\documentclass[preprint, review, 12pt, 3p]{elsarticle}
\usepackage{lscape}
\usepackage{amsmath}
\usepackage[colorlinks=true]{hyperref}
\usepackage{boldline}
\usepackage{dirtytalk}
\usepackage{graphicx}
\usepackage{subfig}
\usepackage{multirow}
\usepackage{lipsum}
\usepackage{booktabs}
\usepackage{multicol}
\usepackage{longtable}
\usepackage{caption}
\usepackage{amsfonts}
\usepackage{makecell}
\usepackage{adjustbox}
\usepackage{tabularx}
\usepackage{floatrow} 
\usepackage[vlines]{tabularht}
\usepackage{pbox}
\usepackage{pdflscape}
\usepackage{mathtools, cuted}
\usepackage{pifont}% http://ctan.org/pkg/pifont
\newcommand{\xmark}{\ding{55}}%

\newcommand*\colourcheck[1]{%
  \expandafter\newcommand\csname #1check\endcsname{\textcolor{#1}{\ding{52}}}%
}
\colourcheck{blue}
\colourcheck{green}
\colourcheck{red}

\newcommand\dunderline[3][-2pt]{{%
  \sbox0{#3}%
  \ooalign{\copy0\cr\rule[\dimexpr#1-#2\relax]{\wd0}{#2}}}}

\DeclareMathOperator{\EX}{\mathbb{E}}% expected value

\usepackage{nomencl}
\makenomenclature

\usepackage{float}
\floatstyle{plaintop}
\restylefloat{table}

\makeatletter
\newcommand{\thickhline}{%
    \noalign {\ifnum 0=`}\fi \hrule height 1pt
    \futurelet \reserved@a \@xhline
}
\newcolumntype{"}{@{\vrule width 1pt}}
\makeatother

\newlength{\Oldarrayrulewidth}

\usepackage[linesnumbered,ruled,vlined]{algorithm2e}
\SetKwInput{KwInput}{Input}                % Set the Input
\SetKwInput{KwOutput}{Output}              % set the Output

\usepackage[table,xcdraw]{xcolor}

\newsavebox\MBox

\usepackage{amsmath}
\DeclareMathOperator*{\argmax}{argmin}

\DeclareMathOperator*{\argmaxv}{min}
\usepackage{tikz}
\biboptions{square,sort&compress}

\journal{Medical Image Analysis}

\begin{document}

\begin{frontmatter}

\title{Multi-scale, Data-driven and Anatomically Constrained Deep Learning Image Registration for Adult and Fetal Echocardiography}

\author{Md. Kamrul Hasan\fnref{label2}}
\ead{k.hasan22@imperial.ac.uk OR kamruleeekuet@gmail.com}

\author[label2]{Haobo Zhu}
\ead{haobo.zhu18@imperial.ac.uk}

\author{Guang Yang\fnref{label3,label4,label5,label6}}
\ead{g.yang@imperial.ac.uk}

\author[label2]{Choon Hwai Yap\corref{cor1}}
\ead{c.yap@imperial.ac.uk}

\address[label2]{Department of Bioengineering, Imperial College London, London SW7 2AZ, UK}

\address[label3]{Bioengineering Department and Imperial-X, Imperial College London, London W12 7SL, UK}

\address[label4]{National Heart and Lung Institute, Imperial College London, London SW7 2AZ, UK}

\address[label5]{Cardiovascular Research Centre, Royal Brompton Hospital, London SW3 6NP, UK}

\address[label6]{School of Biomedical Engineering \& Imaging Sciences, King's College London, London WC2R 2LS, UK}

\cortext[cor1]{Corresponding author}

\newpage

\begin{abstract}
Temporal image registration of echocardiography images is a basic building block that can be used for important clinical quantifications such as cardiac motion estimation, myocardial strain measurements, and stroke volume quantifications. Deep learning image registration (DLIR) is a promising approach as it can be consistently accurate and precise, requiring substantially lower computational time and showing promising results in past implementations. We propose that a greater focus on the warped moving image's anatomic plausibility and image quality can support robust DLIR performance. Further, past implementations have focused on adult echocardiography, and there is an absence of DLIR implementations for fetal echocardiography. We propose a framework combining three strategies for DLIR for both fetal and adult echo: (1) an anatomic shape-encoded loss to preserve physiological myocardial and left ventricular anatomical topologies in warped images; (2) a data-driven loss that is trained adversarially to preserve good image texture features in warped images; and (3) a multi-scale training scheme of a data-driven and anatomically constrained algorithm to improve accuracy. Our experiments show that the shape-encoded loss and the data-driven adversarial loss are strongly correlated to good anatomical topology and image textures, respectively. They improve different aspects of registration performance in a non-overlapping way, justifying their combination. We show that these strategies can provide excellent registration results in both adult and fetal echocardiography using the publicly available CAMUS adult echo dataset and our private multi-demographic fetal echo dataset, despite fundamental distinctions between adult and fetal echo images. Our approach also outperforms traditional non-DL gold standard registration approaches, including Optical Flow and Elastix. Registration improvements could also be translated to more accurate and precise clinical quantification of cardiac ejection fraction, demonstrating a potential for translation. The source code of our data-driven and anatomically constrained DLIR methods will be made publicly available at \url{https://github.com/kamruleee51/DdC-AC-DLIR}.

\end{abstract}


\end{frontmatter}

\tableofcontents
% \listoffigures
% \listoftables

\section{Introduction}
\label{Introduction}
\subsection{Background and Motivation}
Accurate estimations of cardiac motion from clinical echocardiography images are essential and can aid in the evaluation of cardiac function \citep{claus2015tissue} and the detection of dysfunction in conditions such as cardiomyopathy \citep{popovic2008association}, chemotherapy toxicity \citep{balter2007imaging}, and infraction \citep{seo2013computing}. Motion estimation is also a good way to obtain myocardial strains with accuracy on par with speckle tracking \citep{heyde2012elastic}, and it can be used to estimate stroke volume and ejection fraction (EF) accurately. The extracted motion information is also important for biomechanical simulations of cardiac function to elucidate intracardiac flow and myocardial (MYO) stresses \citep{wong2023fluid, ong2021biomechanics}. However, echo measurements, especially fetal echo, have precision issues due to subjective manual inputs, inter-observer variability, and vendor-specific ultrasound machines and processing algorithms. For instance, clinical measurements of fetal left ventricular (LV) strains can vary more than one-fold between reputable groups \citep{van2020fetal}. It is thus essential to improve cardiac motion estimation algorithms that could be achieved by temporal echo image registration \citep{wiputra2020cardiac}, and machine learning is an ideal way to do so.

Cardiac motion estimation is typically achieved via non-rigid pair-wise temporal echo image registration, coupled with regularizations. Such pair-wise image registration determines the spatial transformation that will enable optimally aligned pixel- or voxel-wise correspondence between two images from different time points, thereby retrieving the deformations of the heart between these two time points. Methodologies previously established for registration included free-form deformation \citep{de2012temporal}, demons \citep{thirion1998image}, and optical flow \citep{beauchemin1995computation}, and regularizations such as cyclic cardiac motion and spatial consistency have previously been proposed \citep{wiputra2020cardiac}. However, these traditional approaches take substantial computational time, especially with high dimensionality and extensive regularizations, which can be improved with deep learning (DL)-based approaches.

Recently, DL-based image registration (DLIR) showed the ability to perform accurate registration that effectively avoids ill-conditioned and highly non-convex cost functions \citep{balakrishnan2019voxelmorph, chen2021deep}. DLIR can be supervised or unsupervised. For example, \citet{dosovitskiy2015flownet, rohe2017svf, sokooti2017nonrigid, yang2017quicksilver} developed supervised DLIR, which are easy to train and have lower computational costs. However, obtaining the ground-truth motion field for training is often challenging, and non-DL approaches to seeking ground-truths are often imperfect. For this reason, investigators have attempted synthetic images and ground truths, for example, \citet{ostvik2021myocardial}. For echocardiography images, however, despite advances in generating synthetic images, they may still have differences from natural clinical images. For this reason, unsupervised approaches are of interest, for example, those by \citet{mansilla2020learning, balakrishnan2019voxelmorph, ali2019conv2warp, hu2018label, xu2019deepatlas}, which use constraints enforcing a similarity between the moved and fixed images. Many of such works are based on the VoxelMorph algorithm \citep{balakrishnan2019voxelmorph}, which is successful due to additional considerations of label shape matching between fixed and warped moving images, demonstrating that a focus on the physical plausibility of the moving image after deformation is important to DLIR.

In echocardiography, a few strategies have been proposed for DLIR, including supervised training based on an optical flow estimator \citep{ostvik2021myocardial}, unsupervised patch-based MLP and transformer networks \citep{wang2022unsupervised}, and unsupervised networks utilizing advanced multi-scale correlation \citep{fan2022searchmorph}. The strategy of improving performance by imposing physiological and physical plausibility on the warped moving image, such as employed by VoxelMorph, has not been fully explored. Further, VoxelMorph imposes an overall dice loss, thus imposing pixel-based local constraints, and has not utilized global latent variables for constraints. We hypothesize that greater emphasis on the quality of the warped image can improve DLIR performance. We have investigated this here.

Further, the majority of current DLIR work has used adult echocardiography datasets. Cardiac motion estimation and myocardial strain measurements are equally important for fetal cardiology, which is important to detect abnormalities and guide management and treatment \citep{vasciaveo2023fetal}. Fetal echo images are substantially different from postnatal echo images. The fetal heart is smaller, and transducers are separated further from the heart by maternal tissues, leading to more challenging image quality issues. Consequentially, fetal echo measurements have serious issues with precision \citep{van2020fetal}. It is unclear if our DLIR approaches work well for fetal images. Thus, we investigate both fetal and adult echo images here.

Here, we propose a new framework for echocardiography DLIR focused on the physical and physiological plausibility of warped moving images. Our framework involves enforcing the anatomic topology of the heart in warped images, as well as warped image quality, using a multi-scale training approach. We show that a focus on the warped image quality is sufficient for robust performance compared to current DLIR reports and can be successful for both adult and fetal echo.

\subsection{Related Works}
\label{Related_DLIR_Works}
\subsubsection{Non-Echocardiographic DLIRs}
Several past DLIR studies estimate the spatial transformation parameters under the supervision of the ground-truth deformation vector fields \citep{chen2021deep}. Conventional registration techniques \citep{yang2016fast, cao2017deformable, fan2019birnet} and simulated images with known ground-truth fields \citep{rohe2017svf, sokooti2017nonrigid, eppenhof2018pulmonary} are two common methods for obtaining such ground truths. However, the use of conventional registration as ground truth may imply that the DLIR will have accuracy limited by that of the conventional approach, while synthetic images may have differences from natural clinical images.

Unsupervised DLIRs require additional constraints or regularizations for network training \citep{balakrishnan2019voxelmorph, mansilla2020learning, lian2022cocyclereg, kim2021cyclemorph, yan2018adversarial, he2022deformable, hu2018label, dey2021generative}.
\citet{balakrishnan2019voxelmorph} developed VoxelMorph, which uses an unsupervised DLIR regularized by a segmentation-aware auxiliary loss (dice score) to enforce anatomical plausibility in the deformed images. The authors experimentally demonstrated that this improves registration results for brain MR images compared to conventional methods. Similar auxiliary loss concepts were utilized in \citep{hu2018label} for T2-weighted transrectal MR and ultrasound scans.
VoxelMorph's idea was further extended by \citet{he2022deformable} and \citet{mansilla2020learning}. \citet{he2022deformable} proposed a DLIR method based on the attention-guided fusion of multi-scale deformation fields by having a separate segmentation network for regions of interest to distinguish them from the uninterested areas and successfully tested it on chest X-ray images. \citet{mansilla2020learning} added a new loss function to VoxelMorph's segmentation-aware loss to enforce global non-linear representations of image anatomy from the segmentation mask, which was also tested on chest X-ray images. \citet{xu2019deepatlas} proposed DeepAtlas to perform both weakly supervised registration and semi-supervised segmentation for knee and brain 3D MR images.

Further, \citet{kim2021cyclemorph} and \citet{lian2022cocyclereg} combined Cycle-GAN \citep{zhu2017unpaired} and VoxelMorph \citep{balakrishnan2019voxelmorph} for image registration, using bidirectional deformation to retain cycle consistency and implicit regularization to improve performance, validating the approach with 2D facial expression images, 3D brain MR images, and multi-phase liver CT images. Adversarial constraints were also utilized by \citet{yan2018adversarial}, \citet{dey2021generative}, and \citet{mahapatra2018deformable} for MR and transrectal ultrasound scan (TRUS) image fusion, brain MR images, and retinal and cardiac MR images, respectively. Here, an adversarially-trained network was used to classify between the moved and fixed images \citep{goodfellow2020generative}. Such adversarial training is based on image intensities and their spatial properties, which allows for improved image texture in deformed intensity images. \citet{czolbe2023semantic} developed an unsupervised DLIR network for 2D and 3D MR images of different modalities, using dataset-specific learned spatial invariance features rather than raw intensity images to estimate the loss value. They further showed that deep semantic features can provide DLIR improvements.

\subsubsection{DLIRs for Echocardiography}
A number of supervised and unsupervised DLIRs have been applied to echocardiography \citep{ostvik2021myocardial, wang2022unsupervised, haukom2019basal, wei2020temporal, fan2022searchmorph}.
\citet{ostvik2021myocardial} proposed a supervised PWC-Net \citep{sun2018pwc}-based framework for motion estimation, which is based on a feature pyramid extractor and an optical flow estimator, and showed that the resulting myocardial strains compared well with truths and current commercial machine quantifications.
\citet{wang2022unsupervised} proposed an unsupervised registration method employing a patch-based MLP and transformer network. The patch-based learning features from moving and fixed images were fitted to the cross-feature block for block matching. MLP and/or Swin Transformer were utilized as cross-feature blocks.
\citet{wei2020temporal} proposed co-learning of registration and segmentation using a 3D UNet structure and showed that registration and segmentation benefited from each other.
\citet{fan2022searchmorph} presented an unsupervised multi-scale correlation iterative registration network (SearchMorph) with a correlation layer to improve feature relevance and a correlation pyramid to offer multi-scale relevance information. They additionally developed a deformation field iterator to let the search module register details and large deformations through the model.

\subsection{Our Contributions}
Existing DLIRs applied to echocardiography typically focus on a match between the warped moving image and the fixed image and do not focus on constraints involving the quality of the warped moving images or the anatomic plausibility of the heart in the warped image. Anatomic constraints are considered by Voxelmorph, but the constraints are pixel-wise and localized instead of relying on global latent variables. For these reasons, we propose a DLIR framework with a more robust preservation of the anatomic topology of the heart and the quality of the image after registration warping. Further, to date, DLIR has not been applied to fetal echo images. We thus test our framework on both adult and fetal echo. Our specific contributions in this article are:

\begin{itemize}
    \item We propose a DLIR framework that adds on three strategies to the vanilla DLIR to ensure robust DLIR performance: (1) an anatomic shape-encoded loss to preserve physiological myocardial and cardiac anatomical topologies after warping; (2) a data-driven loss that is trained adversarially to maintain good image texture after warping and to preserve a high degree of perceptual similarity between the fixed and warped intensity images; and (3) a multi-scale training scheme of data-driven and anatomically constrained algorithm to improve accuracy.

    \item We demonstrate that there are non-overlapping benefits in concurrently using the shape-encoded loss and the data-driven adversarial loss, as the former is correlated and likely responsible for improved anatomical topology, while the latter is correlated to and likely responsible for better image textures in warped moving images.
    
    \item We demonstrate that our approach has robust performance in both adult and fetal echocardiography, despite essential differences between the two.
    
    \item We demonstrate that our proposed framework offers robust performance in comparison to other DLIR methods and can support accurate clinical measurements such as ejection fraction.
    
\end{itemize}

The remainder of the article is prepared as follows: Section~\ref{Proposed_DLIR} details the proposed DLIR incorporating different regularizations, including their implementations. The experimental setup, training protocols, evaluation criteria, and datasets are described in Section~\ref{Experimental_setup_and_Datasets}. Section~\ref{Results} describes the experimental results with critical observations and a complete ablation study, demonstrating the advantages of different integral components and their non-overlapping benefits. The use of registration for estimating ejection fraction has been shown in Section~\ref{Estimating_EF}. Finally, Section~\ref{Critical_observation} concludes the article.

\section{Methods}
\label{Proposed_DLIR}
\begin{figure*}[!ht]
\caption{The proposed DLIR model combines different loss functions ($\mathcal{L}$), such as unsupervised intensity image-based loss ($\mathcal{L}_{us}$) (in block A), anatomical shape-, size-, and orientation-based loss ($\mathcal{L}_{gac}$) (in block B), and data-driven adversarial loss ($\mathcal{L}_{DdC}$) (in block C) through a discriminator. The sum of those losses teaches and constrains the deformation field to minimize the differences between fixed and moving images and/or masks to generate more anatomically plausible and realistic intensity and mask images after the deformation.}
\label{fig:RegistrationBlock}
\centering
\includegraphics[width=1.0\textwidth]{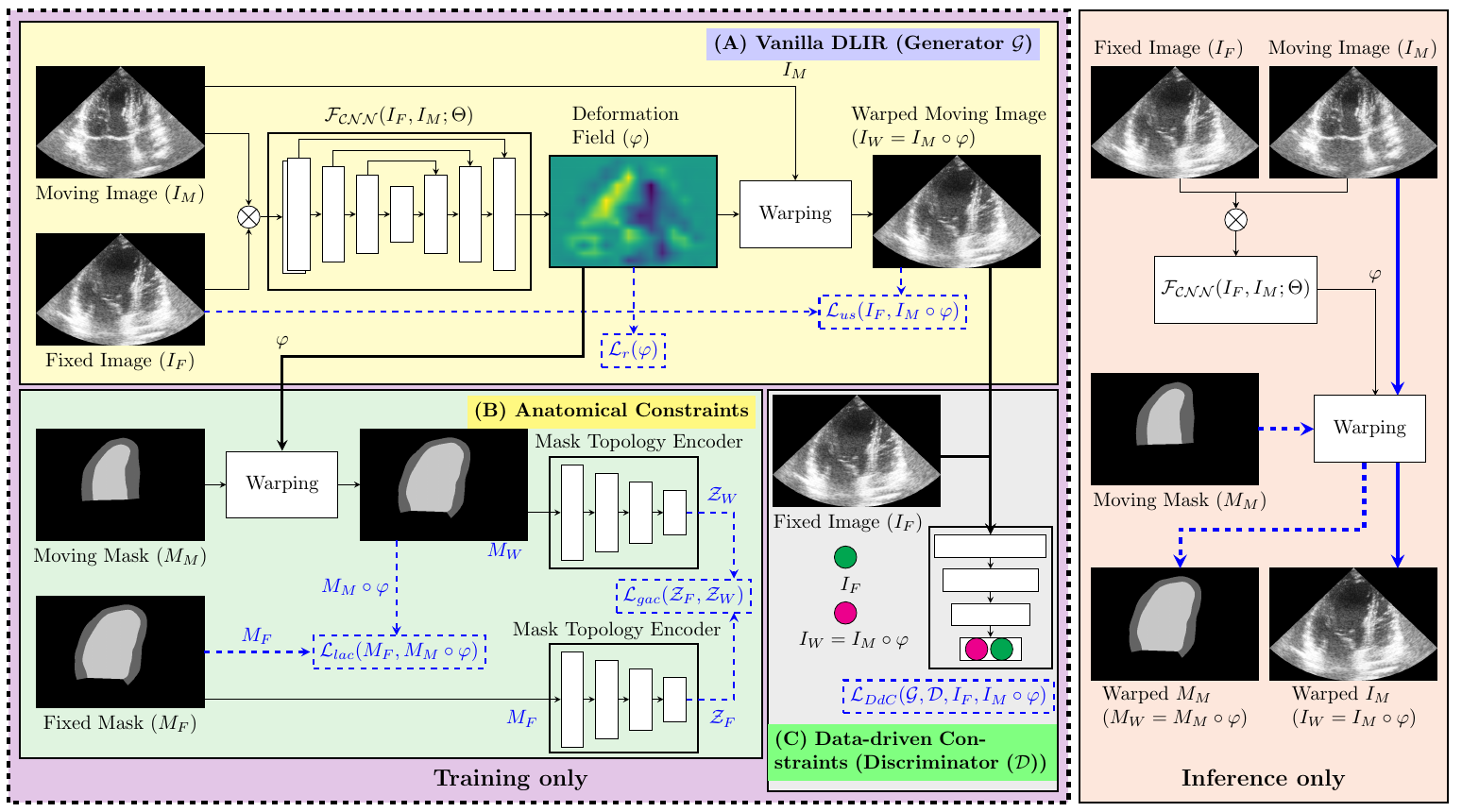}
\end{figure*}

Fig.~\ref{fig:RegistrationBlock} illustrates the proposed DLIR pipeline. The training module consists of the Vanilla DLIR (VanDLIR) (block A in Fig.~\ref{fig:RegistrationBlock}) enhanced with three specific strategies: (1) anatomical constraints (block B in Fig.~\ref{fig:RegistrationBlock}), (2) data-driven constraints (block C in Fig.~\ref{fig:RegistrationBlock}), and (3) multi-scale training.
In contrast, the inference module (Fig.~\ref{fig:RegistrationBlock} right panel) requires only a pair of intensity images to predict a deformation field (DDF), which can be used to warp the intensity and mask moving images to match the fixed images.

\subsection{Basic Framework}
\label{V_DLIR}
The basic form of a DLIR is the unsupervised VanDLIR network, which consists of a CNN-based deformation network ($\mathcal{F_{CNN}}$), containing six layers with (128, 256, 512, 1024, 2048, and 4096) channels in each layer, and a differentiable warping module. $\mathcal{F_{CNN}}$ predicts a deformation field (DDF), $u(x) = \mathcal{F_{CNN}}(I_F, I_M;\Theta)$, where $\Theta$ is a $\mathcal{F_{CNN}}$'s training parameters and $u(x): \mathcal{R}^N \mapsto \mathcal{R}^N$, where $N$ is an image dimension. The warping transformation ($\varphi$) is thus defined as $\varphi(x) = x + u$.

\begin{figure*}[!ht]
\centering
\includegraphics[width=0.8\textwidth]{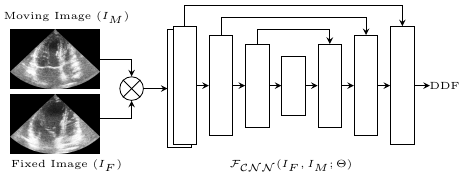}
\caption{The $\mathcal{F_{CNN}}$ network structure for generating a DDF from the inputted intensity image pair ($I_F$ and $I_M$), which follows that UNet \citep{ronneberger2015u} and DeepReg \citep{fu2020deepreg} structures.}
\label{fig:dcnnBlock}
\end{figure*}

For generating DDF from the inputted intensity image pair ($I_F$ and $I_M$), we used MONAI's \citep{cardoso2022monai} RegUNet\footnote{\url{https://docs.monai.io/en/stable/networks.html\#regunet}\label{RegUNet}}, as illustrated in Fig.~\ref{fig:dcnnBlock}, which exploits the benefits of the skip connections from the UNet \citep{ronneberger2015u} to recover the spatial information lost in the encoder pooling.
$\mathcal{F_{CNN}}$ processes the two intensity images, and generates a 2-channel filter map representing the 2D deformation field (3 channels for 3D images), and has a random initialization. For warping the moving image and mask, we used MONAI's \citep{cardoso2022monai} Warp\footnote{\url{https://docs.monai.io/en/stable/networks.html\#warp}} block ($I_W=I_M\circ \varphi$ and $M_W=M_M\circ \varphi$).

The registration between fixed ($I_F$) and moving ($I_M$) images is formulated as an optimization problem, as in Eq.~\ref{main_eqn}, which uses pixel displacement fields to represent the spatial transformation, 
% \begin{strip}
\begin{equation}
\label{main_eqn}
\begin{aligned}
        \widehat{\Theta} = \argmax\limits_{\Theta}\Big\{ \mathcal{L}_{us} (I_F, I_M \circ \varphi) + \lambda_r \times \mathcal{L}_r(\varphi) + \lambda_{lac} \times \mathcal{L}_{lac} (M_F, M_M \circ \varphi) \\ + \lambda_{gac} \times \mathcal{L}_{gac} (M_F, M_M \circ \varphi) + \lambda_{DdC} \times \mathcal{L}_{DdC} (\mathcal{G}, \mathcal{D}, I_F, I_M \circ \varphi) \Big\}, 
\end{aligned}
\end{equation}
% \end{strip}
\noindent where $\mathcal{L}_{us}$ is the mutual information (MI) (dis)similarity measure between $I_F$ and $I_W (= I_M \circ \varphi$), defined as in Eq.~\ref{similarity}.
\begin{equation}
\label{similarity}
        \mathcal{L}_{us} (I_F, I_W=I_M \circ \varphi) = -\sum_{i_f,i_w}p(i_f,i_w) \log \frac{p(i_f,i_w)}{p(i_f)p(i_w)},
\end{equation}
\noindent where $p(i_f)$ represents the probability that a pixel (or voxel) in image $I_F$ has the intensity of $i_f$, $p(i_w)$ represents the same probability for image $I_W$, and $p(i_f,i_w)$ represents the joint probability distribution of the intensities of two images, $I_F$ and $I_W$. The regularization loss (a bending energy loss \citep{fu2020deepreg}), $\mathcal{L}_r$ on $u(x)$ is a smooth regularization on the warping transformations ($\varphi$), and $\lambda_r\geq 0$ is the smoothness weight.

$\mathcal{L}_{lac}$ is the local anatomic similarity constrain, $\mathcal{L}_{gac}$ is the global anatomic similarity constraint, and $\mathcal{L}_{DdC}$ is the adversarial data-driven image similarity constrain, while $\lambda_{lac}$, $\lambda_{gac}$, and $\lambda_{DdC}$ are their respective weights ($\geq 0$). These loss terms are explained below.

\subsection{Anatomic Constraints (AC)}
\label{AC_DLIR}
Several previous studies \citet{balakrishnan2019voxelmorph, czolbe2023semantic, mansilla2020learning, hu2018label, xu2019deepatlas, he2022deformable} show that the VanDLIR often fails to align anatomical correspondences between $I_F$ and $I_W$, and produces anatomically non-plausible shape, size, and orientation of the moved masks. For this reason, we inserted anatomic constraints during training (block B in Fig.~\ref{fig:RegistrationBlock}). First, we modeled a local anatomic similarity loss as proposed in VoxelMorph by \citet{balakrishnan2019voxelmorph},
\begin{equation}
\label{reg_form_v3}
\begin{aligned}
\mathcal{L}_{lac} (M_F, M_M \circ \varphi) =  \frac{2}{K} \sum_{\mathcal{C}=1}^{K}\frac{|M^{\mathcal{C}}_F \cap (M^{\mathcal{C}}_M \circ \varphi)|}{|M^{\mathcal{C}}_F| + |M^{\mathcal{C}}_M \circ \varphi|},
\end{aligned}
\end{equation}
\noindent where $\mathcal{C}\forall K$ is a number of anatomical regions. Here, K=2, as we modeled the LV chamber and myocardium regions. However, this is unlikely to ensure that the warped masks will have sufficient anatomical accuracy. As such, we further applied a variational autoencoder (VAE) \citep{kingma2019introduction}, $\mathcal{EN}$, to encode the anatomic topology. This shape encoder is trained to convert the input LV chamber and myocardium mask ($I$) into latent vectors and to decode them to reconstruct the same mask ($I'$) closely. The lower-dimensional latent representation ($\mathcal{Z}$) retains significant global information of shape, size, and location. A global anatomic similarity loss, $0\leq \mathcal{L}_{gac}\leq 1$, is utilized to enforce an anatomic similarity between $I_F$ and $I_W$,
\begin{equation}
\label{reg_form_v4}
\begin{aligned}
\mathcal{L}_{gac} (M_F, M_M \circ \varphi) = \|\mathcal{Z}_F-\mathcal{Z}_{W=M \circ \varphi}\|_{2}^2,
\end{aligned}
\end{equation}

To train the VAE network (illustrated in~\ref{Topological_Shape_Encoder} (Table~\ref{tab:Shape_Encoder})), we propose using a hybrid loss function, a combination of DSC and structural similarity index (SSIM) \citep{zhao2019correlation}, rather than the traditional distribution-based cross-entropy loss function to estimate the reconstruction error. The segmentation-aware DSC measures the pixel-level similarity values between $I$ and $I'$, whereas SSIM analyzes the structural differences between them using a $N\times N$ pixel window. This compound loss aims to produce a better-reconstructed mask. The overall loss function ($\mathcal{L}_{VAE}$) is defined as, 
% \begin{strip}
\begin{equation}
\label{VAE_loss}
\begin{aligned}
\mathcal{L}_{VAE} = \frac{2}{K} \sum_{\mathcal{C}=1}^{K}\frac{|I^{\mathcal{C}} \cap I'^{\mathcal{C}}|}{|I^{\mathcal{C}}| + |I'^{\mathcal{C}}|} + \Bigg( \frac{2\mu_I\mu_{I'}+C_1}{\mu_I^2+\mu_{I'}^2+C_1} \Bigg)\Bigg( \frac{2\sigma_{II'}+C_2}{\sigma_I^2+\sigma_{I'}^2+C_2}\Bigg)-\sum_{i\in N}p(I_i)\log \Bigg(\frac{p(I_i)}{q(I'_i)}\Bigg),
\end{aligned}
\end{equation}
% \end{strip}

\noindent where $\mathcal{C}\forall K$ is a number of anatomical organs (which is 2 in our experiments, for the LV chamber and myocardium). $C_1$ and $C_2$ are empirically selected scalar parameters; $\mu_I$ and $\mu_{I'}$ are the mean values of a neighborhood around $I$ and $I'$ with the individual variances of $\sigma_I^2$ and $\sigma_{I'}^2$, respectively; and $\sigma_{II'}^2$ is their covariance (see details in \citep{zhao2019correlation}).
The third term in Eq.~\ref{VAE_loss} is the Kullback-Leibler divergence loss \citep{doersch2016tutorial}, which forces the latent vector, $\mathcal{Z}$, to approximate the normal distribution ($\mathcal{N}(0,I)$). $q(I')$ is the reconstructed distribution of the mask, while $p(I)$ is the actual distribution of the mask. In our DLIR framework, we use the encoder of our pre-trained VAE encoder to extract the latent vector ($\mathcal{Z} \in \mathcal{R}^d$) for a given mask ($\mathcal{Z} = \mathcal{EN} (I)$), which are then used in Eq.~\ref{reg_form_v4} as topological shape features.

\subsection{Adversarial Data-driven Constraint (DdC)}
\label{AAAC_DLIR}
In addition, a data-driven constraint was imposed. Similarity constraints imposed by $\mathcal{L}_{US}$ enforce local, pixel-wise similarity and tend to work well with well-aligned images with highly correlated intensities \citep{czolbe2023semantic}, but not with low contrast, noisy images with ambiguous matches, which are typical of echo images. They further lack neighborhood context, such as texture attributes involving neighboring pixels' correlation and consistency. Enforcing texture similarity is known to give superior network performance and preserve the perceptual similarity of images \citep{yan2018adversarial, zhang2018unreasonable, hou2017deep}. To enforce texture similarity, we incorporate a discriminator ($\mathcal{D}$) in block C of Fig.~\ref{fig:RegistrationBlock}, using a classifier of the warped image output from the VanDLIR block. This data-driven loss function is,
\begin{equation}
\label{discriminator_loss}
\begin{aligned}
\mathcal{L}_{DdC} = \Big<\mathcal{D}(I_F),\mathcal{D}(I_M\circ\varphi)\Big> \\
=\EX_{I_F}[\log \mathcal{D}(I_F)] + \EX_{I_W=I_M\circ \varphi}[\log (1-\mathcal{D(G}(I_M)))]
\end{aligned}
\end{equation}
The loss function is enforced with adversarial learning \citep{goodfellow2020generative} by considering the VanDLIR or AC-DLIR as a generator $\mathcal{G}$, and is trained concurrently with VanDLIR or AC-DLIR. The training procedure for $\mathcal{G}$ maximizes the probability of $\mathcal{D}$ making a mistake in such a way that $\mathcal{D}$'s gain is $\mathcal{G}$'s loss, and vice versa. As the training goes on, both $\mathcal{G}$ and $\mathcal{D}$ are iteratively updated. The feedback from $\mathcal{D}$ via data-driven loss $\mathcal{L}_{DdC}$ will be used to improve $\mathcal{G}$ so that eventually, $\mathcal{G}$ will be well-trained to generate $I_W$ close to $I_F$.

The discriminator ($\mathcal{D}$) is an image classifier that classifies images into fixed or moved echo images. It’s architecture is demonstrated in ~\ref{Topological_Shape_Encoder} (Table~\ref{tab:impl_Discriminator}). The discriminator follows the ResNet structure \citep{he2016deep} but with modifications. Instead of batch normalization in ResNet, we use instance normalization \citep{ulyanov2016instance} followed by the $10\,\%$ dropout layer in each residual block. Instance normalization normalizes across each channel instead of normalizing across input features in a training example. In contrast to batch normalization, the instance normalization layer is also implemented during testing due to the mini-batches independence. Further, we replaced the rectified linear unit (ReLU) activation in ResNet with a leaky ReLU \citep{xu2015empirical} in order to avoid discarding potentially vital information due to negative input values.

\subsection{Multi-Scale Training}
\label{MAAC_DLIR}
Coarse-to-fine multi-scale training has shown effectiveness in many previous studies for flow estimation or image registration \citep{zhu2021test, sokooti2017nonrigid, jiang2020multi, shao2022multi, de2019deep} by giving the network more contextual information and sensitivity of features at varying size scales and avoiding local minima \citep{jiang2020multi,sokooti2017nonrigid}. We implemented this for all networks within the three blocks in Fig.~\ref{fig:RegistrationBlock}. During the training phase, trained parameters ($\Theta$) from the lower scale, i.e., resolutions, are used to initialize the training for the following higher scale.

\subsection{DLIR Configuration Summary}
\label{Configurations}
Our proposed approach is a combination of various strategies and can be summarized in table~\ref{tab:DLIR_configurations}. 
{\renewcommand{\arraystretch}{1.0} 
\begin{table*}[!ht]
\fontsize{8pt}{8pt}\selectfont
\caption{Summary of different model configurations to validate the role of different integral components in our proposed DLIR.}
\label{tab:DLIR_configurations}
\begin{tabular}{lcccc}
\hline
\begin{tabular}[c]{@{}l@{}}Model \\ configurations\end{tabular} & \begin{tabular}[c]{@{}l@{}}Anatomical constraints (AC) \\ (W/O shape encoding)\end{tabular} & \begin{tabular}[c]{@{}l@{}}Anatomical constraints (AC) \\ (W/ shape encoding)\end{tabular} & \begin{tabular}[c]{@{}l@{}} Data-driven\\ constraints (DdC)\end{tabular} & \begin{tabular}[c]{@{}l@{}}Multi-scale \\ learning (MS)\end{tabular} \\ \hline
VanDLIR                                                        &    \xmark                                                                &       \xmark                                                              &               \xmark                                                 &                                        \xmark                         \\
VoxelMorph \citep{balakrishnan2019voxelmorph}                                                    &        \checkmark                                                            &        \xmark                                                             &              \xmark                                                  &                       \xmark                                          \\
AC-DLIR                                                         &        \checkmark                                                            &       \checkmark                                                              &        \xmark                                                        &             \xmark                                                    \\
DdC-DLIR                                                       &   \xmark                                                                 &      \xmark                                                               &                      \checkmark                                          &                   \xmark                                              \\
DdC-AC-DLIR                                                      &    \checkmark                                                                &             \checkmark                                                        &        \checkmark                                                        &                              \xmark                                   \\
MS-DdC-AC-DLIR                                                    &     \checkmark                                                               &            \checkmark                                                         &        \checkmark                                                        &                   \checkmark                                              \\ \hline
\end{tabular}
\end{table*}}

\section{Experimental Setup and Datasets}
\label{Experimental_setup_and_Datasets}

\subsection{Elastix and Optical Flow Control Experiments}
\label{Elastix_and_Optical_low}
Non-deep learning registration approaches are performed for controlled comparisons. The Elastix \citep{klein2009elastix} registration is performed using the Python SimpleITK-Elastix package\footnote{\url{https://simpleelastix.github.io/}} with the settings suggested by \citet{chan2021full}, which was initially initialized by a rigid Euler transformation followed by an affine and b-spline transformation. Advanced mean squares, transform bending energy penalty, and advanced mattes mutual information are the penalty functions. The dark regions outside the conical field of view of the ultrasound images are masked out to allow image deformation beyond these boundaries. The grid was configured to $16\times 16$ for image warping. 

For optical flow (OF) \citep{beauchemin1995computation} implementation, we have used the iterative Lucas-Kanade solver \citep{lucas1981iterative} at each level of the image pyramid. The skimage registration package\footnote{\url{https://scikit-image.org/docs/stable/api/skimage.registration.html}} has been used for this OF implementation.

\subsection{Training Protocol}
\label{Training_Protocol}
All the DLIR networks were implemented using Python in the Pytorch\footnote{\url{https://pytorch.org/get-started/locally/}} \citep{paszke2019pytorch} and MONAI\footnote{\url{https://docs.monai.io/en/stable/apps.html}} frameworks \citep{cardoso2022monai}. The batch size for training DLIRs was eight, for a total of one hundred epochs. All DLIRs were trained to utilize an Adam optimizer without AMSGrad, with a learning rate of 0.0002, betas of 0.9 and 0.999, and a weight decay rate of 0. For multi-scale analysis, all intensity images were resized to $32\times 32$, $64\times 64$, $128\times 128$, $256\times 256$, and $512\times 512$ using bicubic interpolation, whereas these sizes were obtained from the manual MYO and LV masks using nearest neighbor interpolation. The mask labels for the background, MYO, and LV chambers were 0, 1, and 2, and the intensity images were normalized to the range 0–1.
During the validation phase, we set the weighting factors to $\lambda_r =1$, $\lambda_{lac}=2$, $\lambda_{gac}=2$, and $\lambda_{DaC}=0.001$ by trial-and-error. The machine (running on Ubuntu 20.04 LTS) used for the experiments was equipped with 4 Nvidia\textsuperscript{\tiny\textregistered} GeForce RTX\textsuperscript{\tiny\textregistered} 3090Ti PCIe 3.0 GPUs with 24GB GDDR6X memory with Intel\textsuperscript{\tiny\textregistered} Xeon\textsuperscript{\tiny\textregistered} W-2295 x18 @ 3.00GHz CPU and 128 GB of memory.

\subsection{Evaluation Criterion}
\label{Evaluation_Criterion}
The quality of intensity image deformation is measured using mean squared error (MSE) and learned perceptual image patch similarity (LPIPS) \citep{zhang2018unreasonable}, considering $I_F$ and $I_W$ (see Eq.~\ref{eq:metrics}). The mean dice similarity coefficient (DSC) and Hausdorff distance (HD) \citep{aydin2021usage} are used to evaluate the quality of mask image deformation, considering $M_F$ and $M_W$ (see Eq.~\ref{eq:metrics}). 
% \begin{strip}
\begin{equation} 
\label{eq:metrics}
\begin{split}
MSE= \frac{1}{N \times M}\displaystyle\sum_{i=1}^{N} \displaystyle\sum_{j=1}^{M} (I_{Fij}-I_{Wij})^2, \\
DSC= \frac{1}{N}\displaystyle\sum_{i=1}^{N} \frac{2\times |M_{Fi} \cap M_{Wi}|}{|M_{Fi}| + |M_{Wi}|}, \\
HD= \frac{1}{2\times N}\displaystyle\sum_{i=1}^{N} \Big(\frac{1}{M_F}\displaystyle\sum_{x\in M_F} \argmaxv\limits_{y\in M_W} d(x,y) + \frac{1}{M_W}\displaystyle\sum_{y\in M_W} \argmaxv\limits_{x\in M_F} d(x,y)\Big), 
\end{split}
\end{equation}
% \end{strip}
\noindent where $M$, $N$, and $d$ denote the number of pixels, sample numbers, and shortest distance, respectively.
The MSE estimates the local pixel-level similarity, while LPIPS measures the perceptual similarity and computes the similarity between the activations of two image patches for some pre-defined network (VGG \citep{simonyan2014very} in our experiments). A low LPIPS score means that image patches are perceptually similar.
The DSC and HD quantify the amount of similarity between $M_F$ and $M_W$ and edge roughness (or irregularity), respectively.

For the clinical evaluation \citep{leclerc2019deep, xue2022improved, sfakianakis2023gudu}, we use the correlation between the true and predicted EFs from the end-diastolic (ED) and end-systolic (ES) volumes. The CAMUS dataset (explained below) contains true ED and ES volumes and EFs, and an assumption was made presuming that the number of pixels of the LV area correlates with chamber volume in accordance with Simpson’s rule, as previously proposed \citep{sfakianakis2023gudu}. Consequently, using Eq.~\ref{eq:clin_metrics}, the predicted ED- and ES-LV volumes (in ml) and corresponding EF are calculated.
\begin{equation} 
\label{eq:clin_metrics}
\begin{split}
LV_{EDV-pred}= LV_{EDV-true}\times \frac{LV_{EDV-pred-pxls}}{LV_{EDV-true-pxls}}, \\
LV_{ESV-pred}= LV_{ESV-true}\times \frac{LV_{ESV-pred-pxls}}{LV_{ESV-true-pxls}}, \\
LV_{EF-pred} = 1- \frac{LV_{ESV-pred}}{LV_{EDV-pred}},
\end{split}
\end{equation}
where $LV_{EDV-pred-pxls}$ and $LV_{EDV-true-pxls}$ are the predicted pixel numbers, i.e., LV ED predicted and true areas, respectively. Similarly, $LV_{ESV-pred-pxls}$ and $LV_{ESV-true-pxls}$ are for the ESs. $LV_{EDV-true}$ and $LV_{ESV-true}$ are the true ED and ES volumes that are provided in the CAMUS dataset \cite{leclerc2019deep}.

\begin{figure}[!ht]
\centering
\includegraphics[width=0.9\textwidth]{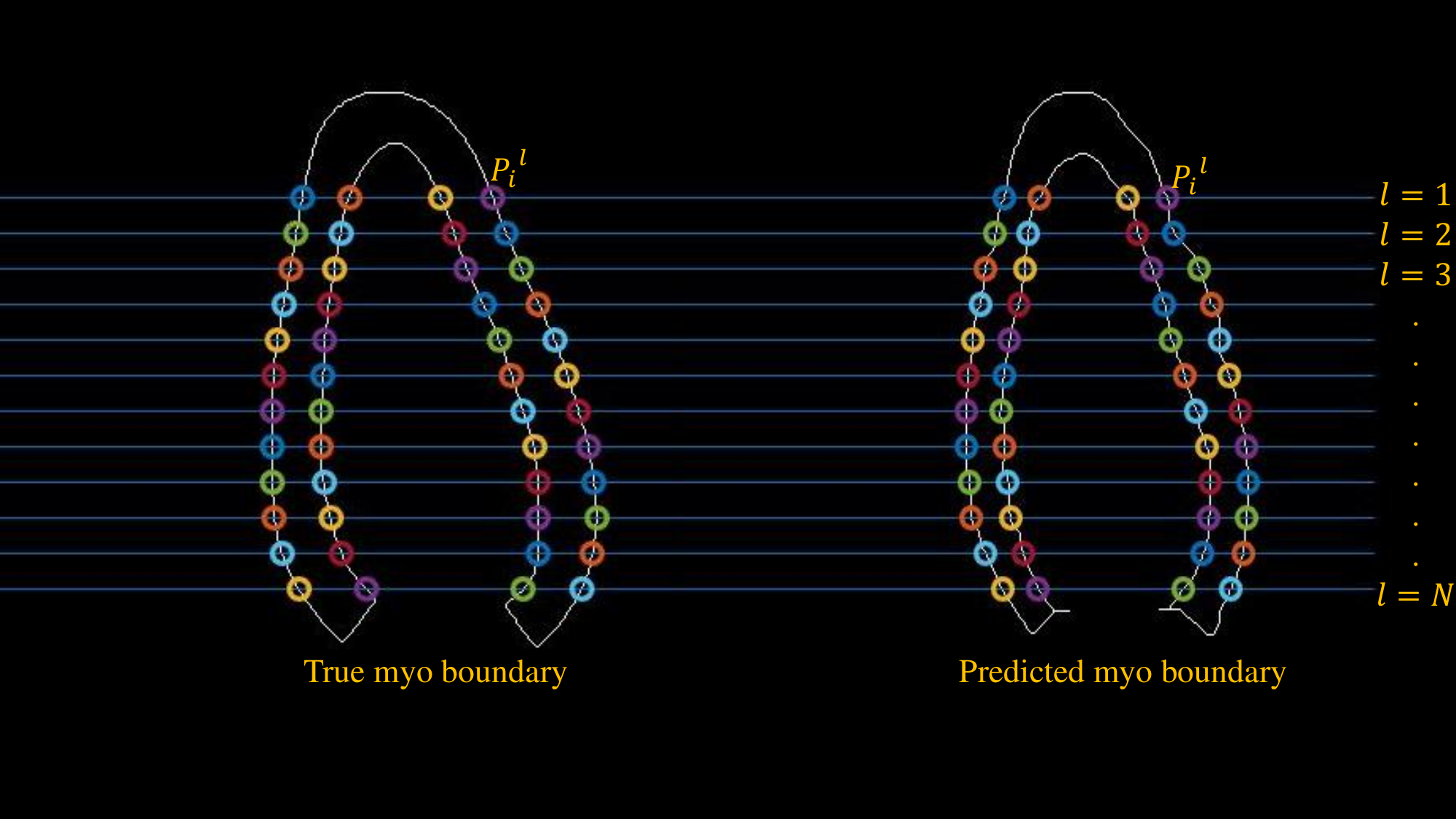}
\caption{The proposed variance estimation of the MYO thickness to measure its thickness uniformity (\textit{TU}) from the true and predicted MYO boundaries.}
\label{fig:Var}
\end{figure}

In addition to DSC and HD, we propose a new metric for assessing the anatomic plausibility of the MYO, which measures the MYO's thickness uniformity (TU) by estimating the thickness variance, as illustrated in Fig.~\ref{fig:Var}.
To calculate TU, we consider N number of horizontal lines ($l$) and determine the intersection points ($P_{i}^l$) between MYO's inner and outer boundaries (see Fig.~\ref{fig:Var}). Each true and predicted MYO boundary returns four ($C=4$) spatial coordinates ($i\forall C$) for each line $l$. Then, the estimated true and predicted TUs are estimated from the N-dimensional distance vector $d_N$ using Eq.~\ref{eq:TU}. 
\begin{equation}
\label{eq:TU}
\begin{split}
d_N= \Bigg[\frac{1}{2} \big(|d_{R}^l(P_{i=1}^l,P_{i=2}^l)| + |d_{L}^l(P_{i=3}^l,P_{i=4}^l)| \big)\Bigg]_{N\times 1},\\
TU= \EX[d_{N}^2]-\EX[d_N]^2
\end{split}
\end{equation}
where $d_{R}^l$ and $d_{L}^l$ are the $l^{th}$ distances between a pair of spatial coordinates (MYO's right- and left-side pairs, i.e., $(P_{1}^l,P_{2}^l)$ and $(P_{3}^l,P_{4}^l)$). The smaller value of $TU$ denotes the less thickness variance in the MYO boundary.

\subsection{Experimental Datasets}
\label{Datasets}
Comprehensive ablation studies are conducted using two different datasets: the publicly available adult echo dataset (CAMUS \citep{leclerc2019deep}) and our private fetal echo dataset. 

\subsubsection{CAMUS Adult Dataset} 
This adult dataset includes 500 patients with an apical two-chamber view (A2C) and an apical four-chamber view (A4C). It was acquired from GE Vivid E95 ultrasound scanners (GE Vingmed Ultrasound, Horten, Norway), with a GE M5S probe, with a pixel resolution of $0.154 m \times 0.154 m$. Further descriptions of the CAMUS dataset can be found in \citep{leclerc2019deep}.

\subsubsection{Private Fetal Dataset} 
\begin{figure*}[!ht]
    \centering
   \includegraphics[width=16.0cm]{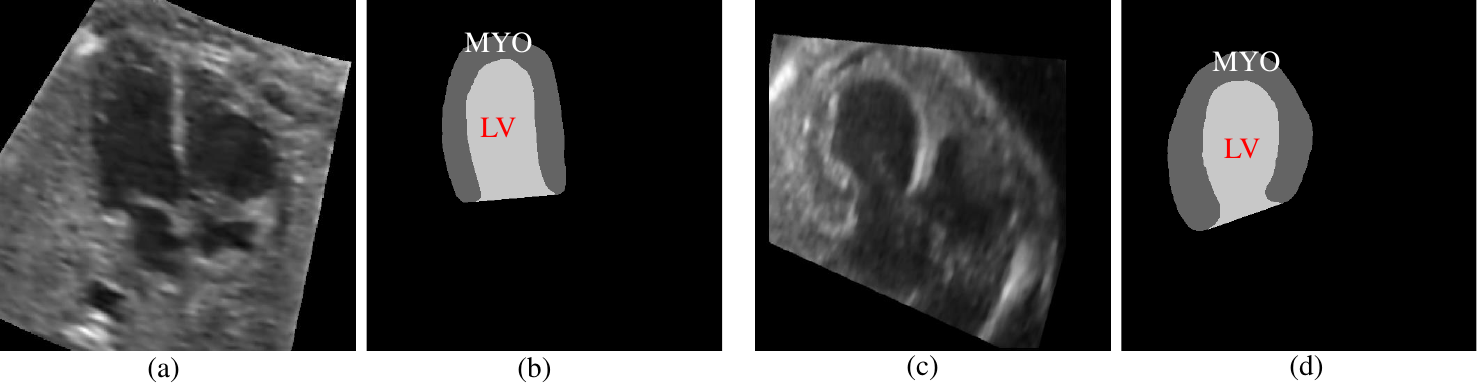} 
    \caption{Examples of manually segmented LV and MYO for our private fetal echo dataset. (a) is for healthy patients with a corresponding mask in (b), and (c) is for patients with the disease with a corresponding mask in (d).}
    \label{fig:example}
\end{figure*}

This fetal dataset is a collection of 4D (3D over time) fetal echocardiography images, consisting of 105 2D echo videos obtained at various planes from the 3D echo videos of 15 patients. Images were acquired with the GE Volusion 730 ultrasound machine with the RAB $4-8 L$ transducer and have an in-plane image resolution of $0.95 \mu m \times 0.90 \mu m$ (see details in \citep{wiputra2020cardiac}). The dataset consists of fifteen healthy cases and nine disease cases (with fetal aortic stenosis), with the training dataset involving thirteen healthy and seven disease cases (5813 2D images and masks) and the validation set containing two of them each (795 2D images and masks). The dataset was labeled manually at the ES and ED time points of each 2D video, followed by a temporal registration to generate the masks of the other time points using a validated cardiac motion estimation algorithm \citep{chan2021full, wiputra2020cardiac}, which imposes a spatial b-spline of temporal Fourier registration over pair-wise Elastix image registration. Fig.~\ref{fig:example} shows two examples of our private fetal echo dataset. 

\section{Experimental Results}
\label{Results}
\begin{figure*}[!ht]
\caption{Example of qualitative results of the deformed intensity ($I_W=I_M\circ \varphi$) and mask ($M_W=M_M\circ \varphi$) images generated by different registration models for CAMUS's A2C (first four columns) and A4C (last four columns) views. The fixed and moving examples are also displayed in the first two rows for comparison.}
\label{fig:qualitativeResults}
\centering
\includegraphics[width=1.0\textwidth]{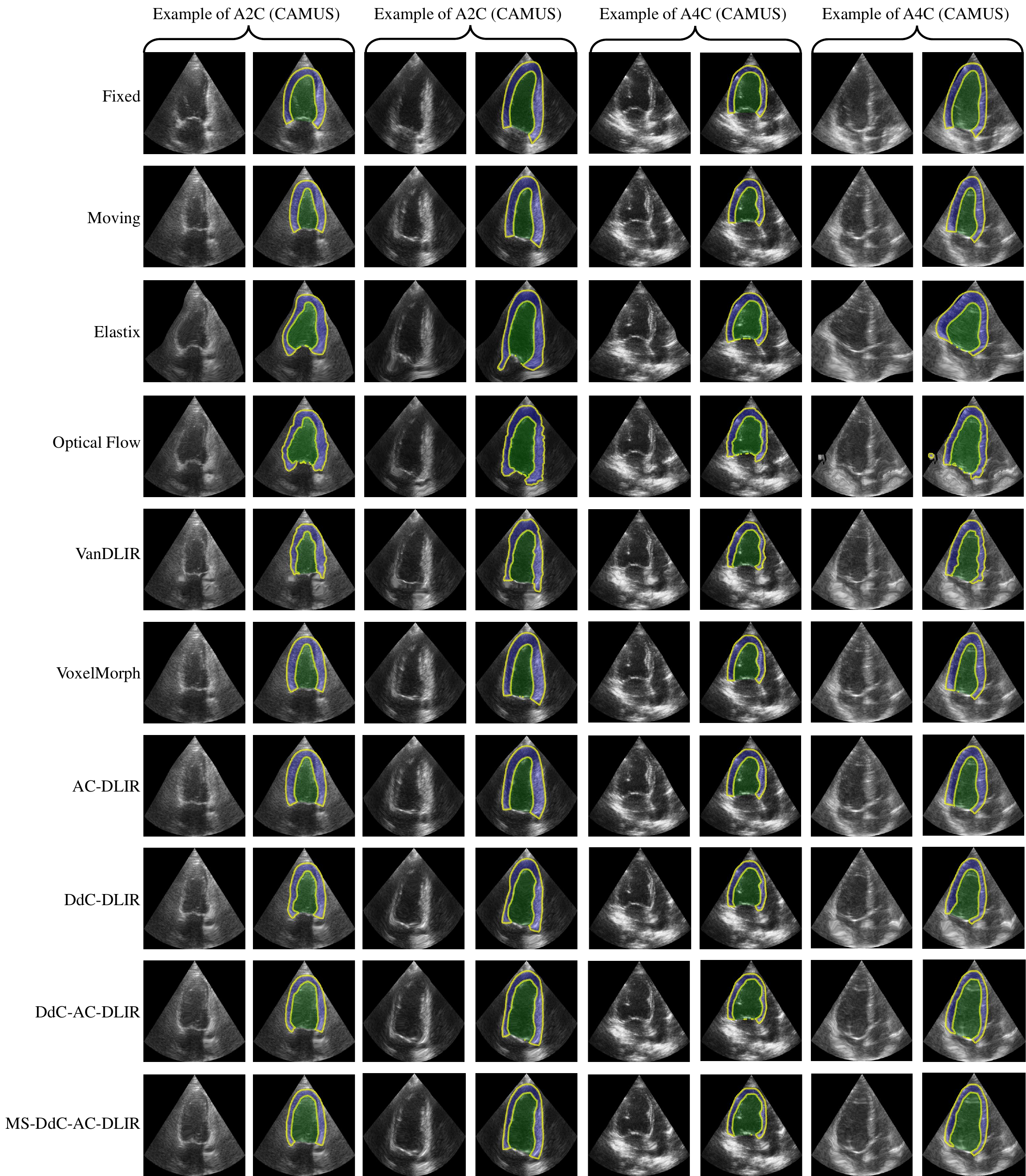}
\end{figure*}

\begin{figure*}[!ht]
\centering
\includegraphics[width=1.0\textwidth]{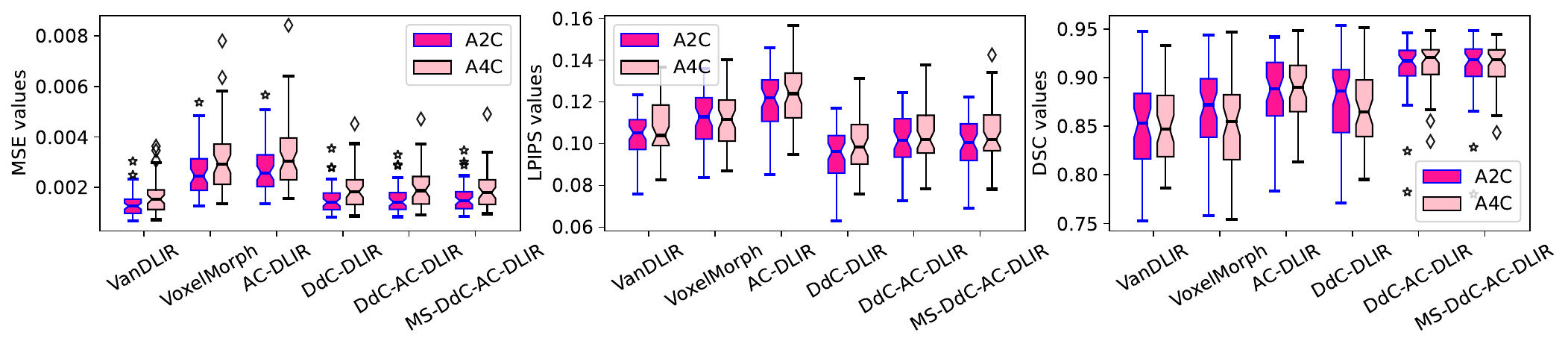}
\caption{Demonstration of the non-overlapping benefits of anatomical and data-driven constraints, as well as the effects of their combination. We use the CAMUS A2C and A4C views to make this figure.}
\label{fig:ANOVA_DSC_LPIPS_MSE}
\end{figure*}

In Sections~\ref{VanDLIR_Results}, \ref{AC_wo_shape}, \ref{DdC_on_VanDLIR}, and \ref{MS_Learning}, we undertake extensive ablation experiments, where we systematically remove specific components to reveal the effects of various components in our proposed DLIR method and to show that our proposed approach gives the best results. The performance of our ablation experiments can be summarized in Table~\ref{tab:Registration_CAMUS_ES_ED}, Fig.~\ref{fig:qualitativeResults}, and Fig.~\ref{fig:ANOVA_DSC_LPIPS_MSE}. These experiments utilized the ED time frame as the fixed ($I_F$ and $M_F$) image and the ES time frame as the moving ($I_M$ and $M_M$) image from the adult CAMUS dataset for both A2C and A4C views. In Section~\ref{Effect_of_Augmentation}, we demonstrate that image augmentation further improves this best-performing DLIR model. Finally, in Section~\ref{Temporal_Registration}, we proceed to test the registration algorithm for a wider range of image pairs between the first time frame and all other time frames and to test the algorithm for both adult echo images and fetal echo images.
{\renewcommand{\arraystretch}{2.3} 
\begin{table*}[!ht]
\fontsize{6pt}{6pt}\selectfont
\caption{Registration results of the CAMUS adult echo dataset from different techniques, employing ES (moving, $I_M$ and $M_M$) and ED (fixed, $I_F$ and $M_F$) images of cardiac A2C and A4C views and demonstrating the non-overlapping benefits of data-driven and anatomical constraints. All the metrics are estimated using fixed images (and masks) and warped moving images (and masks). The DSC and HD are the averages of backgrounds, MYO, and LV metrics, whereas the class-wise metrics of MYO and LV are shown in Table~\ref{tab:Additional_ES_ED} of~\ref{Additional_Results}. Bold fonts denote the best-performing metrics for the A2C echo view, while the best-performing metrics for the A2C view are underlined.}
\label{tab:Registration_CAMUS_ES_ED}
\begin{tabular}{lccccccc}
\hline
\multicolumn{2}{l}{\textbf{Methods}}    & \textbf{DSC ($\uparrow$)} & \textbf{HD $(mm)$ ($\downarrow$)}  & $\mathbf{TU^\ddag}\,(mm)$ & \textbf{MSE ($\downarrow$)} & \textbf{LPIPS ($\downarrow$)} & \textbf{Time ($\downarrow$)} \\ \hline

\multirow{2}{*}{N/A} & A2C  & $0.8024\pm0.0485$  & $5.73\pm1.62$ & $-$ & $0.0026\pm0.0010$ & $-$ & \multirow{2}{*}{$-$} \\
                  & \cellcolor[HTML]{EFEFEF}A4C & \cellcolor[HTML]{EFEFEF}$0.7923 \pm 0.0492$ & \cellcolor[HTML]{EFEFEF}$5.96 \pm 1.64$ & \cellcolor[HTML]{EFEFEF}$-$ & \cellcolor[HTML]{EFEFEF}$0.0031 \pm 0.0014$ & \cellcolor[HTML]{EFEFEF}$-$ &                   \\ \hline
                  
\multirow{2}{*}{Elastix \citep{klein2009elastix}} & A2C & $0.8825 \pm 0.0477$ & $5.23 \pm 2.63$  & $3.83\pm 6.03$ & $0.0033\pm 0.0019$  & $0.2264 \pm 0.0495$  & \multirow{2}{*}{$4270$ ms} \\
                  & \cellcolor[HTML]{EFEFEF}A4C & \cellcolor[HTML]{EFEFEF}$0.8892 \pm 0.0388$ & \cellcolor[HTML]{EFEFEF}$4.75 \pm 2.69$  & \cellcolor[HTML]{EFEFEF}$2.92\pm 3.21$ & \cellcolor[HTML]{EFEFEF}$0.0049 \pm 0.0030$ & \cellcolor[HTML]{EFEFEF}$0.2388\pm 0.0501$ &                   \\ \hline
                  
\multirow{2}{*}{OF \citep{lucas1981iterative} } & A2C & $0.8980 \pm 0.0323$ & $4.48 \pm 1.61$ & $3.14\pm 3.16$ & $0.0590 \pm 0.0145$ & $0.1098\pm 0.0151$ & \multirow{2}{*}{$937$ ms} \\
                  & \cellcolor[HTML]{EFEFEF}A4C & \cellcolor[HTML]{EFEFEF}$0.8995 \pm 0.0358$ & \cellcolor[HTML]{EFEFEF}$4.48 \pm 3.0$ & \cellcolor[HTML]{EFEFEF}$2.75\pm 3.91$ & \cellcolor[HTML]{EFEFEF}$0.0588 \pm 0.0139$ & \cellcolor[HTML]{EFEFEF}$0.1202\pm 0.0314$ &                   \\ \hline
                  
\multirow{2}{*}{VanDLIR} & A2C & $0.8511 \pm 0.0456$  & $5.44  \pm 1.70$ & $2.76 \pm 2.11$ & $\mathbf{0.0013 \pm 0.0005}$ & $0.1049 \pm 0.0113$ & \multirow{2}{*}{85 ms} \\
                  & \cellcolor[HTML]{EFEFEF}A4C & \cellcolor[HTML]{EFEFEF}$0.8484 \pm 0.0421$ & \cellcolor[HTML]{EFEFEF}$5.51 \pm 1.71$ & \cellcolor[HTML]{EFEFEF}$2.17 \pm 1.92$ & \cellcolor[HTML]{EFEFEF}\dunderline{1pt}{$0.0017 \pm 0.0007$} & \cellcolor[HTML]{EFEFEF}$0.1083 \pm 0.0132$&                   \\ \hline
                  
\multirow{2}{*}{VoxelMorph \citep{balakrishnan2019voxelmorph}} & A2C & $0.8698 \pm 0.0408$  & $4.66  \pm 1.36$ & $1.80 \pm 1.90$ & $0.0026 \pm 0.0009$ & $0.1120 \pm 0.0126$ & \multirow{2}{*}{82 ms} \\
                  & \cellcolor[HTML]{EFEFEF}A4C & \cellcolor[HTML]{EFEFEF}$0.8520 \pm 0.0448$ & \cellcolor[HTML]{EFEFEF}$4.71 \pm 1.54$ & \cellcolor[HTML]{EFEFEF}$1.54 \pm 1.58$ & \cellcolor[HTML]{EFEFEF}$0.0031 \pm 0.0014$ & \cellcolor[HTML]{EFEFEF}$0.1115 \pm 0.0135$ &                   \\ \hline
                  
\multirow{2}{*}{AC-DLIR} & A2C & $0.8854 \pm 0.0360$  & $4.13  \pm 1.30$ & $1.85 \pm 2.06$ & $0.0027 \pm 0.0010$ & $0.1204 \pm 0.0141$ & \multirow{2}{*}{91 ms} \\
                  & \cellcolor[HTML]{EFEFEF}A4C & \cellcolor[HTML]{EFEFEF}$0.8867 \pm 0.0353$ & \cellcolor[HTML]{EFEFEF}$3.82 \pm 1.33$ & \cellcolor[HTML]{EFEFEF}$1.64 \pm 1.56$ & \cellcolor[HTML]{EFEFEF}$0.0033 \pm 0.0014$ & \cellcolor[HTML]{EFEFEF}$0.1239 \pm 0.0151$ &                   \\ \hline
                  
\multirow{2}{*}{DdC-DLIR} & A2C & $0.8726 \pm  0.0447$  & $4.78  \pm 1.97$ & $1.90 \pm 1.41$ & $0.0015 \pm 0.0006$ & $0.0951 \pm 0.0126$ & \multirow{2}{*}{92 ms}  \\
                  & \cellcolor[HTML]{EFEFEF}A4C & \cellcolor[HTML]{EFEFEF}$0.8669 \pm 0.0407$ & \cellcolor[HTML]{EFEFEF}$5.35 \pm 1.95$ & \cellcolor[HTML]{EFEFEF}$1.34 \pm 1.28$ & \cellcolor[HTML]{EFEFEF}$0.0019 \pm 0.0008$ & \cellcolor[HTML]{EFEFEF}\dunderline{1pt}{$0.1005 \pm 0.0138$} &                   \\ \hline
                  
\multirow{2}{*}{DdC-AC-DLIR} & A2C & $\mathbf{0.9107 \pm 0.0282}$  & $\mathbf{3.43  \pm 1.32}$ & $1.73 \pm 2.37$ & $0.0016 \pm 0.0005$ & $0.1024 \pm 0.0126$ & \multirow{2}{*}{92 ms} \\ 
                  & \cellcolor[HTML]{EFEFEF}A4C  & \cellcolor[HTML]{EFEFEF}\dunderline{1pt}{$0.9136 \pm 0.0238$} & \cellcolor[HTML]{EFEFEF}$3.99 \pm 1.08$ & \cellcolor[HTML]{EFEFEF}$1.16 \pm 1.13$ & \cellcolor[HTML]{EFEFEF}$0.0020 \pm 0.0008$ & \cellcolor[HTML]{EFEFEF}$0.1046 \pm 0.0139$ &                   \\ \hline
                  
\multirow{2}{*}{MS-DdC-AC-DLIR} & A2C & $0.9105 \pm 0.0293$  & $3.50  \pm 1.32$ & $\mathbf{1.36 \pm 1.49}$ & $0.0016 \pm 0.0006$ & $\mathbf{0.1001 \pm 0.0127}$ & \multirow{2}{*}{93 ms} \\
                  & \cellcolor[HTML]{EFEFEF}A4C & \cellcolor[HTML]{EFEFEF}$0.9126 \pm 0.0216$ & \cellcolor[HTML]{EFEFEF}\dunderline{1pt}{$3.34 \pm 1.20$} & \cellcolor[HTML]{EFEFEF}\dunderline{1pt}{$0.98 \pm 0.98$} & \cellcolor[HTML]{EFEFEF}$0.0019 \pm 0.0008$ & \cellcolor[HTML]{EFEFEF}$0.1054 \pm 0.0139$ &                   \\ \hline
\multicolumn{8}{p{15cm}}{\textbf{$^\ddag$An accurate $TU$ should be as close as the fixed MYO's TU, which is 0.76 mm for adults in A2C and 0.72 mm for adults in A4C views.}} 
\end{tabular}
\end{table*}}

\subsection{Non-Deep Learning Approaches Compared to VanDLIR}
\label{VanDLIR_Results}
Without any registration (N/A), the DSCs between ED and ES time frames are 0.8024 and 0.7923 for CAMUS A2C and A4C views, respectively, while their HD are 5.73 mm and 5.96 mm (Table~\ref{tab:Registration_CAMUS_ES_ED}). This is thus the no-registration control. All registration algorithms we tested have performances above this, with higher DSC and lower HD between the warped moving image and the fixed image. The average $TU$ of the fixed images is 0.76 mm and 0.72 mm for A2C and A4C views, and for some algorithms, the warped moving image has $TU$ close to this baseline. In terms of MSE, the baseline no registration control is 0.0026 and 0.0031 for A2C and A4C views, but not every algorithm improves on this. 

Between the two gold-standard non-deep learning algorithms, Elastix and OF, which underwent tuning, we observe that OF outperformed Elastix in terms of DSC, HD, $TU$, and LPIPS, but not MSE. Better DSC, HD, and $TU$ indicate that OF has a better overall warped myocardium shape, which is also observable in Fig.~\ref{fig:qualitativeResults}. A better LPIPS demonstrates that there is a better perceived visual similarity between the warped moving image and the fixed image. However, OF’s MSE is very high, indicating insufficient similarity in local pixel-level intensity between warped moving and fixed images. On the other hand, Elastix performs very poorly in LPIPS, showing significant perceived differences between warped moving and fixed images. 

Comparing VanDLIR to these gold-standard non-deep learning methods, VanDLIR had poorer results in terms of DSC and HD, demonstrating that its warping produces poorer myocardium and LV chamber shapes with lower anatomic plausibility and uneven topology. The greater $TU$ of VanDLIR than Elastix and OF indicates a more uniform MYO thickness. Its visual perception quality, LPIPS, is also better than Elastix and OF. However, VanDLIR has a very low MSE, indicating an excellent local-pixel intensity match between moving and fixed images. Since the VanDLIR is designed to seek only local intensity matching and does not enforce matching of neighboring pixels or global attributes, it is unsurprising that MSE is over-emphasized and very low at the expense of other performance indicators. 

For all three methods, VanDLIR, Elastix, and OF, the shapes and edges of the myocardium and LV are very irregular and coarse (see Fig.~\ref{fig:qualitativeResults}). Their $TU$ is several times higher than the fixed image $TU$  of 0.76 mm. It is thus important to add anatomic regularization to improve anatomic plausibility and topology.

\subsection{Benefits of Anatomic Constraints}
\label{AC_wo_shape}
Our results show that the addition of shape constraints indeed improved registration performance (see Table~\ref{tab:Registration_CAMUS_ES_ED}). First, we investigate adding an auxiliary local loss function $\mathcal{L}_{lac}$ (Eq.~\ref{reg_form_v3}) to VanDLIR, as proposed in VoxelMorph \citep{balakrishnan2019voxelmorph}, and find that this improves DSC, HD, and $TU$ by $2.2\,\%$, $14.3\,\%$, and $34.8\,\%$ for the A2C view and $0.4\,\%$, $14.7\,\%$, and $29.0\,\%$ for the A4C view, respectively, all of which are statistically significant ($p\ll 0.001$). 
In Fig.~\ref{fig:qualitativeResults}, the improved thickness uniformity and anatomy overlapping between fixed and warped masks can be seen in comparison with VanDLIR. However, LPIPS and MSE worsened, and visual similarity is poorer than VanDLIR (see Fig.~\ref{fig:qualitativeResults} and Fig.~\ref{fig:ANOVA_DSC_LPIPS_MSE}). 
This is again unsurprising, as $\mathcal{L}_{lac}$ in VoxelMorph seeks only the local pixel-level similarity between the warped and fixed masks. It does not enforce matching neighboring pixels' intensity and/or global semantic attributes.

For further anatomical plausibility improvement, we pursued a second shape constraint on the latent space of the VAE for the masks (the AC-DLIR approach). This gives the network global anatomical knowledge and topology, enabling more stringent matching of fixed and warped moving masks. The pre-training of the VAE was successful, producing high DSC between input and output myocardial and LV masks in both the CAMUS and fetal echo datasets, and high visual similarity between input and output masks (see~\ref{Additional_Results} for Fig.~\ref{fig:VAE_DSC} and Fig.~\ref{fig:VAE_masks}). This demonstrates that the latent space design is sufficient to capture the global attributes of both the myocardium and LV chamber masks.

Results show that the AC-DLIR improved on VanDLIR more than VoxelMorph (see Fig.~\ref{fig:ANOVA_DSC_LPIPS_MSE}), where DSC, HD, and $TU$ have been improved by substantial margins, which are statistically significant ($p\ll 0.001$). The DSC and HD of AC-DLIR also outperformed VoxelMorph by $1.8\,\%$ and $11.4\,\%$ for the A2C view and $4.1\,\%$ and $18.9\,\%$ for the A4C view, respectively, while $TU$ and MSE were very close to VoxelMorph. 
A noteworthy $33.0\,\%$ for A2C and $24.4\,\%$ for A2C improvements in $TU$ from VanDLIR indicates much better warped anatomy, which can be confirmed by visual inspection of Fig.~\ref{fig:qualitativeResults}.

\begin{figure*}[!ht]
\caption{Normalized $\mathcal{L}_{gac}$ and $\mathcal{L}_{lac}$ losses for showing the effect of incorporating $L_2$ loss ($\mathcal{L}_{gac}$) in our proposed DLIR for the anatomical plausibility and realistic deformation. Non-plausible and plausible examples are selected from the VanDLIR and our proposed DLIR, respectively.}
\label{fig:L2_effect}
\centering
\includegraphics[width=0.9\textwidth]{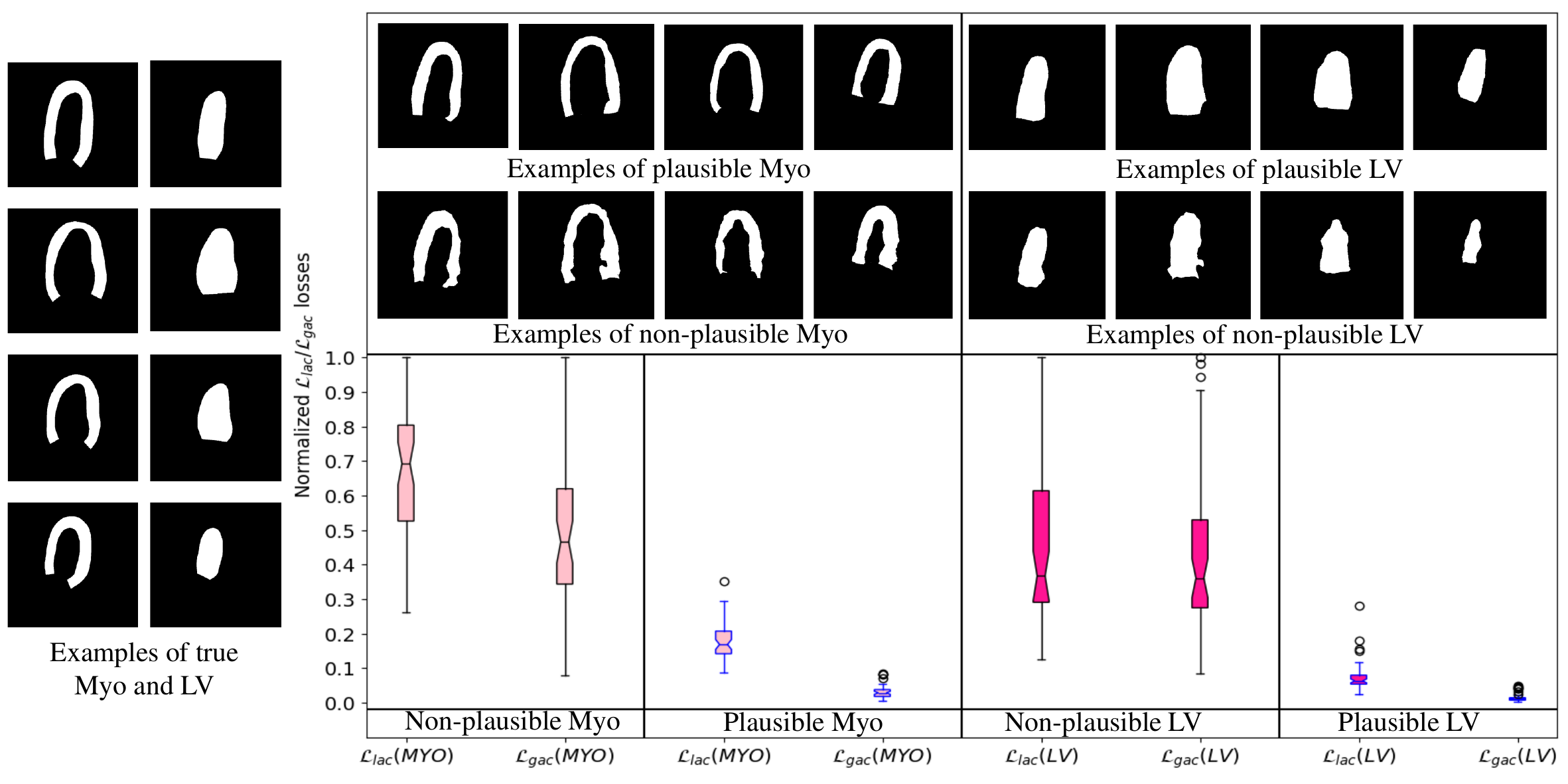}
\end{figure*}

\begin{figure*}[!ht]
\centering
\subfloat[HD vs. $\mathcal{L}_{gac}$]{\includegraphics[width=5cm]{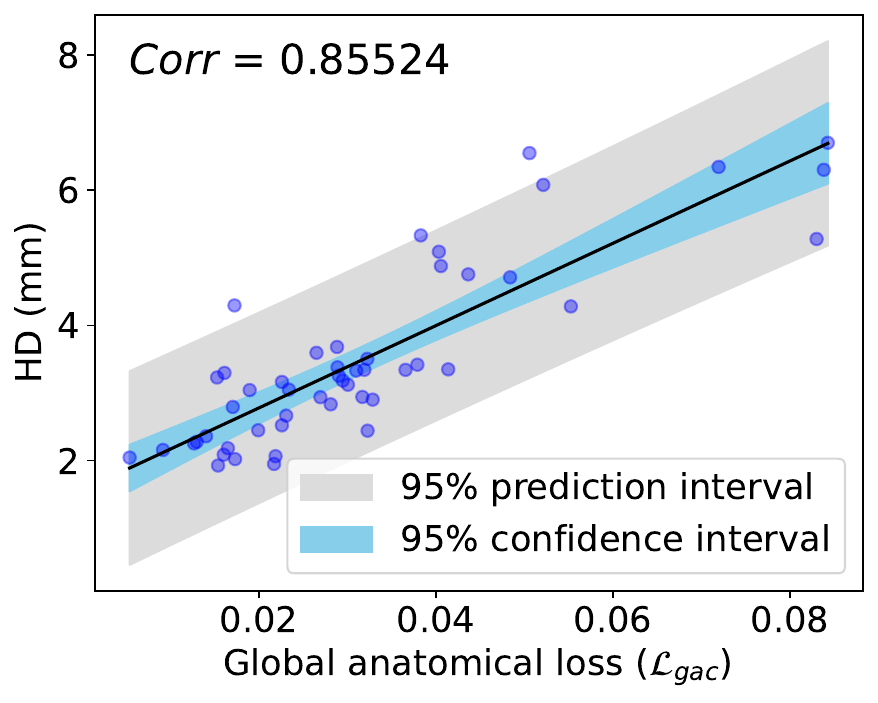}} \hspace{0.2 cm}
\subfloat[HD vs. $\mathcal{L}_{lac}$]{\includegraphics[width=5cm]{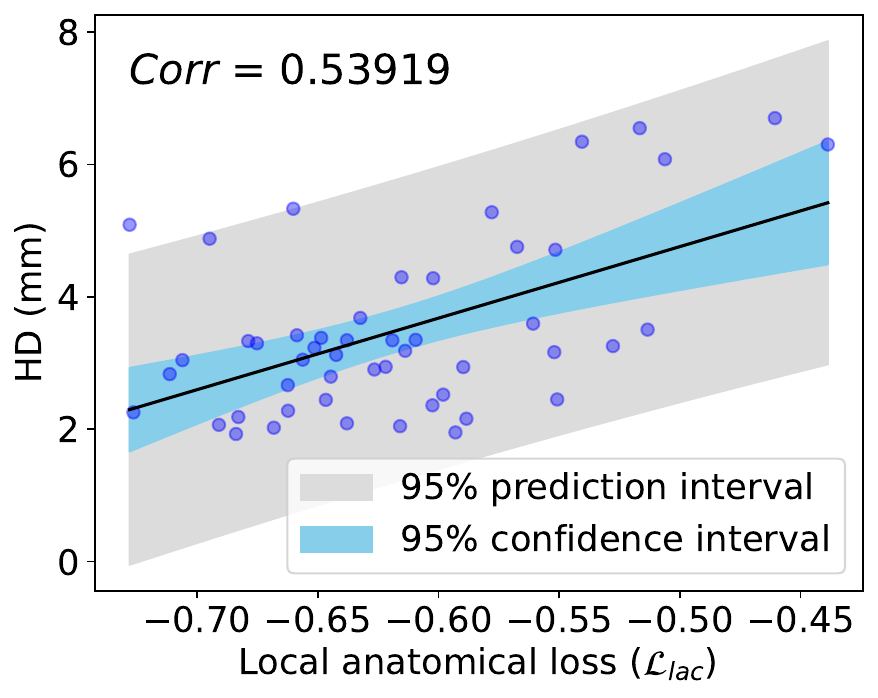}} \hspace{0.2 cm}
\subfloat[HD vs. $\mathcal{L}_{DdC}$]{\includegraphics[width=5cm]{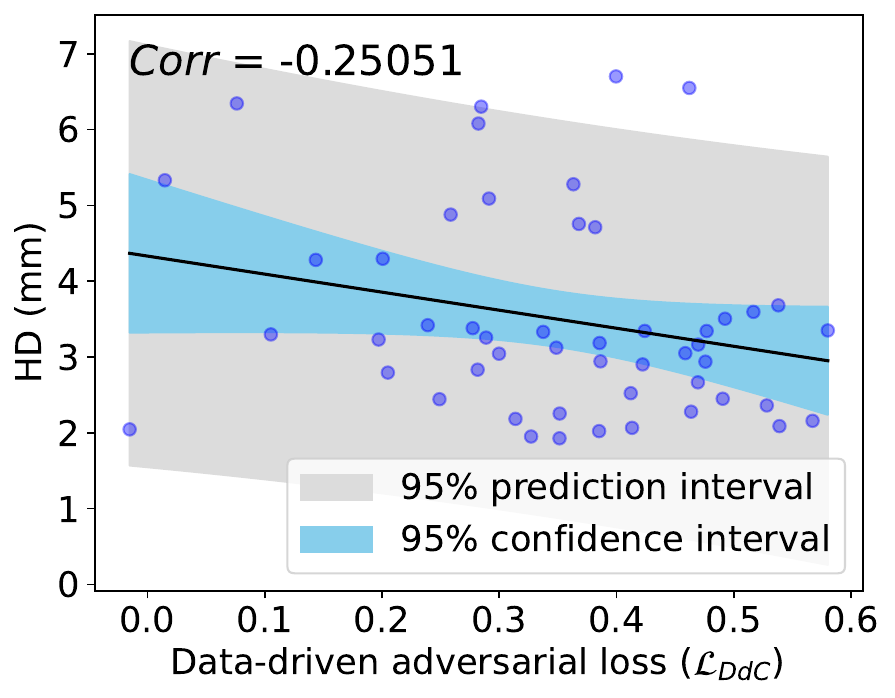}} \\
\subfloat[LPIPS vs. $\mathcal{L}_{gac}$]{\includegraphics[width=5cm]{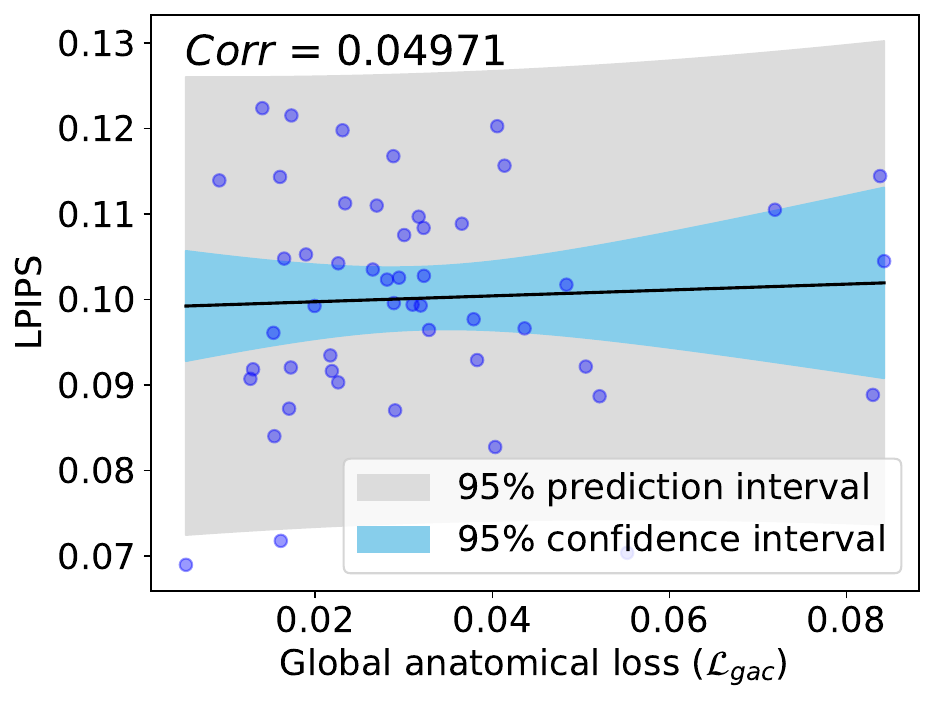}} \hspace{0.2 cm}
\subfloat[LPIPS vs. $\mathcal{L}_{lac}$]{\includegraphics[width=5cm]{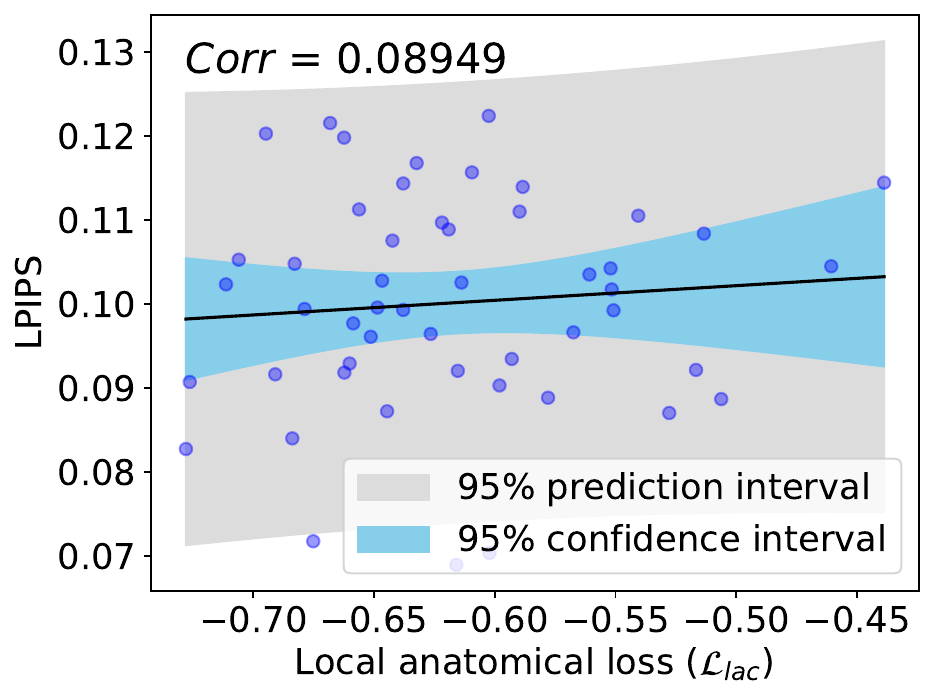}} \hspace{0.2 cm} 
\subfloat[LPIPS vs. $\mathcal{L}_{DdC}$]{\includegraphics[width=5cm]{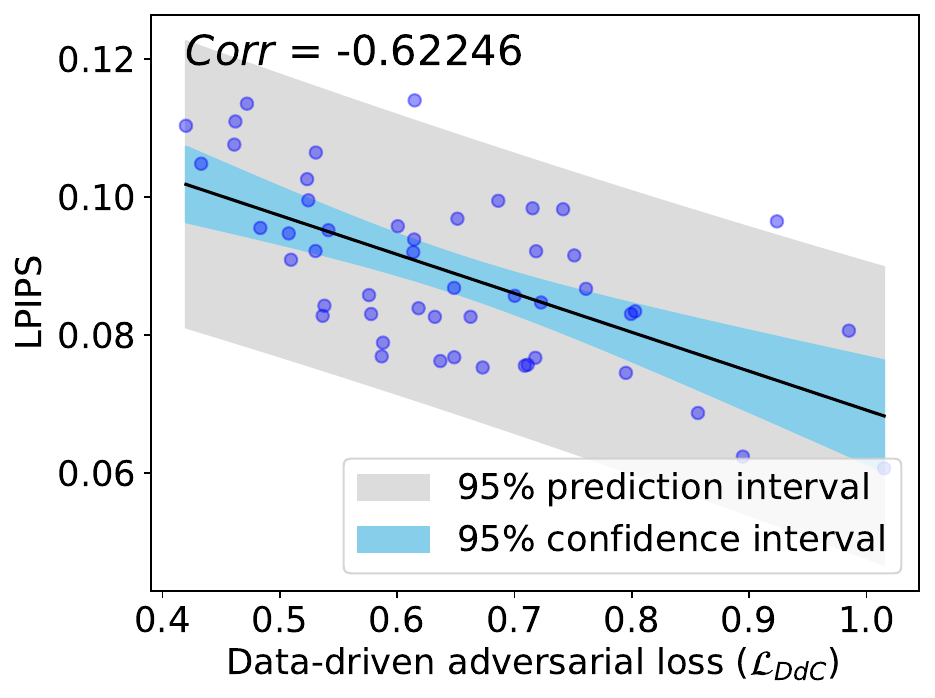}} 
\caption{Degree of correlation between two anatomical losses ($\mathcal{L}_{lac}$ and $\mathcal{L}_{gac}$) and data-driven loss ($\mathcal{L}_{DdC}$) with anatomical measures (HD) and visual similarity measures (LPIPS). The negative correlations in (c) and (f) indicate that when $\mathcal{L}_{DdC}$ is high, the discriminator becomes more confused about classifying $I_F$ and $I_W$, i.e., they are similar, yielding better LPIPS and HD.}
\label{fig:correlation_regression}
\end{figure*}

Visual inspection in Fig.~\ref{fig:L2_effect} also revealed that the low $\mathcal{L}_{gac}$ loss function correlated well with qualitatively physiologic-looking masks.
At the same time, correlation analysis shows that $\mathcal{L}_{gac}$ loss function has a strong correlation coefficient of 0.855 with HD, which is stronger than the correlation of $\mathcal{L}_{lac}$  or $\mathcal{L}_{DdC}$ with HD (see Fig.~\ref{fig:correlation_regression} (a) and Fig.~\ref{fig:correlation_regression} (b)). This indicates that the global shape constraint in AC-DLIR is important for good warped anatomical shapes and DSC outcomes (see Fig.~\ref{fig:ANOVA_DSC_LPIPS_MSE}). 
However, Fig.~\ref{fig:correlation_regression} (d) and Fig.~\ref{fig:correlation_regression} (e) also show that there are poor correlations between LPIPS and $\mathcal{L}_{gac}$ or $\mathcal{L}_{lac}$. This effect has been reflected in Table~\ref{tab:Registration_CAMUS_ES_ED} as the addition of those two anatomical losses worsened the MSE and LPIPS compared to the VanDLIR (see Fig.~\ref{fig:qualitativeResults} and Fig.~\ref{fig:ANOVA_DSC_LPIPS_MSE}). Thus, although the shape constraints can improve warped image anatomic shapes, they do not play a role in enhancing the perceptual match between the fixed and warped images.  

\subsection{Benefits of Adversarial Data-driven Constraints}
\label{DdC_on_VanDLIR}
To improve the warped image's visual quality, we investigated the addition of the adversarial image classifier network (DdC-DLIR), using a data-driven loss function that maximizes the global semantic correspondence between the fixed and warped moving intensity images. 
Results in Table~\ref{tab:Registration_CAMUS_ES_ED} show that adding DdC to VanDLIR significantly ($p\ll 0.001$) improves all metrics (DSC, HD, $TU$, and LPIPS) except for MSE, which remains similar to VanDLIR. Visually, the image has better similarity than the fixed image, but the warped label does not have a good anatomical shape in many cases (see Fig.~\ref{fig:qualitativeResults}).

\begin{figure*}[!ht]
\centering
\includegraphics[width=0.85\textwidth]{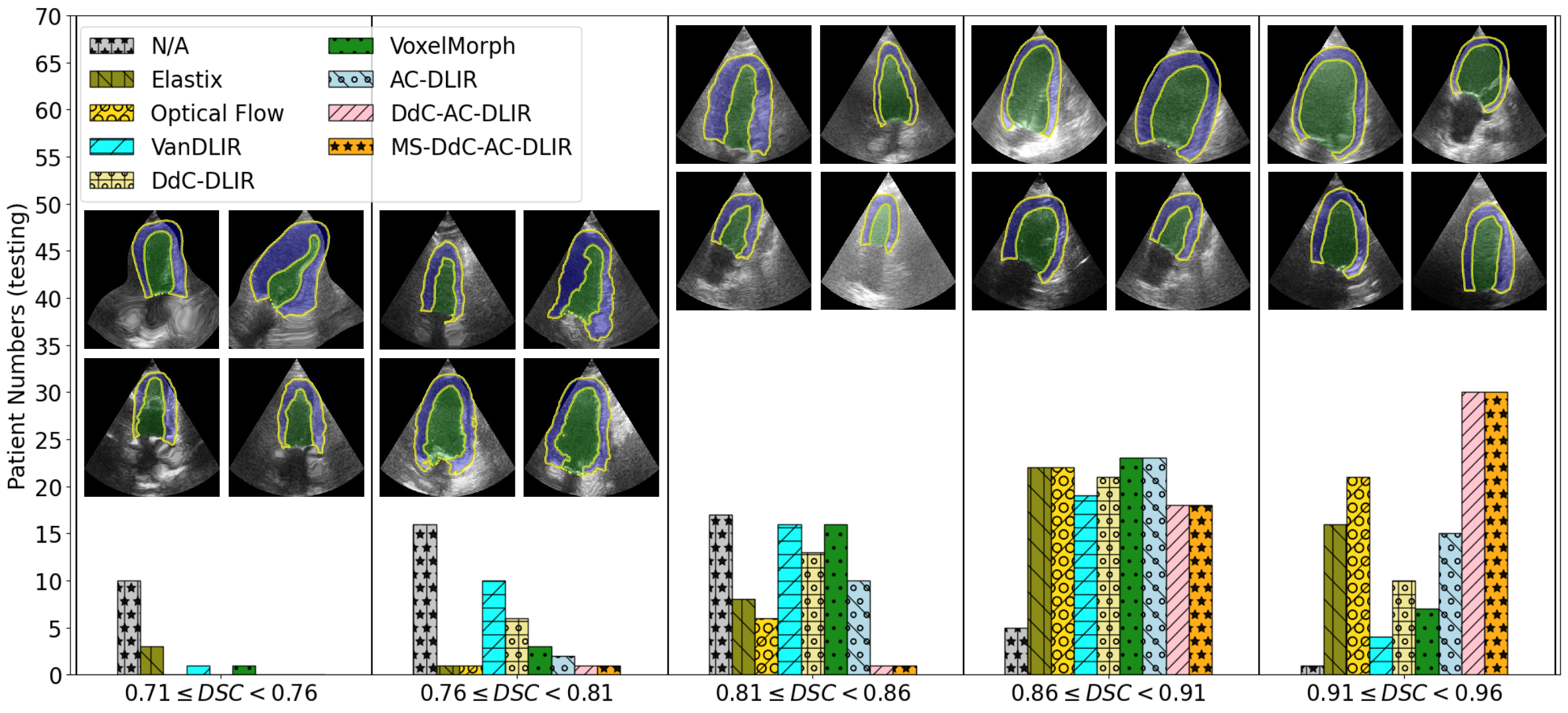}
\caption{Five different groups of obtained DSCs, demonstrating the number of testing patients in each DSC group. This figure is for A2C of CAMUS, and a similar figure for A4C of CAMUS is in Fig.~\ref{fig:Hist_pot_A4C} of~\ref{Additional_Results}.}
\label{fig:Hist_pot_A2C}
\end{figure*}

Compared to AC-DLIR, DdC-DLIR had an almost similar DSC. However, it had lower LPIPS, indicating better visual warped image quality. This suggests that the anatomic constraints imposed in AC-DLIR enforced a better warped cardiac shape, while the semantic correspondence constraint in DdC-DLIR enforced a better visual perception for the image, and the two approaches provide non-overlapping benefits. Correlation analysis in Fig.~\ref{fig:correlation_regression} further confirmed that the $\mathcal{L}_{gac}$ and $\mathcal{L}_{lac}$ loss functions in AC-DLIR correlated better with HD, while the $\mathcal{L}_{DdC}$ loss function in DdC-DLIR correlated better with LPIPS (see Fig.~\ref{fig:correlation_regression} (f)).
We thus investigated the combination of both strategies, DdC-AC-DLIR, to reap the benefits of both approaches (see Fig.~\ref{fig:ANOVA_DSC_LPIPS_MSE}). Results in Table~\ref{tab:Registration_CAMUS_ES_ED} show that the combined approach caused improvements to DSC and $TU$, above AC-DLIR and DdC-DLIR, while other metrics are similar to AC-DLIR and DdC-DLIR. Average DSC has now improved to $>$0.91 for both A2C and A4C, while $TU$ substantially reduces to less than 2.0 mm.
Fig.~\ref{fig:Hist_pot_A2C} demonstrates that when AC-DLIR and DdC-DLIR are combined in DdC-AC-DLIR, $60.0\,\%$ of testing patients have a DSC of 0.91 to 0.96, but only $30.0\,\%$ of patients have DSCs in that range when AC-DLIR is used alone and only $20.0\,\%$ of patients have DSCs in that range when DdC-DLIR is used alone. 
It has also been observed that the DdC-AC-DLIR produces DSCs of $<$0.86 for an extremely small number of image samples, unlike AC-DLIR and DdC-DLIR, again demonstrating the non-overlapping benefits of the two strategies.

\subsection{Effect of Multi-scale Learning}
\label{MS_Learning}
\begin{figure*}[!ht]
\centering
\includegraphics[width=0.85\textwidth]{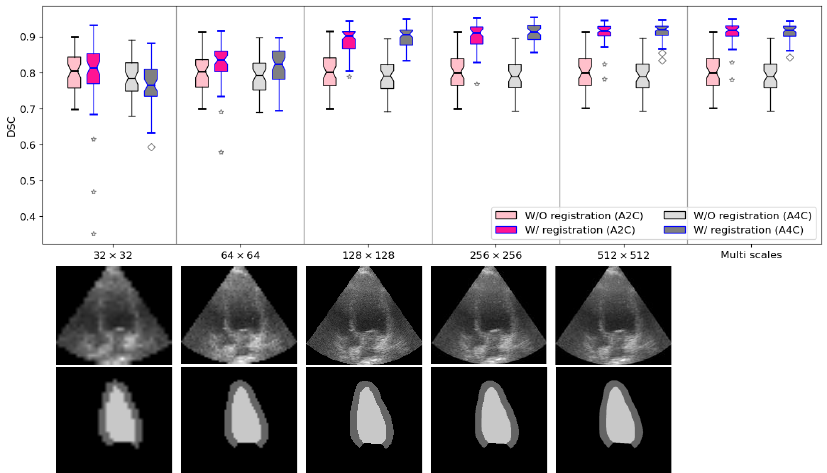}
\caption{Advantages of multi-scale training over single-scale. The registration method barely increases or decreases DSC when image resolution is low; for example, $32\times 32$ and $64\times 64$. Multi-scale training increases other measures (see Table~\ref{tab:Registration_CAMUS_ES_ED}) despite having a similar DSC to single $256\times 256$ and $512\times 512$. CAMUS A2C and A4C views were used to create this diagram.}
\label{fig:Scale}
\end{figure*}

We further investigated the use of a multi-scale learning approach to improve performance. We first tested the performance of DdC-AC-DLIR at various image scales, from $32\times 32$ to $512\times 512$. We find that increasing the image dimensions enhanced the DSC results significantly, as shown in Fig.~\ref{fig:Scale}. 
Next, we utilized a multi-scale training approach that progressively refined the training from coarse-to-fine scales. The network was first trained at the lowest scale, and trained parameters were used to initialize the training for the next image scale until the finest scale was trained. This strategy is named MS-DdC-AC-DLIR. Due to the low resolutions and comparatively poor performance of the $32\times 32$ and $64\times 64$ scales, we started training with $128\times 128$ and refined it to $256\times 256$ and $512\times 512$. 
Table~\ref{tab:Registration_CAMUS_ES_ED}, Fig.~\ref{fig:qualitativeResults}, Fig.~\ref{fig:ANOVA_DSC_LPIPS_MSE}, and Fig.~\ref{fig:Scale} demonstrate the enhancement of DLIR results due to the multi-scale strategy. Compared to the DdC-AC-DLIR trained at $512\times 512$ alone, $TU$ was increased from 1.73 mm to 1.36 in the CAMUS A2C view and from 1.16 mm to 0.98 mm in the CAMUS A4C view. The other performance metrics, like DSC, HD, MSE, and LPIPS, either increased or remained almost similar. These enhancements can be observed visually in Fig.~\ref{fig:qualitativeResults}, which depicts enhanced mask overlap and a marked reduction in MYO border irregularity alongside an increase in perceptual similarity in deformed intensity images ($I_W$), especially for the CAMUS A2C view.
The multi-scale training of DdC-AC-DLIR, as illustrated in Fig.~\ref{fig:Hist_pot_A2C} and Fig.~\ref{fig:Hist_pot_A4C}, not only results in an improvement in MYO's thickness uniformity but also does not result in a reduction in the number of testing patients in the better DSC group.

\subsection{Effect of Data Augmentation}
\label{Effect_of_Augmentation}
The MS-DdC-AC-DLIR has undergone additional testing by incorporating geometric and intensity image augmentations such as motion blur, Gaussian blur, defocusing \citep{hendrycks2019benchmarking}, contrast limited adaptive histogram equalization (CLAHE) \citep{yadav2014contrast}, and horizontal flipping. Examples of augmented images are displayed in Fig.~\ref{fig:augmentations} of~\ref{Additional_Results}. This Aug-MS-DdC-AC-DLIR has led to a significant improvement ($p\ll 0.001$) in the thickness uniformity of MYO, with the $TU$ improving from 1.36 mm to 0.84 mm for A2C and 0.98 mm to 0.58 mm for A4C. The HD has decreased from 3.50 mm to 3.22 mm for A2C and from 3.34 mm to 3.16 mm for A4C, whereas the DSCs have comparable values with and without the addition of augmentations. As a result of these anatomical metric enhancements, the mask now overlaps more effectively, and there has been a significant reduction in the irregularity and coarseness of MYO's inner and outer boundaries. Fig.~\ref{fig:MS_Aug} of~\ref{Additional_Results} illustrates the corresponding qualitative enhancements made to Aug-MS-DdC-AC-DLIR, comparing MS-DdC-AC-DLIR. Despite the improvement of pixel-level intensity differences in terms of MSE from 0.0016 mm to 0.0015 mm for A2C and from 0.0019 mm to 0.0018 mm for A4C, the LPIPS did not improve.

\subsection{Temporal Registration}
\label{Temporal_Registration}
\begin{figure*}[!ht]
    \centering
    \subfloat[\centering Fixed ($t_0$)]{{\includegraphics[width=3.1cm]{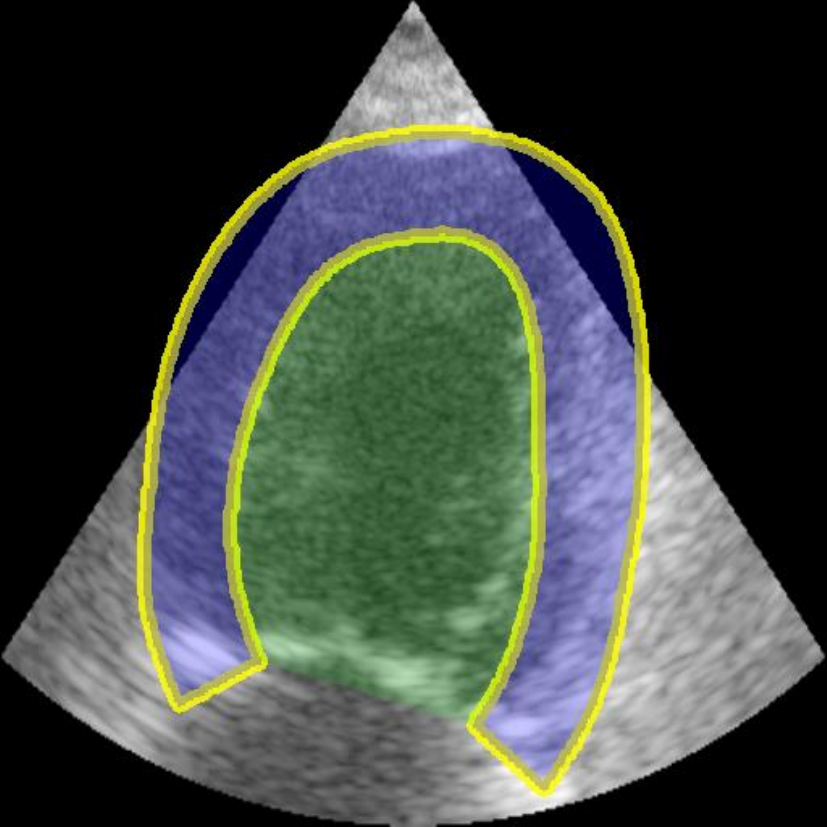} }} 
    \subfloat[\centering Moving ($t_3$)]{{\includegraphics[width=3.1cm]{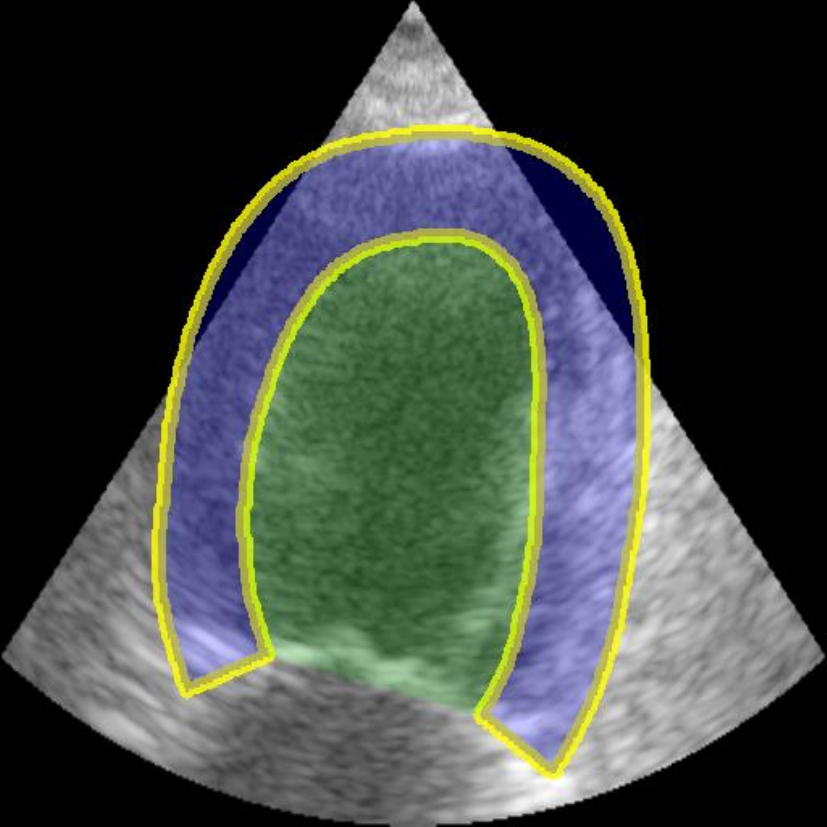} }} 
    \subfloat[\centering Moved ($t_3\mapsto t_0$)]{{\includegraphics[width=3.1cm]{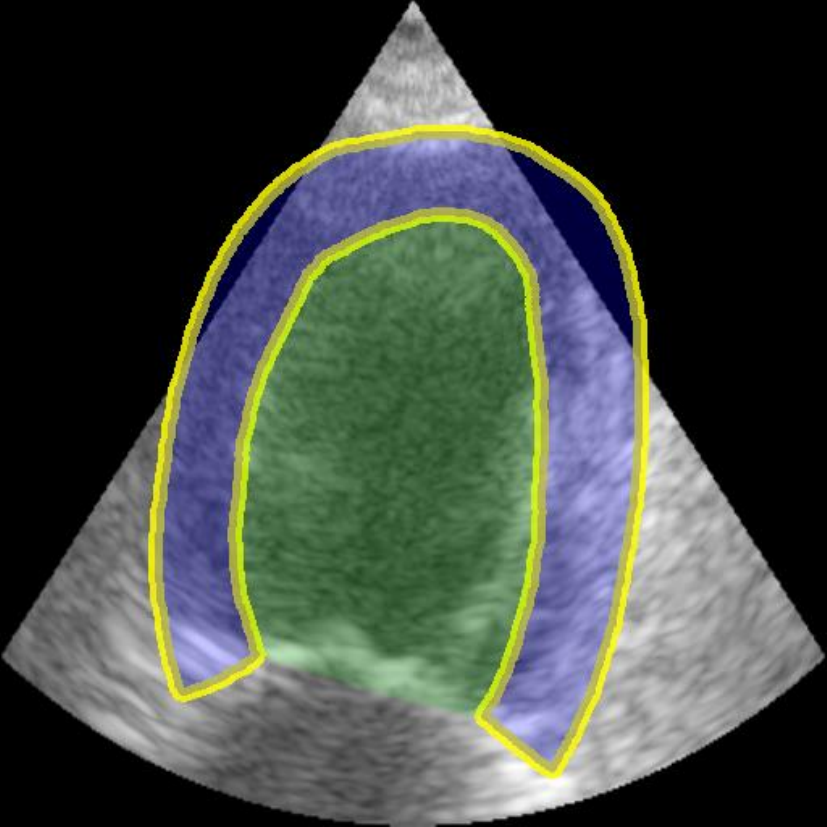} }} 
    \subfloat[\centering Moving ($t_{11}$)]{{\includegraphics[width=3.1cm]{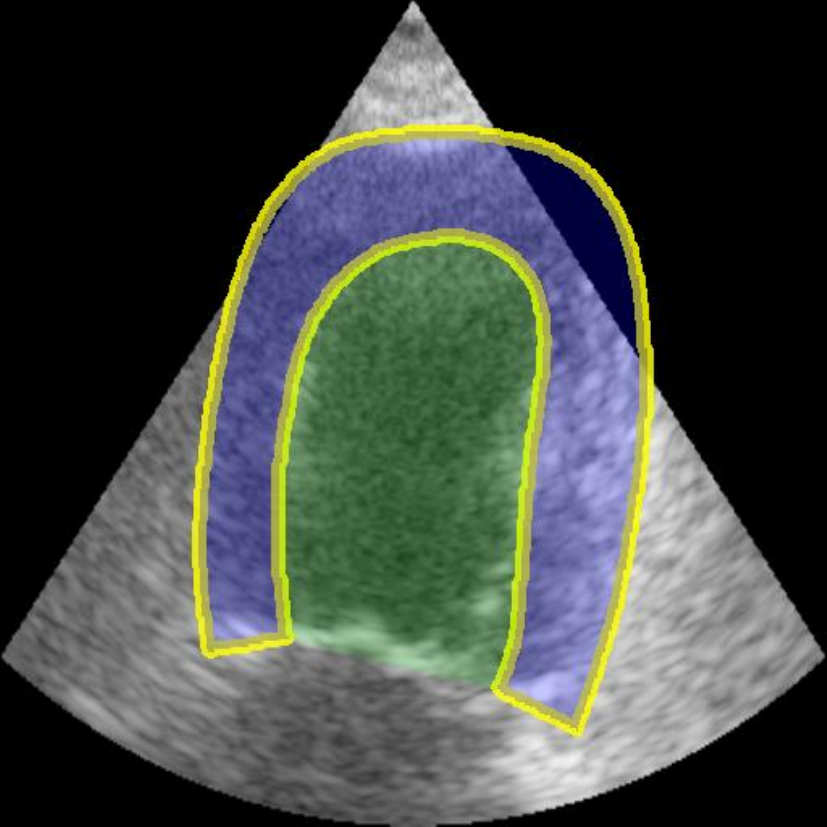} }} 
    \subfloat[\centering Moved ($t_{11}\mapsto t_0$)]{{\includegraphics[width=3.1cm]{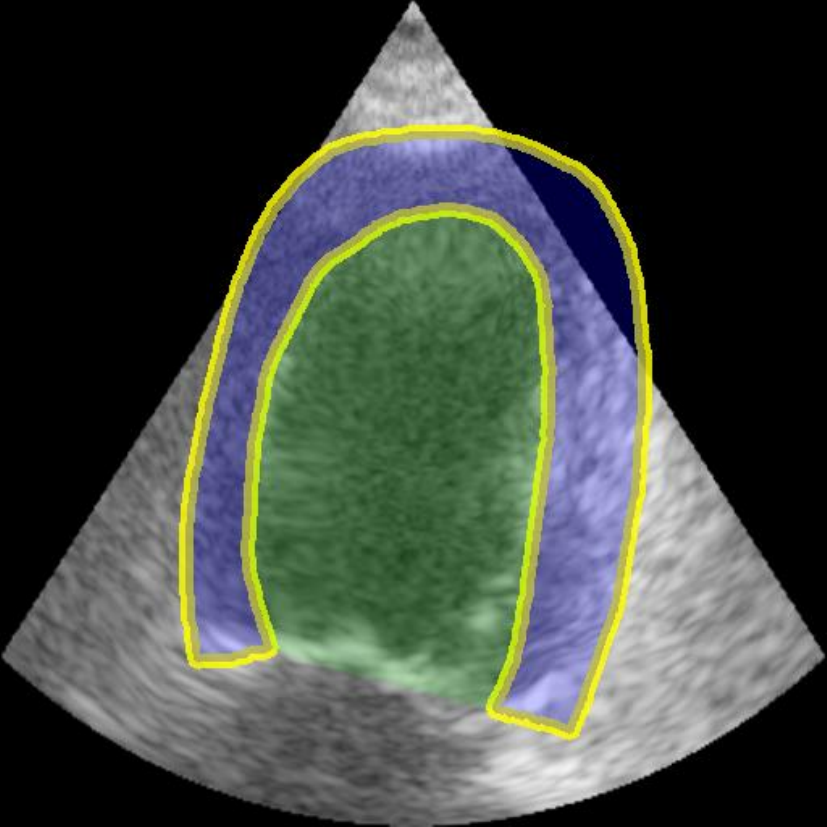} }} \\
    \subfloat[\centering Fixed ($t_0$)]{{\includegraphics[width=3.1cm]{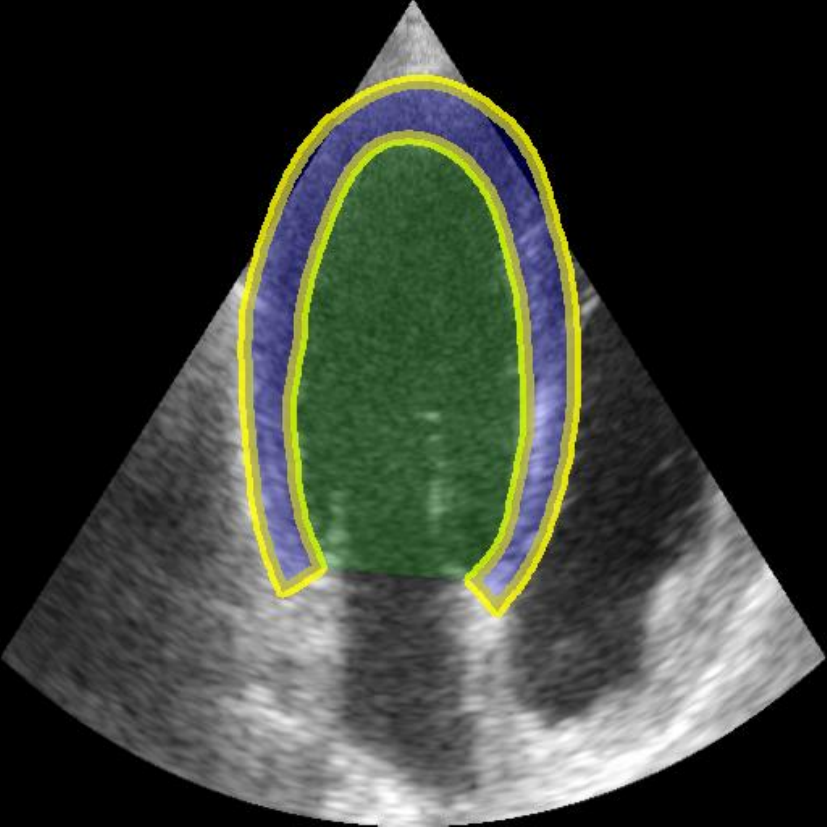} }} 
    \subfloat[\centering Moving ($t_3$)]{{\includegraphics[width=3.1cm]{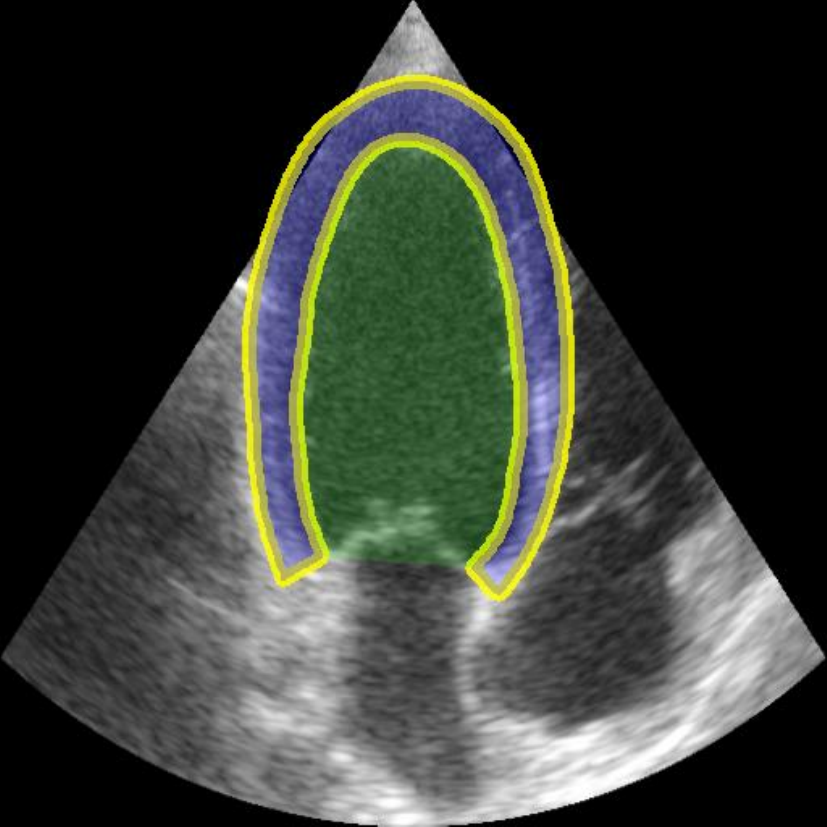} }} 
    \subfloat[\centering Moved ($t_3\mapsto t_0$)]{{\includegraphics[width=3.1cm]{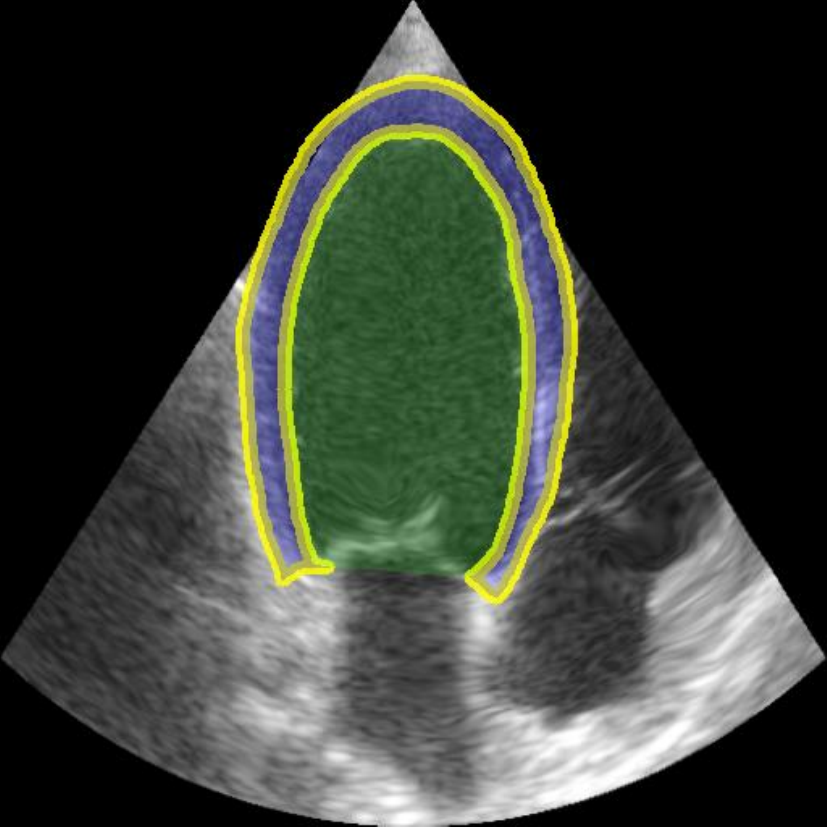} }} 
    \subfloat[\centering Moving ($t_{17}$)]{{\includegraphics[width=3.1cm]{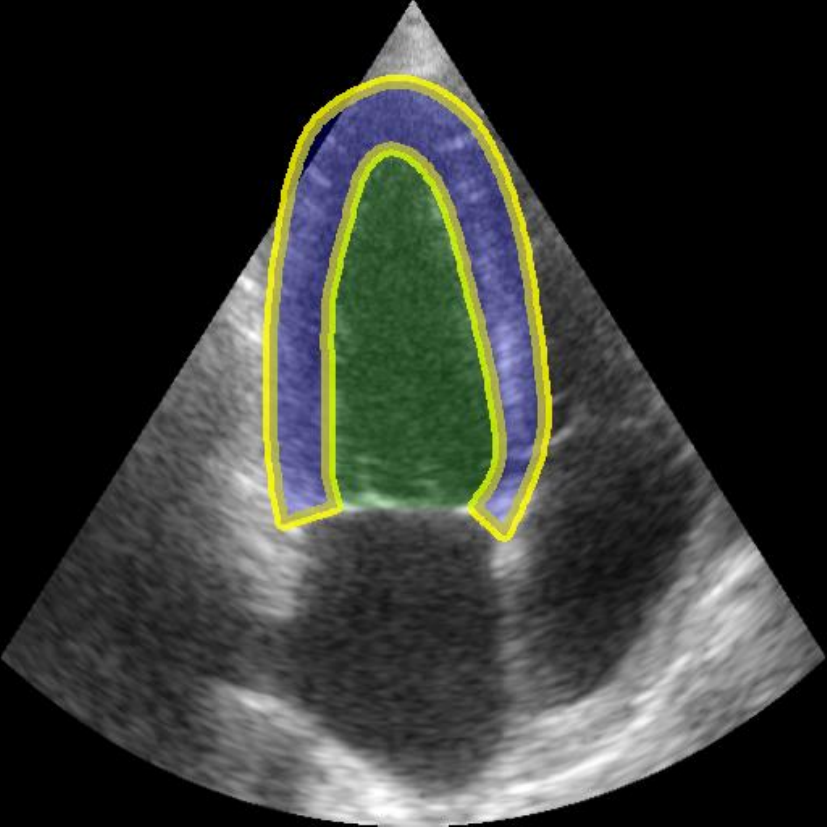} }} 
    \subfloat[\centering Moved ($t_{17}\mapsto t_0$)]{{\includegraphics[width=3.1cm]{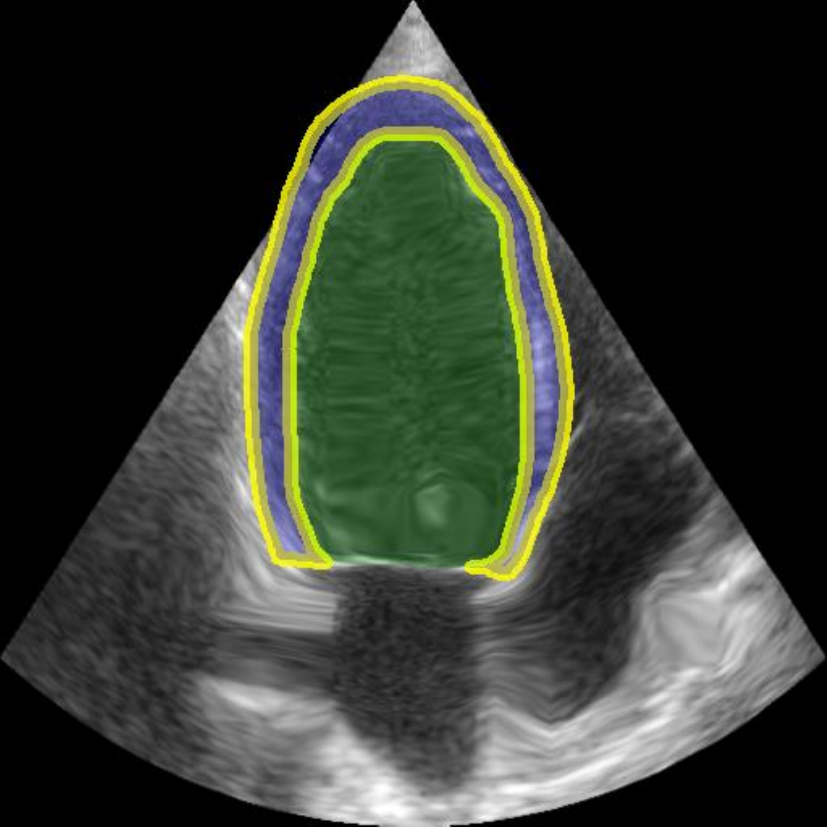} }} \\
    \subfloat[\centering Fixed ($t_0$)]{{\includegraphics[width=3.1cm]{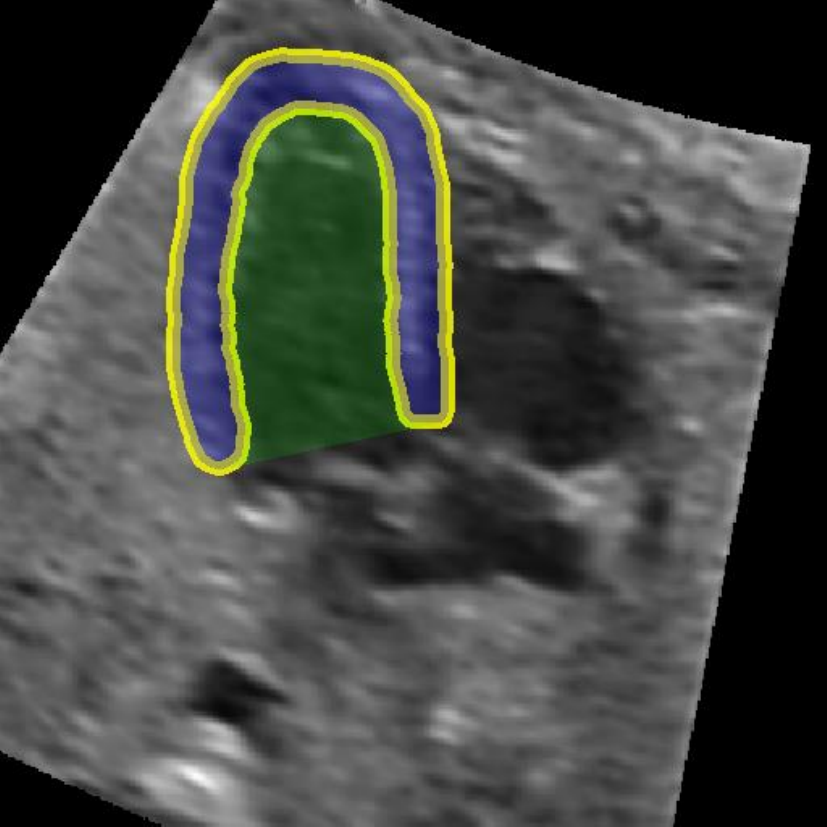} }} 
    \subfloat[\centering Moving ($t_3$)]{{\includegraphics[width=3.1cm]{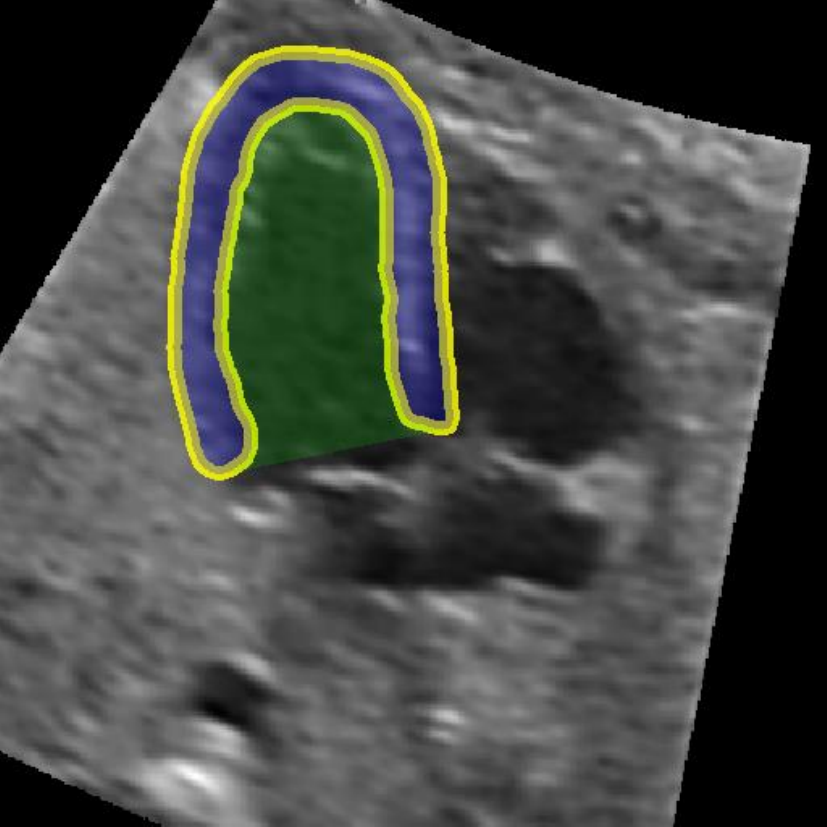} }} 
    \subfloat[\centering Moved ($t_3\mapsto t_0$)]{{\includegraphics[width=3.1cm]{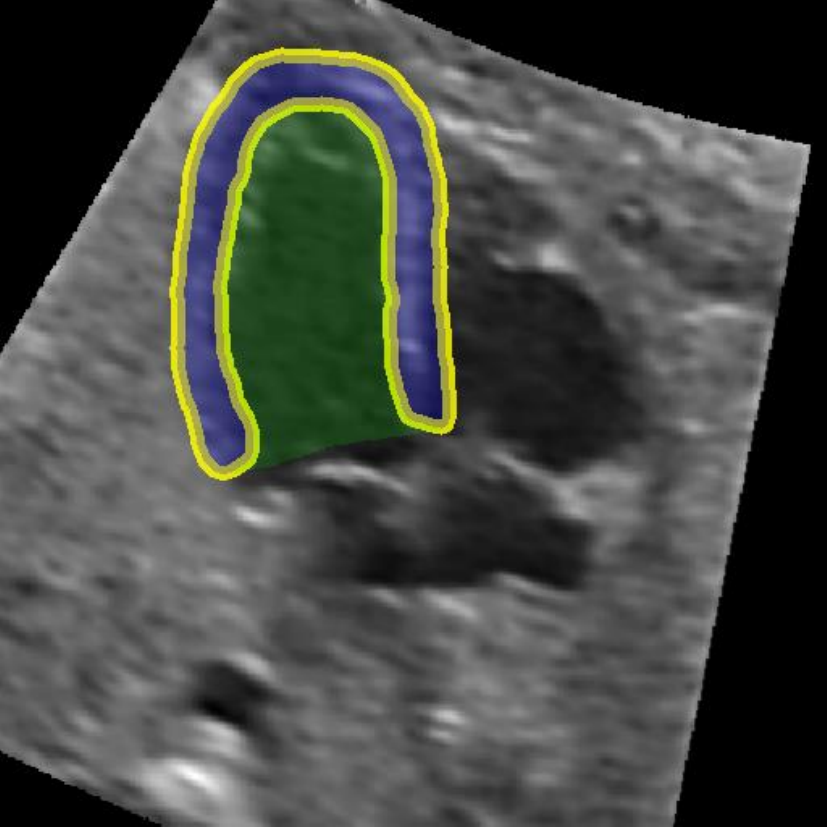} }} 
    \subfloat[\centering Moving ($t_{24}$)]{{\includegraphics[width=3.1cm]{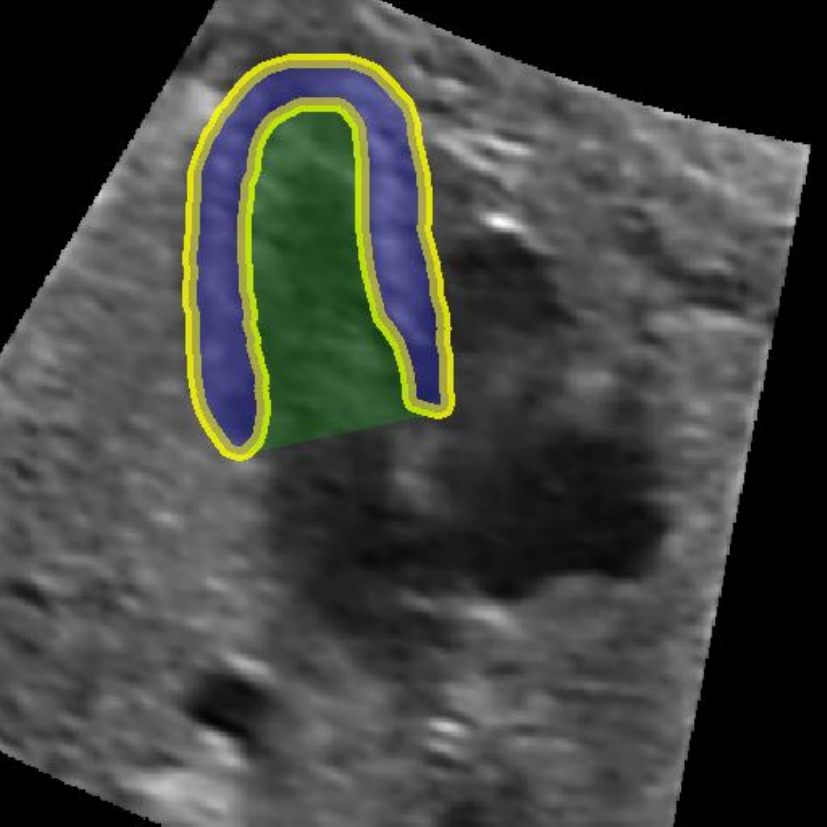} }} 
    \subfloat[\centering Moved ($t_{24}\mapsto t_0$)]{{\includegraphics[width=3.1cm]{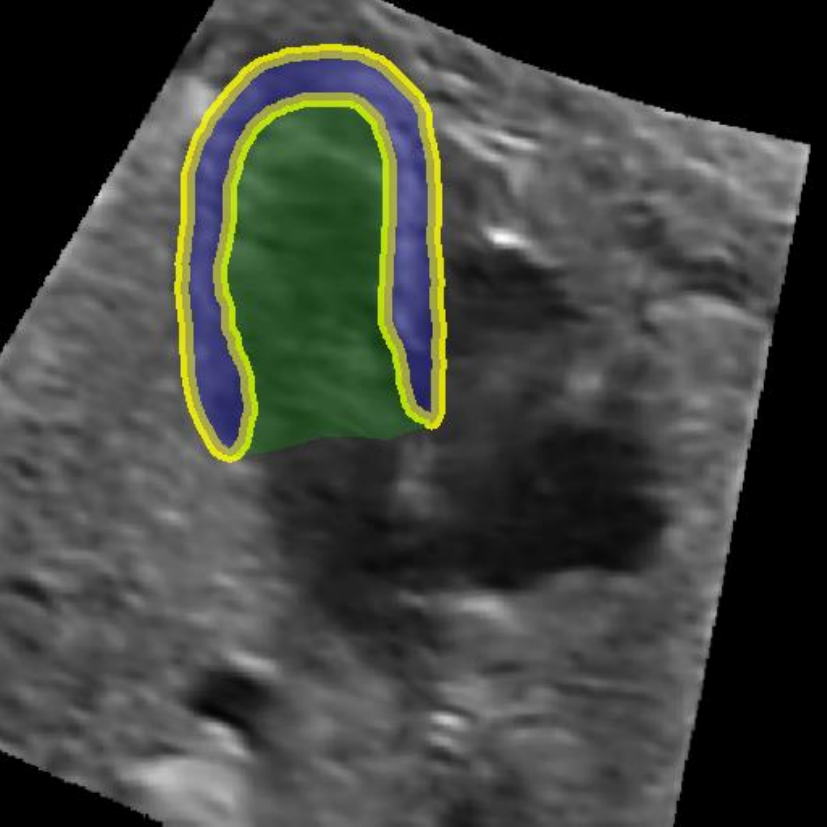} }} \\
    \subfloat[\centering Fixed ($t_0$)]{{\includegraphics[width=3.1cm]{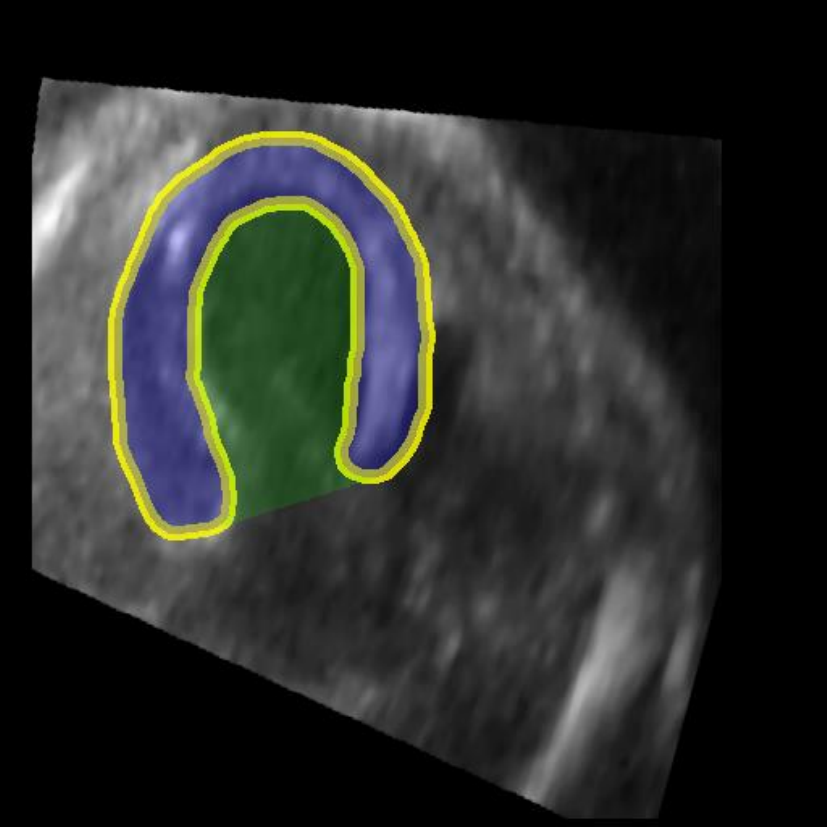} }} 
    \subfloat[\centering Moving ($t_3$)]{{\includegraphics[width=3.1cm]{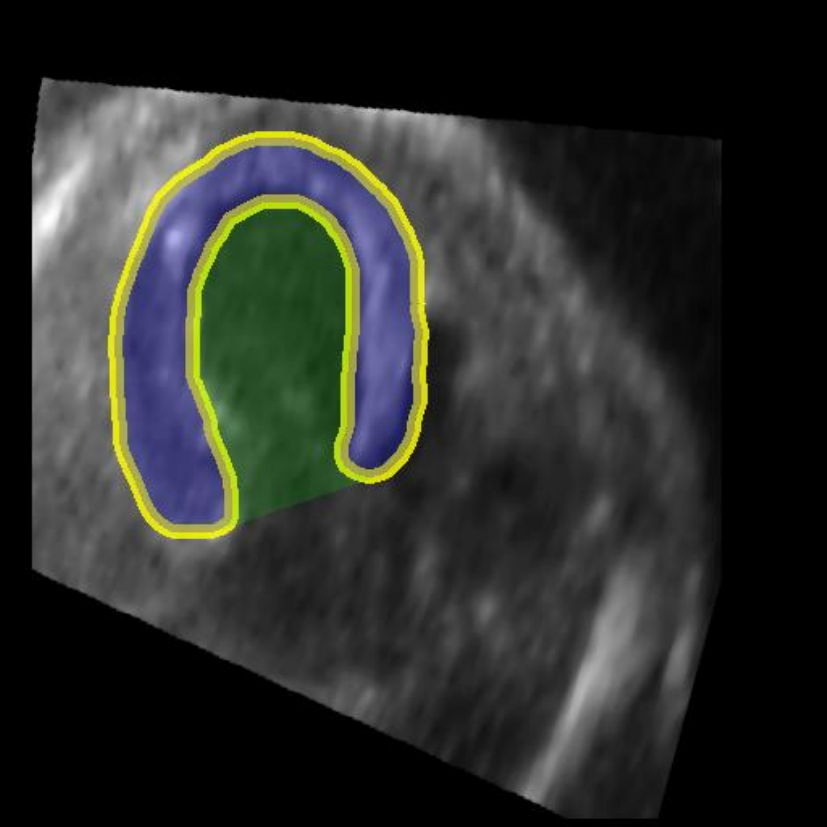} }} 
    \subfloat[\centering Moved ($t_3\mapsto t_0$)]{{\includegraphics[width=3.1cm]{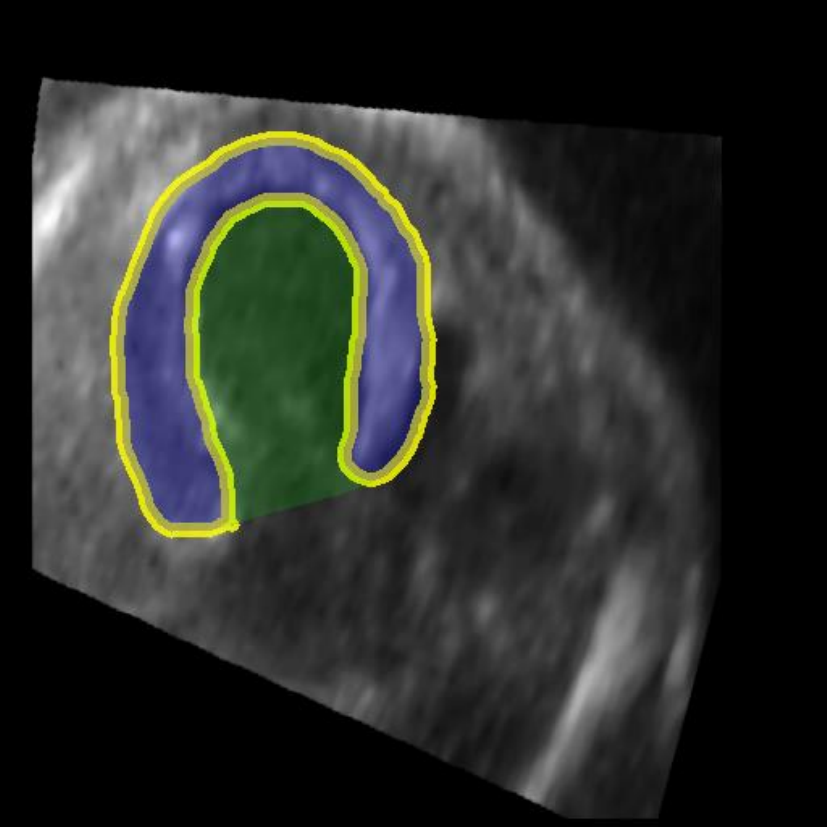} }} 
    \subfloat[\centering Moving ($t_{21}$)]{{\includegraphics[width=3.1cm]{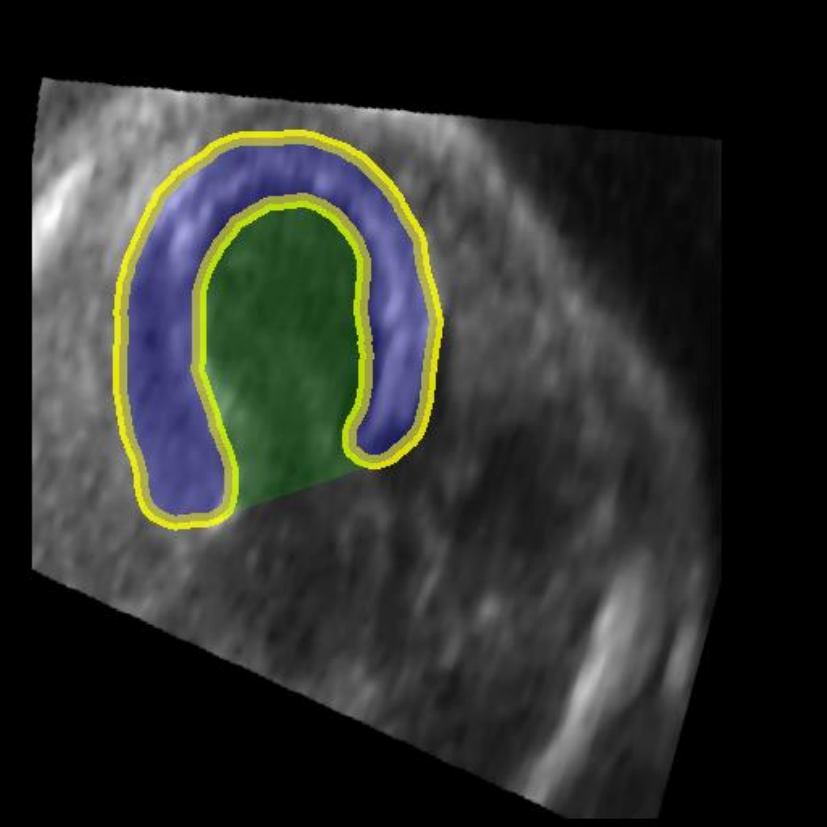} }} 
    \subfloat[\centering Moved ($t_{21}\mapsto t_0$)]{{\includegraphics[width=3.1cm]{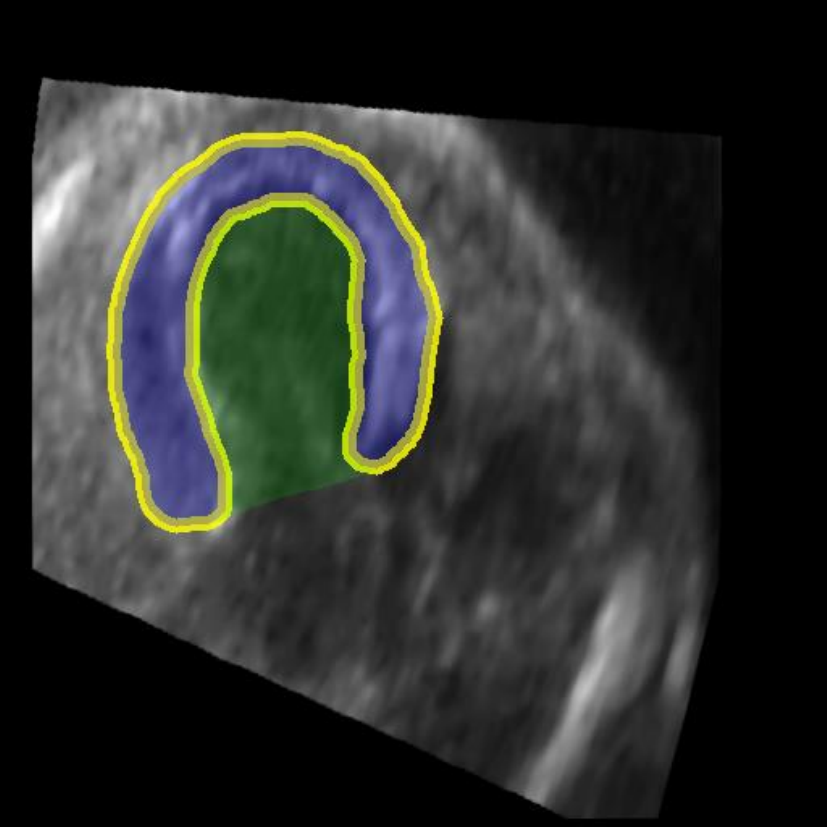} }} 
    \caption{Example of qualitative temporal registration results for the public CAMUS echoes dataset, with the first row for A2C view (see video\footref{CAMUSA2C}) and the second row for A4C view (see video\footref{CAMUSA4C}). The last two rows are for our private fetal echo A4C view (see video\footref{FetalA4C}), where the third row is for the healthy patient and the fourth row is for the diseased patient. Two distinct time positions, one near and one far from the fixed time point, are aligned with it.}
    \label{fig:Adult_Fetal}
\end{figure*}

Next, we proceed to registration over a temporal range. To do this, we fixed the first time point as the fixed image and mask ($I_F(t_0)$ and $M_F(t_0)$). The other time points ($I_F(t_n)$ and $M_F(t_n)$) are registered to this first time point, i.e., $I_W(t_n\mapsto t_0)=I_M(t_n)\circ \varphi$ and $M_W(t_n\mapsto t_0)=M_M(t_n)\circ \varphi$. Results are sown in Table~\ref{tab:TR_reg}, Fig.~\ref{fig:Adult_Fetal}, and Fig.~\ref{fig:DSC_Patients_A2C_A4C}.

{\renewcommand{\arraystretch}{2.3} 
\begin{table*}[!ht]
\fontsize{6pt}{6pt}\selectfont
\caption{Temporal image registration results for adult and fetal echo images, demonstrating the non-overlapping benefits of data-driven and anatomical constraints. All the metrics are estimated using fixed images (and masks) and warped moving images (and masks). The DSC and HD are the averages of backgrounds, MYO, and LV metrics, whereas the class-wise metrics of MYO and LV are shown in Table~\ref{tab:Additional_TR} of~\ref{Additional_Results}. Underlined, double-underlined, and bold fonts denote the best-performing metrics for the A2C view of adult echo, the A4C view of adult echo, and the A4C view of fetal echo, respectively.}
\label{tab:TR_reg}
\begin{tabular}{lccccccc}
\hline
\multicolumn{2}{l}{\textbf{Methods}}    & \textbf{DSC ($\uparrow$)} & \textbf{HD $(mm)$ ($\downarrow$)}  & $\mathbf{TU^\ddag}\,(mm)$  & \textbf{MSE ($\downarrow$)} & \textbf{LPIPS ($\downarrow$)}   \\ \hline

\multirow{3}{*}{N/A} & \textbf{Adult (A2C)} & $0.8942 \pm 0.0635$ & $3.11 \pm 1.89$ & $-$ & $0.0017 \pm 0.0009$ & $-$ \\ \cline{2-7} 
                  & \cellcolor[HTML]{EFEFEF}\textbf{Adult (A4C)} & \cellcolor[HTML]{EFEFEF}$0.8878 \pm 0.0671$ & \cellcolor[HTML]{EFEFEF}$3.21 \pm 1.96$ & \cellcolor[HTML]{EFEFEF}$-$ & \cellcolor[HTML]{EFEFEF}$0.0021 \pm 0.0013$ & \cellcolor[HTML]{EFEFEF}$-$ \\ \cline{2-7} 
                  
                  & \cellcolor[HTML]{C0C0C0}\textbf{Fetal (A4C)} & \cellcolor[HTML]{C0C0C0}$0.9024 \pm 0.0665$ & \cellcolor[HTML]{C0C0C0}$1.20 \pm 0.64$ & \cellcolor[HTML]{C0C0C0}$-$ & \cellcolor[HTML]{C0C0C0}$0.0008 \pm 0.0008$ & \cellcolor[HTML]{C0C0C0}$-$ \\ \hline

\multirow{3}{*}{VanDLIR} & \textbf{Adult (A2C)} & $0.9247 \pm 0.0470$ & $2.83 \pm 1.61$ & $1.04 \pm 1.02$ & \dunderline{1pt}{$0.0011 \pm 0.0006$} & $0.0806 \pm 0.0145$ \\ \cline{2-7} 
                  & \cellcolor[HTML]{EFEFEF}\textbf{Adult (A4C)} & \cellcolor[HTML]{EFEFEF}$0.9190 \pm 0.0490$ & \cellcolor[HTML]{EFEFEF}$3.27 \pm 1.80$ & \cellcolor[HTML]{EFEFEF}$0.80 \pm 0.74$ & \cellcolor[HTML]{EFEFEF}$0.0014 \pm 0.0009$ & \cellcolor[HTML]{EFEFEF}$0.0840 \pm 0.0166$ \\ \cline{2-7} 
                  
                  & \cellcolor[HTML]{C0C0C0}\textbf{Fetal (A4C)} & \cellcolor[HTML]{C0C0C0}$0.9365 \pm 0.0415$ & \cellcolor[HTML]{C0C0C0}$1.32 \pm 0.64$ & \cellcolor[HTML]{C0C0C0}$0.89 \pm 1.01$ & \cellcolor[HTML]{C0C0C0}$0.0008 \pm 0.0008$ & \cellcolor[HTML]{C0C0C0}$0.0828 \pm 0.0339$ \\ \hline

\multirow{3}{*}{AC-DLIR} & \textbf{Adult (A2C)} & $0.9376 \pm 0.0344$ & $2.52 \pm 1.39$ & \dunderline{1pt}{$0.90 \pm 1.02$} & $0.0013 \pm 0.0007$ & $0.0806 \pm 0.0150$ \\ \cline{2-7} 
                  & \cellcolor[HTML]{EFEFEF}\textbf{Adult (A4C)} & \cellcolor[HTML]{EFEFEF}$0.9278 \pm 0.04304$ & \cellcolor[HTML]{EFEFEF}$2.80 \pm 1.57$ & \cellcolor[HTML]{EFEFEF}$0.72 \pm 0.63$ & \cellcolor[HTML]{EFEFEF}$0.0017 \pm 0.0010$ & \cellcolor[HTML]{EFEFEF}$0.0838 \pm 0.0167$ \\ \cline{2-7} 
                  
                  & \cellcolor[HTML]{C0C0C0}\textbf{Fetal (A4C)} & \cellcolor[HTML]{C0C0C0}$0.9425 \pm 0.0358$ & \cellcolor[HTML]{C0C0C0}$1.32 \pm 0.62$ & \cellcolor[HTML]{C0C0C0}$1.01 \pm 1.18$ & \cellcolor[HTML]{C0C0C0}$0.0008 \pm 0.0008$ & \cellcolor[HTML]{C0C0C0}$0.0868 \pm 0.0326$ \\ \hline

\multirow{3}{*}{DdC-AC-DLIR} & \textbf{Adult (A2C)} & $0.9378 \pm 0.0356$ & $2.51 \pm 1.41$ & \dunderline{1pt}{$0.90 \pm 1.09$} & $0.0012 \pm 0.0006$ & \dunderline{1pt}{$0.0803 \pm 0.0146$} \\ \cline{2-7} 
                  & \cellcolor[HTML]{EFEFEF}\textbf{Adult (A4C)} & \cellcolor[HTML]{EFEFEF}$0.9269 \pm 0.0437$ & \cellcolor[HTML]{EFEFEF}$2.73 \pm 1.51$ & \cellcolor[HTML]{EFEFEF}\underline{\underline{$0.68 \pm 0.54$}} & \cellcolor[HTML]{EFEFEF}$0.0017 \pm 0.0011$ & \cellcolor[HTML]{EFEFEF}$0.0847 \pm 0.0166$ \\ \cline{2-7}      
                  & \cellcolor[HTML]{C0C0C0}\textbf{Fetal (A4C)} & \cellcolor[HTML]{C0C0C0}$0.9489 \pm 0.0291$ & \cellcolor[HTML]{C0C0C0}$1.24 \pm 0.57$ & \cellcolor[HTML]{C0C0C0}$1.10 \pm 1.28$ & \cellcolor[HTML]{C0C0C0}$0.0006 \pm 0.0007$ & \cellcolor[HTML]{C0C0C0}$0.0827 \pm 0.0337$ \\ \hline

\multirow{3}{*}{MS-DdC-AC-DLIR} & \textbf{Adult (A2C)} & \dunderline{1pt}{$0.9413 \pm 0.0262$} & \dunderline{1pt}{$2.29 \pm 1.08$} & $1.04 \pm 1.15$ & \dunderline{1pt}{$0.0011 \pm 0.0005$} & $0.0857 \pm 0.0167$ \\ \cline{2-7} 
                  & \cellcolor[HTML]{EFEFEF}\textbf{Adult (A4C)} & \cellcolor[HTML]{EFEFEF}\underline{\underline{$0.9353 \pm 0.0288$}} & \cellcolor[HTML]{EFEFEF}\underline{\underline{$2.38 \pm 1.16$}} & \cellcolor[HTML]{EFEFEF}$0.83 \pm 0.73$ & \cellcolor[HTML]{EFEFEF}$0.0013 \pm 0.0007$ & \cellcolor[HTML]{EFEFEF}$0.0899 \pm 0.0193$ \\ \cline{2-7} 
                  & \cellcolor[HTML]{C0C0C0}\textbf{Fetal (A4C)} & \cellcolor[HTML]{C0C0C0}$\mathbf{0.9523 \pm 0.0261}$ & \cellcolor[HTML]{C0C0C0}$\mathbf{1.17 \pm 0.54}$ & \cellcolor[HTML]{C0C0C0}$\mathbf{1.11 \pm 1.08}$ & \cellcolor[HTML]{C0C0C0}$\mathbf{0.0005 \pm 0.0006}$ & \cellcolor[HTML]{C0C0C0}$\mathbf{0.0826 \pm 0.0304}$ \\ \hline
\multicolumn{7}{p{15cm}}{\textbf{$^\ddag$An accurate $TU$ should be as close as the fixed MYO's TU, which is 0.74 mm for adults in A2C, 0.67 mm for adults in A4C, and 1.18 mm for fetal in A4C views.}} 
\end{tabular}
\end{table*}}
Generally, results were similar to those in Table~\ref{tab:Registration_CAMUS_ES_ED}'s ES to ED registration results. Table~\ref{tab:TR_reg}'s temporal registration results demonstrate progressive improving performance with the sequentially added strategies of anatomic constraints, data-driven constraints, and a multi-scale approach. The progressive improvements are the clearest for DSC, HD, and MSE, where MS-DdC-ACDLIR are the best in ($p\ll 0.001$ compared to VanDLIR). The progressive improvements are also evident for DSC from \ref{Additional_Results} (see Fig.~\ref{fig:DSC_Patients_A2C_A4C}). However, for $TU$ and LPIPS, progressive improvements are not observed. This is likely because our proposed strategies work well for large deformations, and here, many time frames have only small deformations, and the benefits are not as clearly observable.

Fig.~\ref{fig:Adult_Fetal} displays plausible and realistic warped moving images and masks for adults and fetal echo datasets using the MS-DdC-AC-DLIR algorithm.
The figure demonstrates that the proposed DLIR provides a plausible deformed image and mask not only for the closest time point but also for distant moving time points. Supplementary videos show the warped moving image over all the time points, demonstrating reasonable temporal consistency of the warped MYO mask (Video 1 for CAMUS-A2C\footnote{CAMUS-A2C: \url{https://youtu.be/EXTlyImIGgA}\label{CAMUSA2C}}, Video 2 for CAMUS-A4C\footnote{CAMUS-A4C: \url{https://youtu.be/l6ua9Qrc3JE}\label{CAMUSA4C}}, and Video 3 for Fetal-A4C\footnote{Fetal-A4C: \url{https://youtu.be/eGUU-rqWznY}\label{FetalA4C}}). The videos also demonstrate good anatomical plausibility and perceptual realism in the images.

\section{Estimating Ejection Fraction (EF)}
\label{Estimating_EF}
\begin{figure*}[!ht]
\centering
\subfloat[Elastix]{\includegraphics[width=4.85cm]{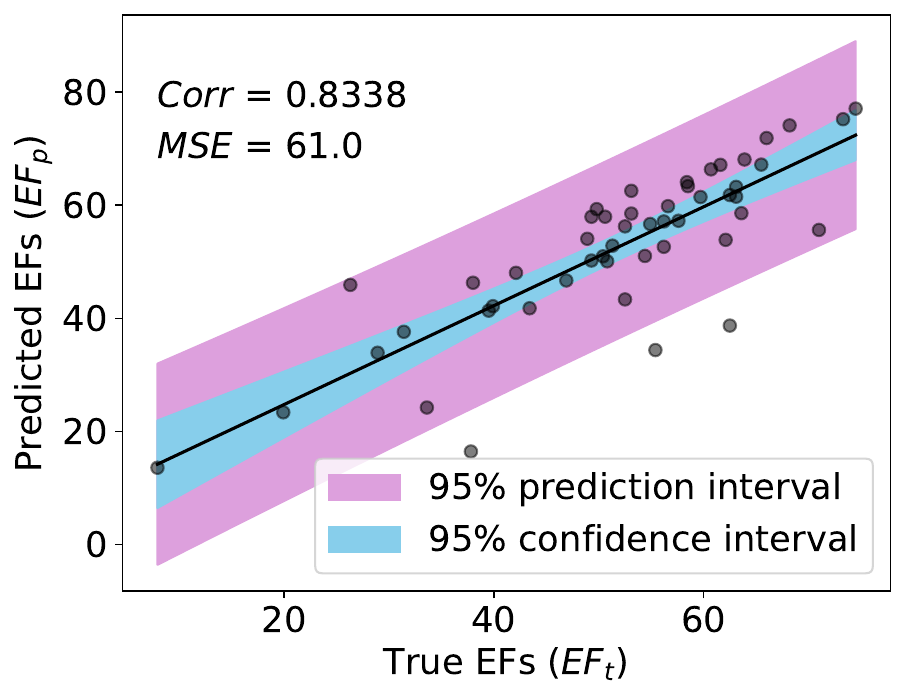}} \hspace{0.15cm}
\subfloat[OF]{\includegraphics[width=4.85cm]{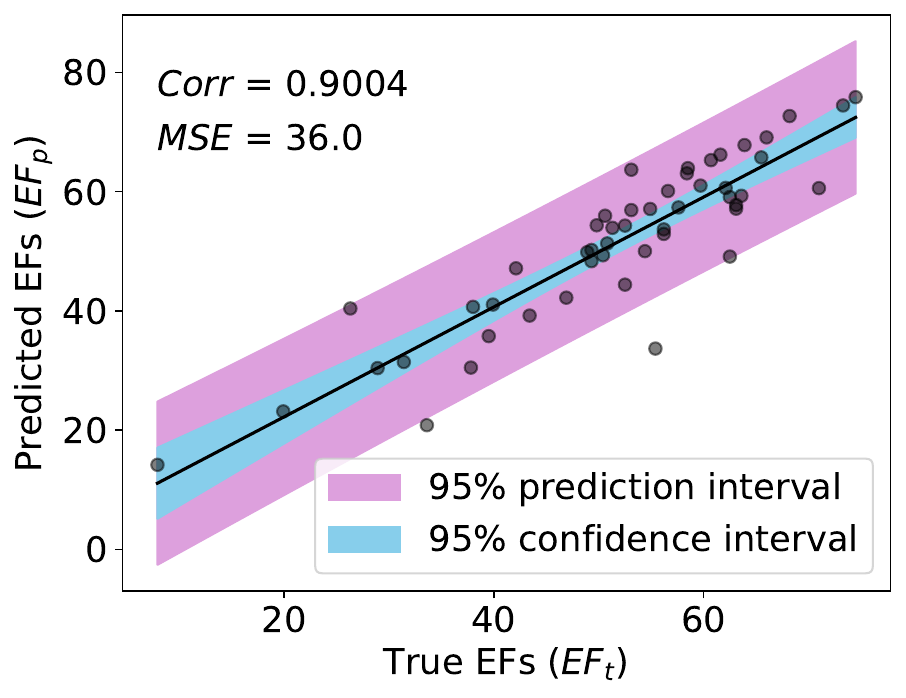}} \hspace{0.15cm}
\subfloat[VanDLIR]{\includegraphics[width=5cm]{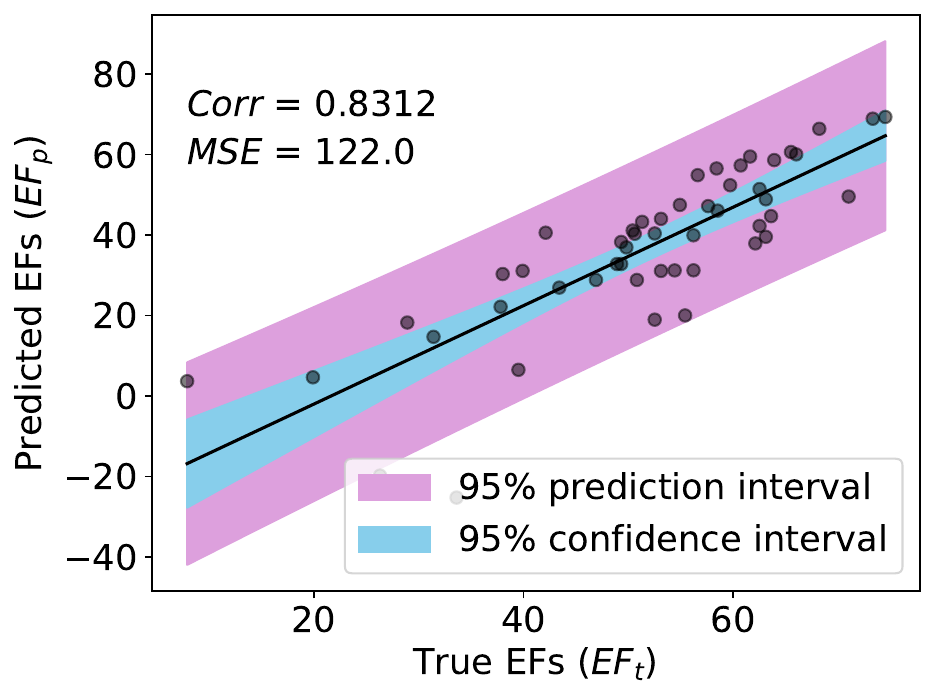}}\\
\subfloat[VoxelMorph]{\includegraphics[width=4.9cm]{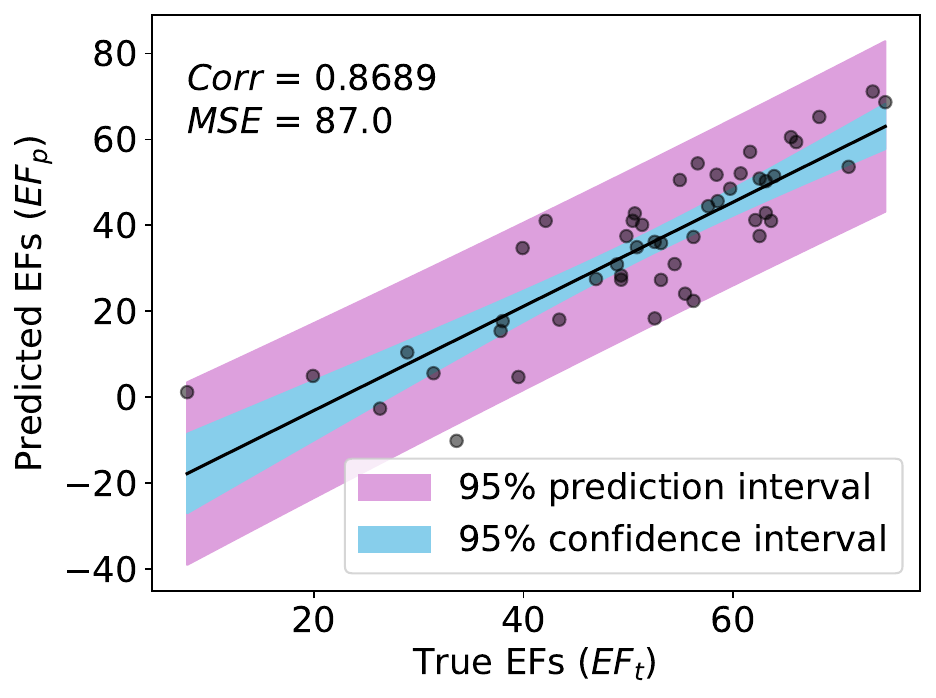}} \hspace{0.15cm}
\subfloat[AC-DLIR]{\includegraphics[width=4.9cm]{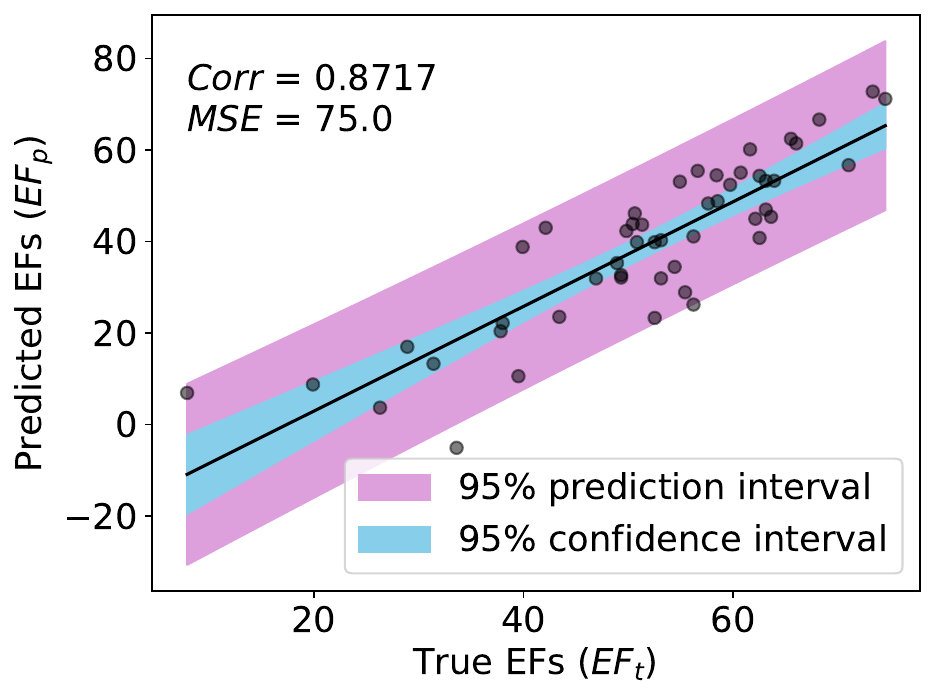}} \hspace{0.15cm}
\subfloat[DdC-DLIR]{\includegraphics[width=4.9cm]{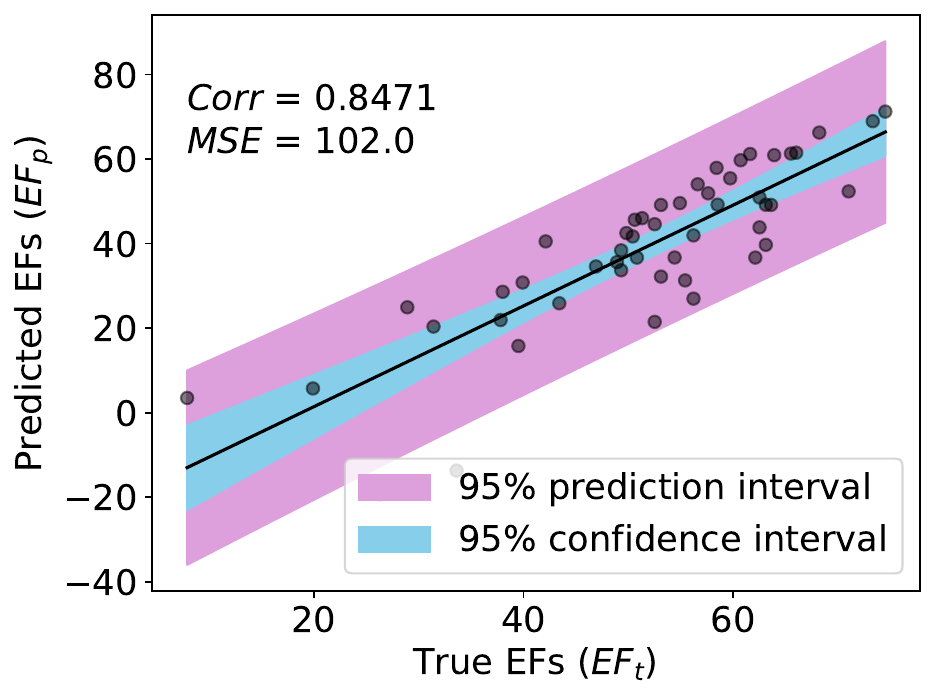}} \\
\subfloat[DdC-AC-DLIR]{\includegraphics[width=4.85cm]{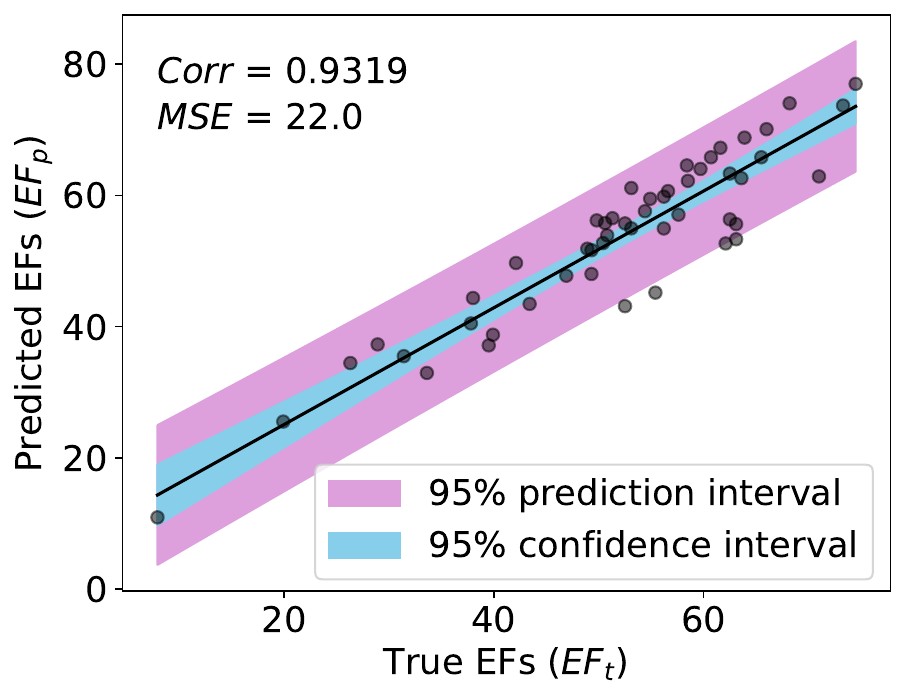}} \hspace{0.15cm}
\subfloat[MS-DdC-AC-DLIR]{\includegraphics[width=4.85cm]{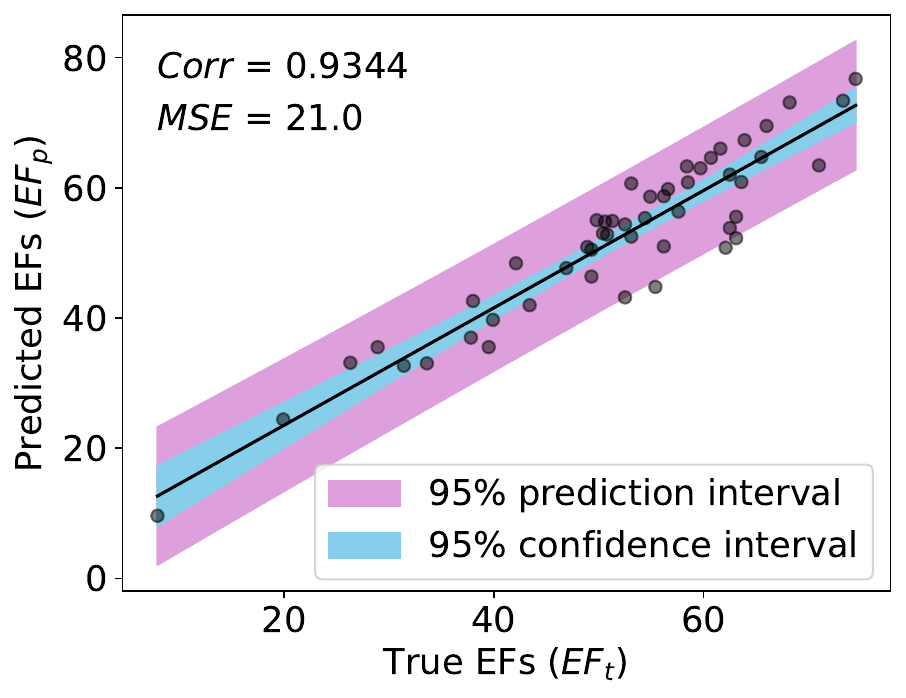}} \hspace{0.15cm}
\subfloat[Aug-MS-DdC-AC-DLIR]{\includegraphics[width=4.85cm]{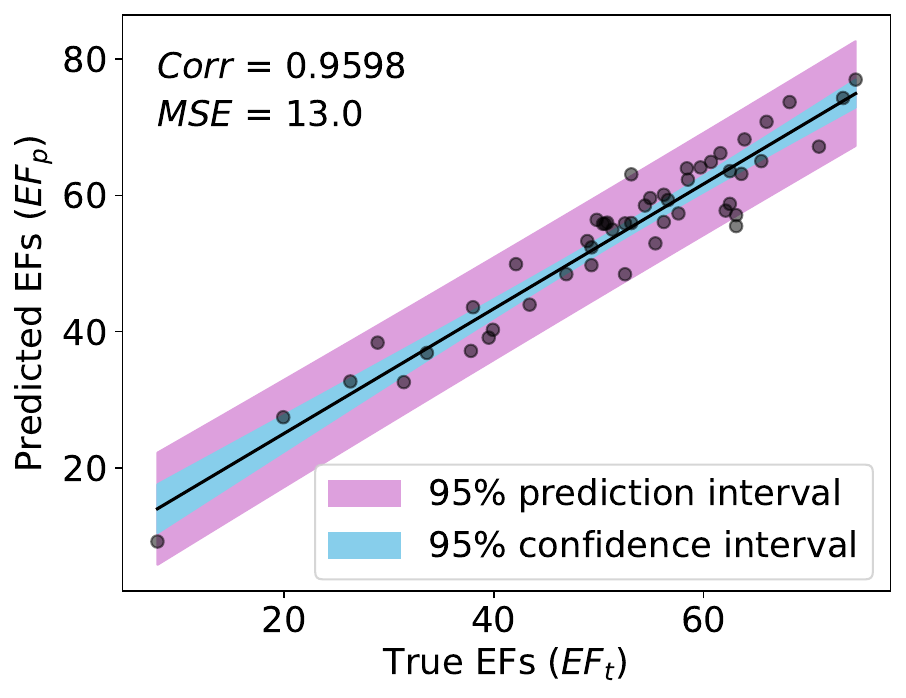}}
\caption{Regression plots between the true and estimated EFs from different registration methods to determine the degree of agreement in terms of the correlation between these two EFs. Those figures were created using the CAMUS A2C view.}
\label{fig:regression}
\end{figure*}

Finally, we evaluate the various DLIR algorithms for their ability to estimate cardiac EF using Eq.~\ref{eq:clin_metrics}. Regression analysis of the predicted and actual EF is shown in Fig.~\ref{fig:regression}.
Here, Elastix and VanDLIR had similar performance in terms of correlation coefficients of around 0.83, but OF achieved 0.90. There is again a progressive enhancement of the correlation coefficient, moving from VanDLIR to the progressive more complex techniques. Interestingly, OF performs better than most of the basic DLIR models, and only DdC-AC-DLIR and its multi-scale and augmented multi-scale versions perform better. This thus provides another validation for our proposed approach.
Additionally, it is observed that VanDLIR provides negative EFs for two out of fifty testing patients, which is physiologically impossible and likely due to poor image quality. Using only local anatomical constraints in VoxelMorph and DdC-DLIR, they remain negative, but adding global anatomical constraints in AC-DLIR results in reduced negative values for one patient. However, with DdC-AC-DLIR and its multi-scale and augmented versions, negative results are resolved. Results thus suggest that superior registration metrics, provided our proposed combination constraints, can translate to better clinical quantifications.

\section{Discussion and Conclusion}
\label{Critical_observation}
Temporal registration of echo images is an important foundational step for extracting clinically relevant features from images. For example, it can be used for estimating stroke volume and ejection fraction, myocardial strains and strain rates, etc. However, the inherent high noise level, low contrast, and limited image resolution, which lead to fuzzy anatomical boundaries, make the registration process challenging, especially for fetuses whose hearts are small and far from the transducer. It is thus important to optimize registration results so that clinical measurements can have manageably low errors and reasonable precision.

Our proposed strategies can help bridge these gaps. We show that the proposed strategies can achieve very robust registration metrics, as demonstrated in Tables~\ref{tab:Registration_CAMUS_ES_ED} and~\ref{tab:TR_reg}, and we show that this, in turn, can translate to better quality clinical measurements, as demonstrated in Fig.~\ref{fig:regression} on EF quantification. These good results are due to our use of a combination of strategies, where the anatomic constraint ensures better-warped anatomy while the adversarial data-driven constraint ensures better visual quality of the warped echo image. As our ablation studies show, this combination provides non-overlapping benefits to achieve excellent final outcomes. These results also demonstrate that such combinations can be a good algorithm design approach.

Our strategy revolves around enforcing high-quality anatomic shapes and image quality in the warped images. The rationale for this is that echo images are generated from ultrasound physics and echogenic material in tissues and are governed by realistic in vivo anatomy and motions; thus, warped images should achieve similarly physiological cardiac shapes and should have realistic echo images. Requiring warped images to have physiological cardiac good echo image textures can be argued to be a way of enforcing physiological deformations and thus a way of gaining accuracy. Traditionally, anatomic constraints are used in segmentation networks, such as those by \citet{oktay2017anatomically}, where anatomy latent spaces are used as constraints for segmentation training. Our study shows that it can also be a valuable strategy to improve registration.

Our Aug-MS-DdC-AC-DLIR approach provides improvements on traditional registration methods that are considered the gold standard, highlighting the benefits of adopting deep learning approaches. The deep learning approach also provides substantial computational time savings. On the same machine, DLIR took a significantly shorter time to compute a single deformation field, approximately 80–95 ms, while Elastix and OF require 4270 ms and 937 ms, respectively.
Our performance compares well with good techniques in the literature. For example, \citet{wei2020temporal} proposed the CLAS approach that concurrently performed segmentation and registration across the whole cycle with the 3D UNet and tested it on the same CAMUS dataset. They find that the mean DSC of A2C and A4C views from object tracking of the LV chamber were 0.923 and 0.903 for end-diastole and end-systole, which are lower than our DSC for the LV chamber of 0.9261 for the A2C view and 0.9360 for the A4C view. \citet{wang2022unsupervised} proposed a patch-based MLP and transformers for registration and reported similar DSC scores for both the LV chamber and MYO region as our study. However, we achieve better HD results, likely due to our anatomical constraints. \citet{fan2022searchmorph} used an unsupervised multi-scale correlation iterative registration network (SearchMorph) with a correlation layer. The author finds that the mean DSCs for MYO and LV are 0.880 and 0.888 for the A2C view and 0.891 and 0.919 for the A4C view, whereas for the same CAMUS dataset, those values from our DLIR are 0.881 and 0.953 for the A2C view and 0.864 and 0.951 for the A4C view.

Thus, in conclusion, a combination of anatomical constraints via a shape encoder, an adversarially trained data-driven constraint, and multi-scale training can produce excellent image registration results for both adult and fetal echocardiography, which can translate to higher-quality clinical measurements.

\section*{Acknowledgement}
Md. Kamrul Hasan was supported by the doctoral training program (DTP) studentship funds of the Engineering and Physical Sciences Research Council (EPSRC) (2022-2025).

Guang Yang was supported in part by the ERC IMI (101005122), the H2020 (952172), the MRC (MC/PC/21013), the Royal Society (IEC/NSFC/211235), the NVIDIA Academic Hardware Grant Program, the SABER project supported by Boehringer Ingelheim Ltd, and the UKRI Future Leaders Fellowship (MR/V023799/1).

\bibliographystyle{model1-num-names}
\bibliography{sample}

\onecolumn
\appendix

\section{Structure of Networks}
\label{Topological_Shape_Encoder}
{\renewcommand{\arraystretch}{1.0} 
\begin{table*}[!ht]
\fontsize{8pt}{8pt}\selectfont
\caption{The architecture of the topological shape encoder to learn the latent vectors of LV and MYO. Conv: Convolution layer with a dilation of 1; BN: Batch normalization for each mini-batch; ReLU: Rectified linear unit; FC: Fully connected layer; TConv: Transpose convolution layer for the upsampling with a dilation of 1; SIG: Sigmoid activation at the output; B: Mini-batch size; and $d$: Latent vector dimension ($\mathcal{Z}^d$).}
\label{tab:Shape_Encoder}
\begin{tabular}{llllll}
\hline
Layer & \begin{tabular}[c]{@{}l@{}}Feature dimension\\ (B $\times$ n\_filters $\times$ height $\times$ width)\end{tabular} & Parameters &Kernel & Stride & Padding \\ \hline
\multicolumn{5}{c}{Encoding mask topology to Latent vector ($\mathcal{Z}^d$) using Encoder ($\mathcal{EN}$) }                                                                                                                                    \\ \hline
Input ($I$)      &    $B\times 2^\dag \times H \times W$  &       $-$                                                &        $-$     &    $-$         &      $-$        \\ \hline
Conv      &      $B\times 8 \times H/2 \times W/2$  & $(3\times 3 \times 2 + 1)\times8$                                                   &  $3\times 3$           &  $2\times 2$           &  $1\times 1$            \\
ReLU      &    $B\times 8 \times H/2 \times W/2$   & $-$                                                    &  $-$            &     $-$         &     $-$          \\ \hline
Conv      &      $B\times 16 \times H/4 \times W/4$ & $(3\times 3 \times 8 + 1)\times16$                                                      &  $3\times 3$           &  $2\times 2$           &  $1\times 1$            \\
BN      &    $B\times 16 \times H/4 \times W/4$  &  $32$                                                    &  $-$            &     $-$         &     $-$          \\
ReLU      &    $B\times 16 \times H/4 \times W/4$    & $-$                                                   &  $-$            &     $-$         &     $-$          \\ \hline

Conv      &      $B\times 32 \times H/8 \times W/8$ & $(3\times 3 \times 16 + 1)\times32$                                                      &  $3\times 3$           &  $2\times 2$           &  $1\times 1$            \\
ReLU      &    $B\times 32 \times H/8 \times W/8$  & $-$                                                     &  $-$            &     $-$         &     $-$          \\ \hline

FC      &   $B\times 128$            &  $(32\times H/8 \times W/8 \times 128)+128$                                            &         $-$          &     $-$        &     $-$             \\ 
ReLU      &   $B\times 128$    &    $-$                                                   &   $-$          &     $-$        &     $-$              \\
FC      &   $B\times 128$   &   $(128\times 128)+128$                                                   &         $-$          &     $-$        &     $-$             \\ 
ReLU      &   $B\times 128$    &     $-$                                                  &   $-$          &     $-$        &     $-$              \\

FC      &   $B\times d$   &      $(128\times d) +d$                                               &         $-$          &     $-$        &     $-$             \\ 
ReLU      &   $B\times d$     &   $-$                                                   &   $-$          &     $-$        &     $-$              \\\hline

\multicolumn{5}{c}{Reconstructing mask topology from Latent vector ($\mathcal{Z}^d$) using Decoder ($\mathcal{DE}$)}                                                                                                                                    \\ \hline

FC      &   $B\times 128$   &      $(128\times d) +128$                                               &         $-$          &     $-$        &     $-$             \\ 
ReLU      &   $B\times 128$     &   $-$                                                   &   $-$          &     $-$        &     $-$              \\\hline

FC      &   $B\times (32\times H/8 \times W/8)$   &       \begin{tabular}[c]{@{}l@{}}$(32\times H/8 \times W/8 \times 128)\,+ $\\ $32\times H/8 \times W/8$\end{tabular}                                               &         $-$          &     $-$        &     $-$             \\ 
ReLU      &   $B\times (32\times H/8 \times W/8)$     &  $-$                                                  &   $-$          &     $-$        &     $-$              \\\hline

TConv      &      $B\times 16 \times H/4 \times W/4$ & $(3\times 3 \times 32 + 1)\times16$                                                      &  $3\times 3$           &  $2\times 2$           &  $1\times 1$            \\
BN      &    $B\times 16 \times H/4 \times W/4$  &  $32$                                                    &  $-$            &     $-$         &     $-$          \\
ReLU      &    $B\times 16 \times H/4 \times W/4$    & $-$                                                   &  $-$            &     $-$         &     $-$          \\ \hline

TConv      &      $B\times 8 \times H/2 \times W/2$ & $(3\times 3 \times 16 + 1)\times8$                                                      &  $3\times 3$           &  $2\times 2$           &  $1\times 1$            \\
BN      &    $B\times 8 \times H/2 \times W/2$  &  $16$                                                    &  $-$            &     $-$         &     $-$          \\
ReLU      &    $B\times 8 \times H/2 \times W/2$    & $-$                                                   &  $-$            &     $-$         &     $-$          \\ \hline

TConv      &      $B\times 2 \times H \times W$ & $(3\times 3 \times 8 + 1)\times2$                                                      &  $3\times 3$           &  $2\times 2$           &  $1\times 1$            \\
ReLU      &   $B\times 2 \times H \times W$     & $-$                                                   &  $-$            &     $-$         &     $-$          \\ \hline

SIG      &   $B\times 2 \times H \times W$     & $-$                                                   &  $-$            &     $-$         &     $-$          \\ \hline
      
Output ($I'$)    &   $B\times 2^\dag \times H \times W$ &  $-$                                              &   $-$          &     $-$        &     $-$         \\ \hline
\multicolumn{5}{l}{$^\dag$This article focuses on LV’s and MYO’s topologies of the adult and fetal hearts.}
\end{tabular}
\end{table*}}

{\renewcommand{\arraystretch}{1.0} 
\begin{table*}[!ht]
\fontsize{7pt}{7pt}\selectfont
\caption{Discriminator's architecture to learn the intensity images' attributes ($I_F$ and $I_W=I_M\circ \varphi$) in adversarial training to retrieve the texture features in the moved intensity images ($I_W=I_M\circ \varphi$). Conv: Convolution layer with a dilation of 1; IN: Instance normalization for each mini-batch; LReLU: Leaky Rectified linear unit; FC: Fully connected layer; SIG: Sigmoid activation at the output; and B: Mini-batch size.}
\label{tab:impl_Discriminator}
\begin{tabular}{llllll}
\hline
Layer & \begin{tabular}[c]{@{}l@{}}Feature dimension\\ (B $\times$ n\_filters $\times$ height $\times$ width)\end{tabular} & Parameters &Kernel & Stride & Padding \\ \hline
\multicolumn{5}{c}{2D feature learning blocks}                                                                                                                                    \\ \hline
Input       &    $B\times 1 \times H \times W$  &       $-$                                                &        $-$     &    $-$         &      $-$        \\ \hline

Conv      &      $B\times 8 \times H/2 \times W/2$  & $(3\times 3 \times 1 + 1)\times8$                                                   &  $3\times 3$           &  $2\times 2$           &  $1\times 1$            \\ \hline

Conv      &      $B\times 8 \times H/2 \times W/2$  & $(3\times 3 \times 1 + 1)\times8$                                                   &  $3\times 3$           &  $2\times 2$           &  $1\times 1$            \\ 
IN      &    $B\times 8 \times H/2 \times W/2$   & $-$                                                    &  $-$            &     $-$         &     $-$          \\

Dropout ($10\,\%$)      &    $B\times 8 \times H/2 \times W/2$   & $-$                                                    &  $-$            &     $-$         &     $-$          \\

LReLU      &    $B\times 8 \times H/2 \times W/2$   & $-$                                                    &  $-$            &     $-$         &     $-$          \\ \hline

Conv      &      $B\times 8 \times H/2 \times W/2$  & $(3\times 3 \times 8 + 1)\times8$                                                   &  $3\times 3$           &  $1\times 1$           &  $1\times 1$            \\ 
IN      &    $B\times 8 \times H/2 \times W/2$   & $-$                                                    &  $-$            &     $-$         &     $-$          \\

Dropout ($10\,\%$)     &    $B\times 8 \times H/2 \times W/2$   & $-$                                                    &  $-$            &     $-$         &     $-$          \\

LReLU      &    $B\times 8 \times H/2 \times W/2$   & $-$                                                    &  $-$            &     $-$         &     $-$          \\ \hline

Conv      &      $B\times 16 \times H/4 \times W/4$  & $(3\times 3 \times 8 + 1)\times16$                                                   &  $3\times 3$           &  $2\times 2$           &  $1\times 1$            \\ \hline

Conv      &      $B\times 16 \times H/4 \times W/4$  & $(3\times 3 \times 8 + 1)\times16$                                                   &  $3\times 3$           &  $2\times 2$           &  $1\times 1$            \\ 
IN      &    $B\times 16 \times H/4 \times W/4$   & $-$                                                    &  $-$            &     $-$         &     $-$          \\

Dropout ($10\,\%$)      &    $B\times 16 \times H/4 \times W/4$   & $-$                                                    &  $-$            &     $-$         &     $-$          \\

LReLU      &    $B\times 16 \times H/4 \times W/4$   & $-$                                                    &  $-$            &     $-$         &     $-$          \\ \hline

Conv      &      $B\times 16 \times H/4 \times W/4$  & $(3\times 3 \times 16 + 1)\times16$                                                   &  $3\times 3$           &  $1\times 1$           &  $1\times 1$            \\ 
IN      &    $B\times 16 \times H/4 \times W/4$   & $-$                                                    &  $-$            &     $-$         &     $-$          \\

Dropout ($10\,\%$)     &    $B\times 16 \times H/4 \times W/4$   & $-$                                                    &  $-$            &     $-$         &     $-$          \\

LReLU      &    $B\times 16 \times H/4 \times W/4$   & $-$                                                    &  $-$            &     $-$         &     $-$          \\ \hline

Conv      &      $B\times 32 \times H/8 \times W/8$  & $(3\times 3 \times 16 + 1)\times 32$                                                   &  $3\times 3$           &  $2\times 2$           &  $1\times 1$            \\ \hline

Conv      &      $B\times 32 \times H/8 \times W/8$  & $(3\times 3 \times 16 + 1)\times 32$                                                   &  $3\times 3$           &  $2\times 2$           &  $1\times 1$            \\ 
IN      &    $B\times 32 \times H/8 \times W/8$   & $-$                                                    &  $-$            &     $-$         &     $-$          \\

Dropout ($10\,\%$)      &    $B\times 32 \times H/8 \times W/8$   & $-$                                                    &  $-$            &     $-$         &     $-$          \\

LReLU      &    $B\times 32 \times H/8 \times W/8$   & $-$                                                    &  $-$            &     $-$         &     $-$          \\ \hline

Conv      &      $B\times 32 \times H/8 \times W/8$  & $(3\times 3 \times 32 + 1)\times 32$                                                   &  $3\times 3$           &  $1\times 1$           &  $1\times 1$            \\ 
IN      &    $B\times 32 \times H/8 \times W/8$   & $-$                                                    &  $-$            &     $-$         &     $-$          \\

Dropout ($10\,\%$)     &    $B\times 32 \times H/8 \times W/8$   & $-$                                                    &  $-$            &     $-$         &     $-$          \\

LReLU      &    $B\times 32 \times H/8 \times W/8$   & $-$                                                    &  $-$            &     $-$         &     $-$          \\ \hline

Conv      &      $B\times 64 \times H/16 \times W/16$  & $(3\times 3 \times 32 + 1)\times64$                                                   &  $3\times 3$           &  $2\times 2$           &  $1\times 1$            \\ \hline

Conv      &      $B\times 64 \times H/16 \times W/16$  & $(3\times 3 \times 32 + 1)\times64$                                                   &  $3\times 3$           &  $2\times 2$           &  $1\times 1$            \\ 
IN      &    $B\times 64 \times H/16 \times W/16$   & $-$                                                    &  $-$            &     $-$         &     $-$          \\

Dropout ($10\,\%$)      &    $B\times 64 \times H/16 \times W/16$   & $-$                                                    &  $-$            &     $-$         &     $-$          \\

LReLU      &    $B\times 64 \times H/16 \times W/16$   & $-$                                                    &  $-$            &     $-$         &     $-$          \\ \hline

Conv      &      $B\times 64 \times H/16 \times W/16$  & $(3\times 3 \times 64 + 1)\times64$                                                   &  $3\times 3$           &  $1\times 1$           &  $1\times 1$            \\ 
IN      &    $B\times 64 \times H/16 \times W/16$   & $-$                                                    &  $-$            &     $-$         &     $-$          \\

Dropout ($10\,\%$)     &    $B\times 64 \times H/16 \times W/16$   & $-$                                                    &  $-$            &     $-$         &     $-$          \\

LReLU      &    $B\times 64 \times H/16 \times W/16$   & $-$                                                    &  $-$            &     $-$         &     $-$          \\ \hline

\multicolumn{5}{c}{Multi-layer perceptron for the classification of $I_F$ and $I_W=I_M\circ \varphi$}                                                                                                                                    \\ \hline

Flatten      &    $B\times (64 \times H/16 \times W/16)$   & $64 \times H/16 \times W/16$                                                    &  $-$            &     $-$         &     $-$  \\ \hline

FC      &    $B\times 1024$   & $64 \times H/16 \times W/16 \times 1024 + 1024$                                                    &  $-$            &     $-$         &     $-$  \\ \hline

FC      &    $B\times 256$   & $1024 \times 256 + 256$                                                    &  $-$            &     $-$         &     $-$  \\ \hline

FC+SIG      &    $B\times 1$   & $256 \times 1 + 1$                                                   &  $-$            &     $-$         &     $-$  \\ \hline
      
Output    &   $B\times 1$ &  $-$                                              &   $-$          &     $-$        &     $-$         \\ \hline
% \multicolumn{5}{l}{$^\dag$This article focuses on LV’s and MYO’s topologies of the adult and fetal hearts.}
\end{tabular}
\end{table*}}

\newpage
\section{Additional Results}
\label{Additional_Results}

{\renewcommand{\arraystretch}{1.5} 
\begin{table*}[!ht]
\fontsize{7pt}{7pt}\selectfont
\caption{Additional registration results of the CAMUS adult echo dataset from different registration techniques, demonstrating the metrics for MYO and LV, where Table~\ref{tab:Registration_CAMUS_ES_ED} shows the average metrics of background, MYO, and LV. All the metrics are estimated using fixed images (and masks) and warped moving images (and masks). Bold fonts denote the best-performing metrics for the A2C echo view, while the best-performing metrics for the A2C view are underlined.}
\label{tab:Additional_ES_ED}
\begin{tabular}{lccccc}
\hline
\multicolumn{2}{l}{\multirow{2}{*}{\textbf{Methods}}} & \multicolumn{2}{c}{\textbf{DSC ($\uparrow$)}} & \multicolumn{2}{c}{\textbf{HD $(mm)$ ($\downarrow$)}} \\ \cline{3-6} 
\multicolumn{2}{c}{}                         & MYO         & LV        & MYO        & LV        \\ \hline
\multirow{2}{*}{Elastix \citep{klein2009elastix}}            & A2C           &      $0.7809 \pm 0.0768$       &    $0.8906 \pm 0.0669$       &   $6.04 \pm 3.10$         &  $5.07\pm 3.47$         \\
                             & \cellcolor[HTML]{EFEFEF}A4C           &  \cellcolor[HTML]{EFEFEF}$0.7792\pm 0.0682$           & \cellcolor[HTML]{EFEFEF}$0.9096\pm 0.0502$          &    \cellcolor[HTML]{EFEFEF}$5.64\pm3.00$        &  \cellcolor[HTML]{EFEFEF}$4.18 \pm 3.17$         \\ \hline
\multirow{2}{*}{OF \citep{lucas1981iterative} }            & A2C           &  $0.8140\pm 0.0456$           &      $0.8990\pm 0.0520$     &    $5.32\pm 1.96$        &     $4.10 \pm 1.93$      \\
                             & \cellcolor[HTML]{EFEFEF}A4C           &  \cellcolor[HTML]{EFEFEF}$0.8029\pm 0.0651$           &   \cellcolor[HTML]{EFEFEF}$0.9116\pm 0.0425$        &  \cellcolor[HTML]{EFEFEF}$5.49\pm 3.42$          & \cellcolor[HTML]{EFEFEF}$4.05 \pm 3.77$           \\ \hline

\multirow{2}{*}{VanDLIR}           & A2C           &     $0.7294\pm 0.0731$        &     $0.8512\pm 0.0719$      &      $6.78\pm2.17$      &   $4.48\pm 1.69$        \\
                             & \cellcolor[HTML]{EFEFEF}A4C           &  \cellcolor[HTML]{EFEFEF}$0.7173\pm 0.0713$           &  \cellcolor[HTML]{EFEFEF}$0.8519\pm 0.0596$          &    \cellcolor[HTML]{EFEFEF}$6.42\pm 2.02$        &   \cellcolor[HTML]{EFEFEF}$5.05\pm 1.68$        \\ \hline
\multirow{2}{*}{VoxelMorph \citep{balakrishnan2019voxelmorph}}            & A2C           &    $0.7838\pm 0.0609$         &    $0.8443 \pm 0.0641$       &   $6.39 \pm 1.92$         &     $3.81\pm 1.30$      \\
                             & \cellcolor[HTML]{EFEFEF}A4C           &  \cellcolor[HTML]{EFEFEF}$0.7316\pm 0.0769$           &  \cellcolor[HTML]{EFEFEF}$0.8448\pm 0.0619$         &     \cellcolor[HTML]{EFEFEF}$5.79 \pm 1.96$       &  \cellcolor[HTML]{EFEFEF}$4.23 \pm 1.48$          \\ \hline
\multirow{2}{*}{AC-DLIR}            & A2C           &   $0.8079\pm 0.0526$          &   $0.8649\pm 0.0581$        &    $5.40 \pm 1.93$        &   $3.49 \pm 1.17$        \\
                             & \cellcolor[HTML]{EFEFEF}A4C           &  \cellcolor[HTML]{EFEFEF}$0.7925 \pm 0.0580$          &    \cellcolor[HTML]{EFEFEF}$0.8832 \pm 0.0517$       &   \cellcolor[HTML]{EFEFEF}$4.77 \pm 1.73$         &   \cellcolor[HTML]{EFEFEF}$3.41\pm 1.27$        \\ \hline
\multirow{2}{*}{DdC-DLIR}            & A2C           &  $0.7704\pm 0.0674$           &    $0.8701\pm 0.0688$        &    $5.80 \pm 2.56$        &   $4.12\pm 1.86$        \\
                             & \cellcolor[HTML]{EFEFEF}A4C           &  \cellcolor[HTML]{EFEFEF}$0.7525\pm 0.0681$           &  \cellcolor[HTML]{EFEFEF}$0.8699\pm 0.0559$         &  \cellcolor[HTML]{EFEFEF}$6.23 \pm 2.24$           &  \cellcolor[HTML]{EFEFEF}$4.82\pm 2.03$         \\ \hline

\multirow{2}{*}{DdC-AC-DLIR}           & A2C           &   $\mathbf{0.8247 \pm 0.0455}$          &    $0.9225\pm 0.0409$       &   $4.51 \pm 1.90$         &  $2.78 \pm 1.37$         \\
                             & \cellcolor[HTML]{EFEFEF}A4C           &  \cellcolor[HTML]{EFEFEF}\dunderline{1pt}{$0.8175 \pm 0.0428$}           &  \cellcolor[HTML]{EFEFEF}$0.9364\pm 0.0303$         &   \cellcolor[HTML]{EFEFEF}$6.53\pm 1.70$         &  \cellcolor[HTML]{EFEFEF}$2.49 \pm 1.15$         \\ \hline
\multirow{2}{*}{MS-DdC-AC-DLIR}           & A2C           &    $0.8229\pm 0.0469$         &    $0.9249 \pm 0.0423$       &   $4.75\pm 2.04$         &  $2.61\pm 1.28$         \\
                             & \cellcolor[HTML]{EFEFEF}A4C           & \cellcolor[HTML]{EFEFEF}$0.8159\pm0.0398$            & \cellcolor[HTML]{EFEFEF}$0.9349\pm 0.0278$          &  \cellcolor[HTML]{EFEFEF}$4.56 \pm 1.78$          &  \cellcolor[HTML]{EFEFEF}$2.47 \pm 1.01$         \\ \hline  

\multirow{2}{*}{Aug-MS-DdC-AC-DLIR}           & A2C           &   $0.8200\pm 0.0467$          &    $\mathbf{0.9261\pm 0.0355}$       &     $\mathbf{4.10\pm 1.69}$       & $\mathbf{2.53\pm 0.97}$         \\
                             & \cellcolor[HTML]{EFEFEF}A4C           &  \cellcolor[HTML]{EFEFEF}$0.8095\pm 0.0390$           &  \cellcolor[HTML]{EFEFEF}\dunderline{1pt}{$0.9360\pm 0.0270$ }        &   \cellcolor[HTML]{EFEFEF}\dunderline{1pt}{$4.12\pm 1.82$}         &   \cellcolor[HTML]{EFEFEF}\dunderline{1pt}{$2.37\pm 0.99$}        \\ \hline    
\end{tabular}
\end{table*}}

{\renewcommand{\arraystretch}{1.5} 
\begin{table*}[!ht]
\fontsize{7pt}{7pt}\selectfont
\caption{Additional temporal image registration results for adult and fetal echo images, demonstrating the metrics for MYO and LV, where Table~\ref{tab:TR_reg} shows the average metrics of background, MYO, and LV. All the metrics are estimated using fixed images (and masks) and warped moving images (and masks). Underlined, double-underlined, and bold fonts denote the best-performing metrics for the A2C view of adult echo, the A4C view of adult echo, and the A4C view of fetal echo, respectively.}
\label{tab:Additional_TR}
\begin{tabular}{lccccc}
\hline
\multicolumn{2}{l}{\multirow{2}{*}{\textbf{Methods}}} & \multicolumn{2}{c}{\textbf{DSC ($\uparrow$)}} & \multicolumn{2}{c}{\textbf{HD $(mm)$ ($\downarrow$)}} \\ \cline{3-6} 
\multicolumn{2}{c}{}                         & MYO         & LV        & MYO        & LV        \\ \hline

\multirow{3}{*}{VanDLIR}           & Adult (A2C)           &  $0.8590\pm 0.0829$           &  $0.9287 \pm 0.0549$         &    $3.72 \pm 2.39$        &     $2.28\pm 1.40$      \\
                             & \cellcolor[HTML]{EFEFEF}Adult (A4C)           &     \cellcolor[HTML]{EFEFEF}$0.8455\pm 0.0883$        &   \cellcolor[HTML]{EFEFEF}$0.9243 \pm 0.0554$        &  \cellcolor[HTML]{EFEFEF}$4.13\pm 2.45$         & \cellcolor[HTML]{EFEFEF}$2.80\pm 1.73$         \\ 
                             & \cellcolor[HTML]{C0C0C0}Fetal (A4C)           &     \cellcolor[HTML]{C0C0C0}$0.8888 \pm 0.0658$        &   \cellcolor[HTML]{C0C0C0}$0.9273 \pm 0.0583$        &   \cellcolor[HTML]{C0C0C0}$1.56 \pm 0.90$         &   \cellcolor[HTML]{C0C0C0}$1.26 \pm0.77$        \\ \hline

\multirow{3}{*}{AC-DLIR}            & Adult (A2C)           &   $0.8717\pm 0.0689$          &   \dunderline{1pt}{$0.9539\pm0.0320$}        &   $3.64\pm 2.20$         &  \dunderline{1pt}{$1.61\pm 1.0$}         \\
                             & \cellcolor[HTML]{EFEFEF}Adult (A4C)           &  \cellcolor[HTML]{EFEFEF}$0.8505\pm 0.0844$            &   \cellcolor[HTML]{EFEFEF}$0.9446\pm 0.0441$        &  \cellcolor[HTML]{EFEFEF}$3.90\pm 2.29$          &  \cellcolor[HTML]{EFEFEF}$2.20 \pm 1.38$          \\ 
                             & \cellcolor[HTML]{C0C0C0}Fetal (A4C)           &     \cellcolor[HTML]{C0C0C0}$0.9011\pm 0.0570$        &    \cellcolor[HTML]{C0C0C0}$0.9319 \pm 0.0512$       &   \cellcolor[HTML]{C0C0C0}$1.68 \pm 0.93$         &  \cellcolor[HTML]{C0C0C0}$1.32 \pm 0.77$         \\ \hline

\multirow{3}{*}{DdC-AC-DLIR}           & Adult (A2C)           &    $0.8738\pm 0.0696$         &   $0.9523 \pm 0.0346$        &   $3.59\pm 2.18$        &  $1.69 \pm 1.09$         \\
                             & \cellcolor[HTML]{EFEFEF}Adult (A4C)            &   \cellcolor[HTML]{EFEFEF}$0.8477 \pm 0.0854$          & \cellcolor[HTML]{EFEFEF}$0.9445 \pm 0.0453$          &  \cellcolor[HTML]{EFEFEF}$3.83 \pm 2.23$         &  \cellcolor[HTML]{EFEFEF}$1.95 \pm 1.33$         \\ 
                             
                             & \cellcolor[HTML]{C0C0C0}Fetal (A4C)           &     \cellcolor[HTML]{C0C0C0}$0.9106 \pm 0.0488$        &   \cellcolor[HTML]{C0C0C0}$0.9411 \pm 0.0383$        &   \cellcolor[HTML]{C0C0C0}$1.57 \pm 0.92$         &  \cellcolor[HTML]{C0C0C0}$1.22 \pm 0.72$        \\ \hline

\multirow{3}{*}{MS-DdC-AC-DLIR}           & Adult (A2C)           &   \dunderline{1pt}{$0.8814 \pm 0.0535$}          &    $0.9529 \pm 0.0241$       &   \dunderline{1pt}{$3.15 \pm 1.78$}         &  $1.77 \pm 0.86$         \\
                             & \cellcolor[HTML]{EFEFEF}Adult (A4C)           &    \cellcolor[HTML]{EFEFEF}\underline{\underline{$0.8638 \pm 0.0584$}}         & \cellcolor[HTML]{EFEFEF}\underline{\underline{$0.9507 \pm 0.0307$}}           &   \cellcolor[HTML]{EFEFEF}\underline{\underline{$3.36 \pm 1.94$}}         &   \cellcolor[HTML]{EFEFEF}\underline{\underline{$1.74 \pm 0.90$}}        \\ 

                             & \cellcolor[HTML]{C0C0C0}Fetal (A4C)           & \cellcolor[HTML]{C0C0C0}$\mathbf{0.9155\pm 0.0457}$            &   \cellcolor[HTML]{C0C0C0}$\mathbf{0.9463\pm 0.0332}$        &   \cellcolor[HTML]{C0C0C0}$\mathbf{1.42\pm 0.84}$         &  \cellcolor[HTML]{C0C0C0}$\mathbf{1.14 \pm 0.68}$         \\ \hline

\end{tabular}
\end{table*}}

\begin{figure}[!ht]
\caption{Training and testing histories of the VAE to learn the latent vectors ($\mathcal{Z}$). This figure demonstrates that the learned $\mathcal{Z}$ can successfully represent the anatomical MYO and LV topology of the fetal and adult datasets, as the DSCs between the inputted and reconstructed masks are high (see Fig.~\ref{fig:VAE_masks}).}
\label{fig:VAE_DSC}
\centering
\includegraphics[width=0.9\textwidth]{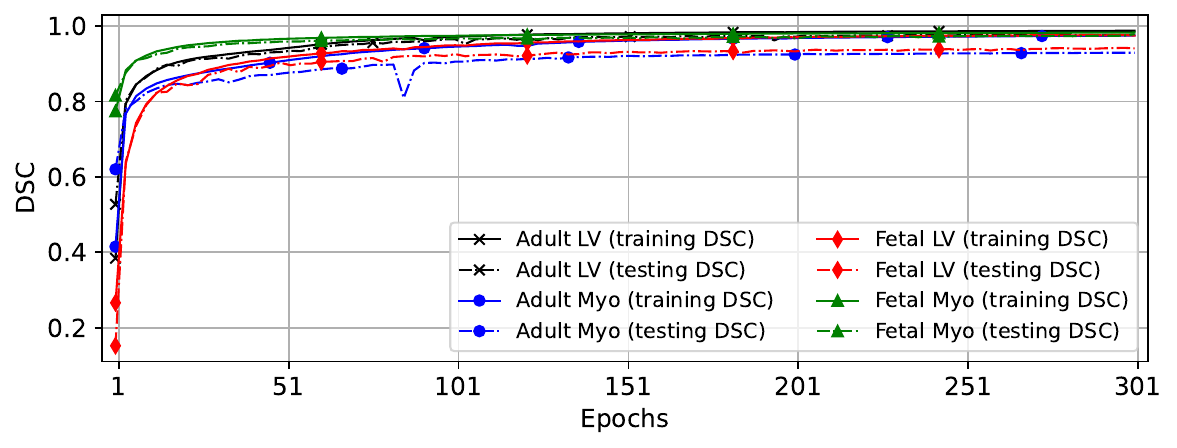}
\end{figure}

\begin{figure}[!ht]
\caption{Qualitative results of the VAE for both adult (two top rows) and fetal (two bottom rows) datasets. The input masks were encoded to the latent vector ($\mathcal{Z}$), and the reconstructed masks were decoded from the $\mathcal{Z}$. This figure illustrates that the inputted and reconstructed masks follow a similar anatomical topology, ensuring the latent vector's quality to represent the global attributes of MYO and LV.}
\label{fig:VAE_masks}
\centering
\includegraphics[width=0.95\textwidth]{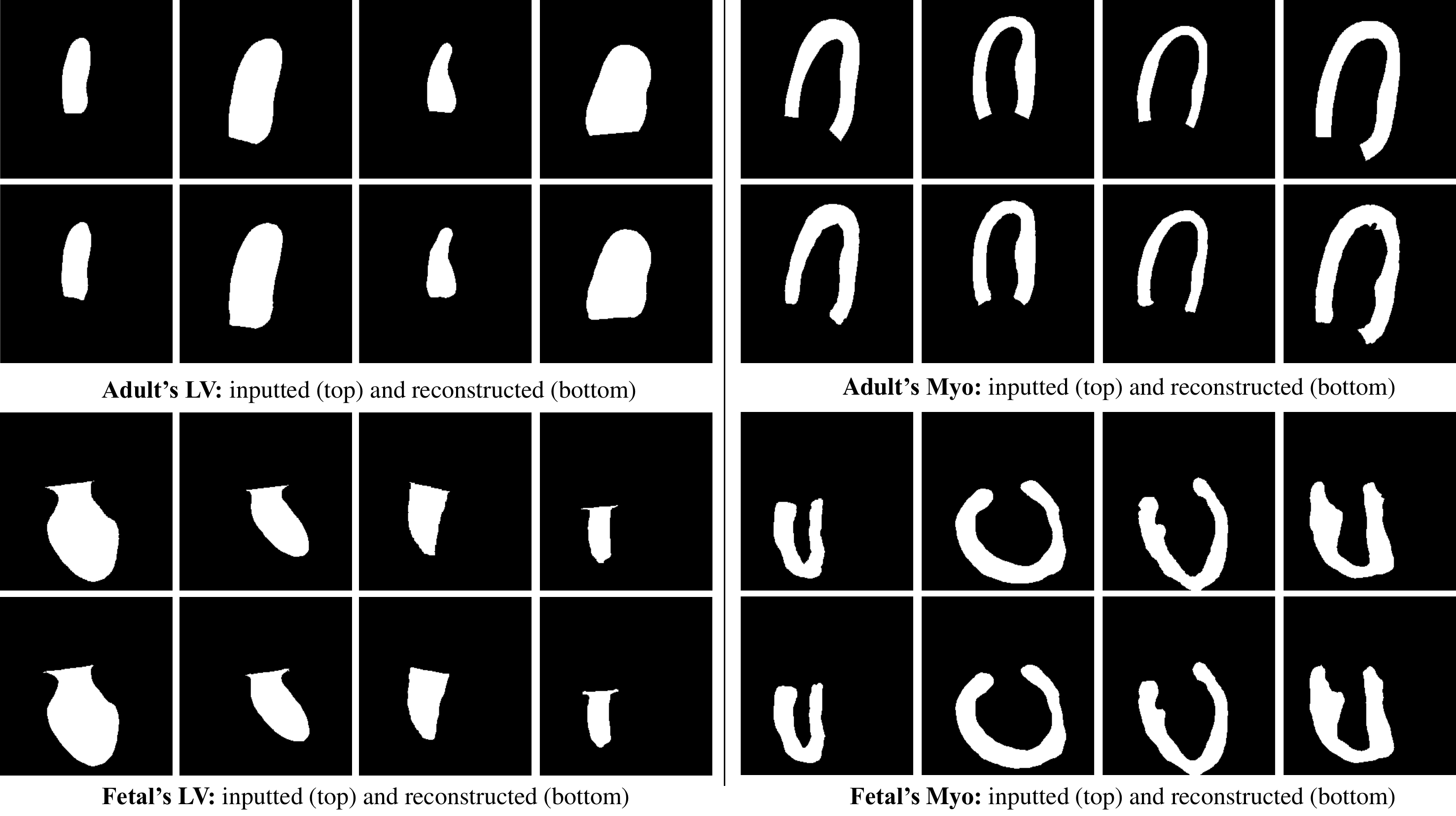}
\end{figure}

\begin{figure}[!ht]
\centering
\includegraphics[width=0.85\textwidth]{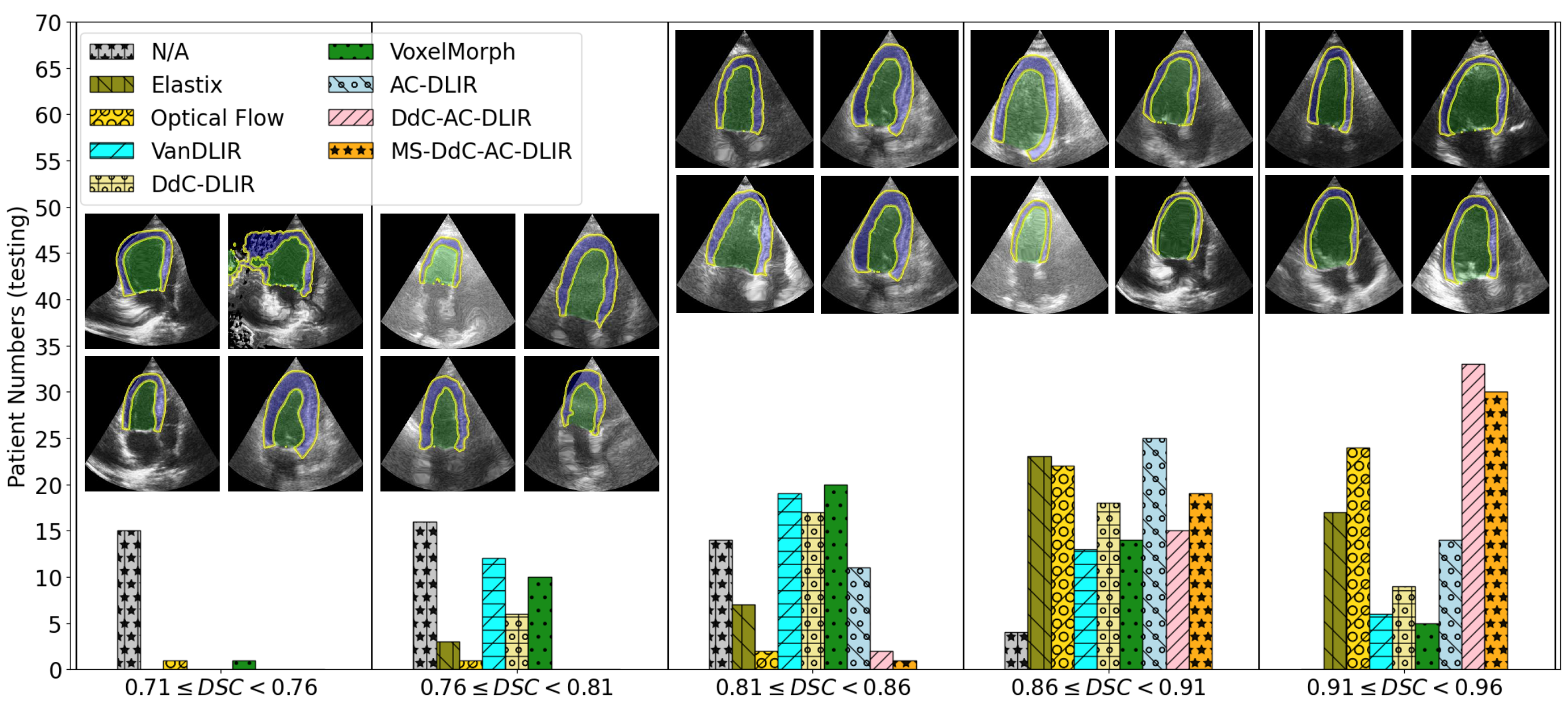}
\caption{Five different groups of obtained DSCs, demonstrating the number of testing patients for A4C of CAMUS in each DSC group. Similar results for A2C of CAMUS are in Fig.~\ref{fig:Hist_pot_A2C}.}
\label{fig:Hist_pot_A4C}
\end{figure}

\begin{figure*}[!ht]
    \centering
    \includegraphics[width=1.0\textwidth]{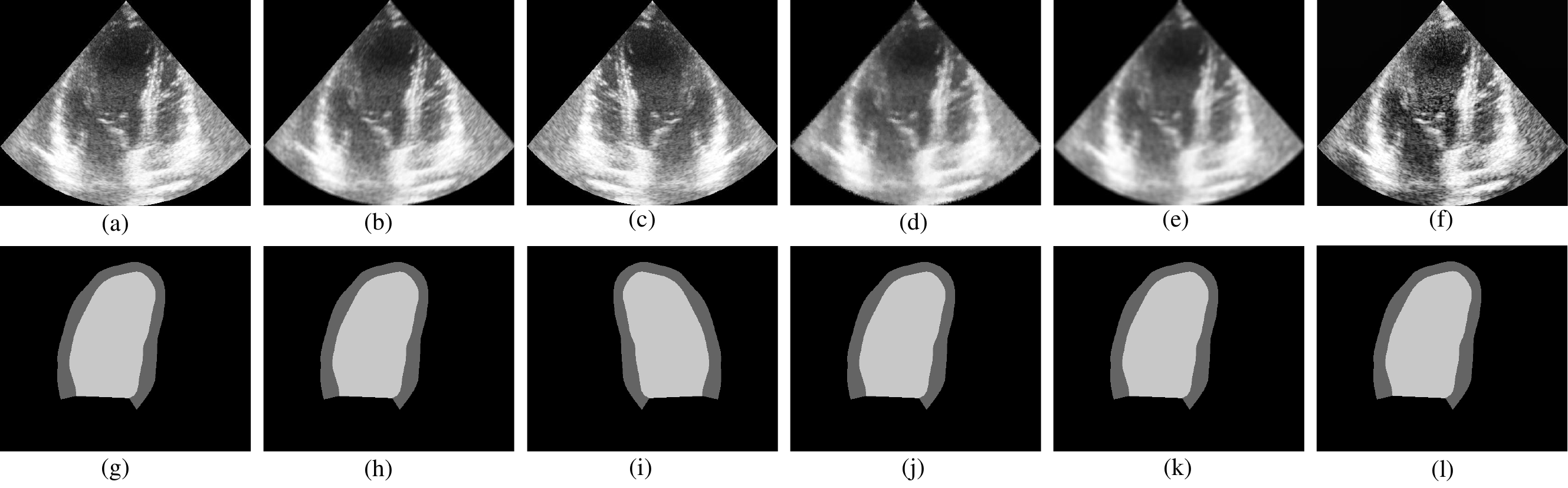} 
    \caption{Examples of image augmentations showing the intensity images with the corresponding masks, where (a,g): original pair, (b,h): motion blurred pair, (c,i): horizontal flipped pair, (d,j): Gaussian blurred pair, (e,k): defocused \citep{hendrycks2019benchmarking} pair, and (f,l): CLAHE \citep{yadav2014contrast} pair.}
    \label{fig:augmentations}
\end{figure*}

\begin{figure*}[!ht]
\caption{Sample of qualitative results of the warped image and mask using the proposed MS-DdC-AC-DLIR and Aug-MS-DdC-AC-DLIR for A2C (top two rows) and A4C (bottom two rows) views.}
\label{fig:MS_Aug}
\centering
\includegraphics[width=1.0\textwidth]{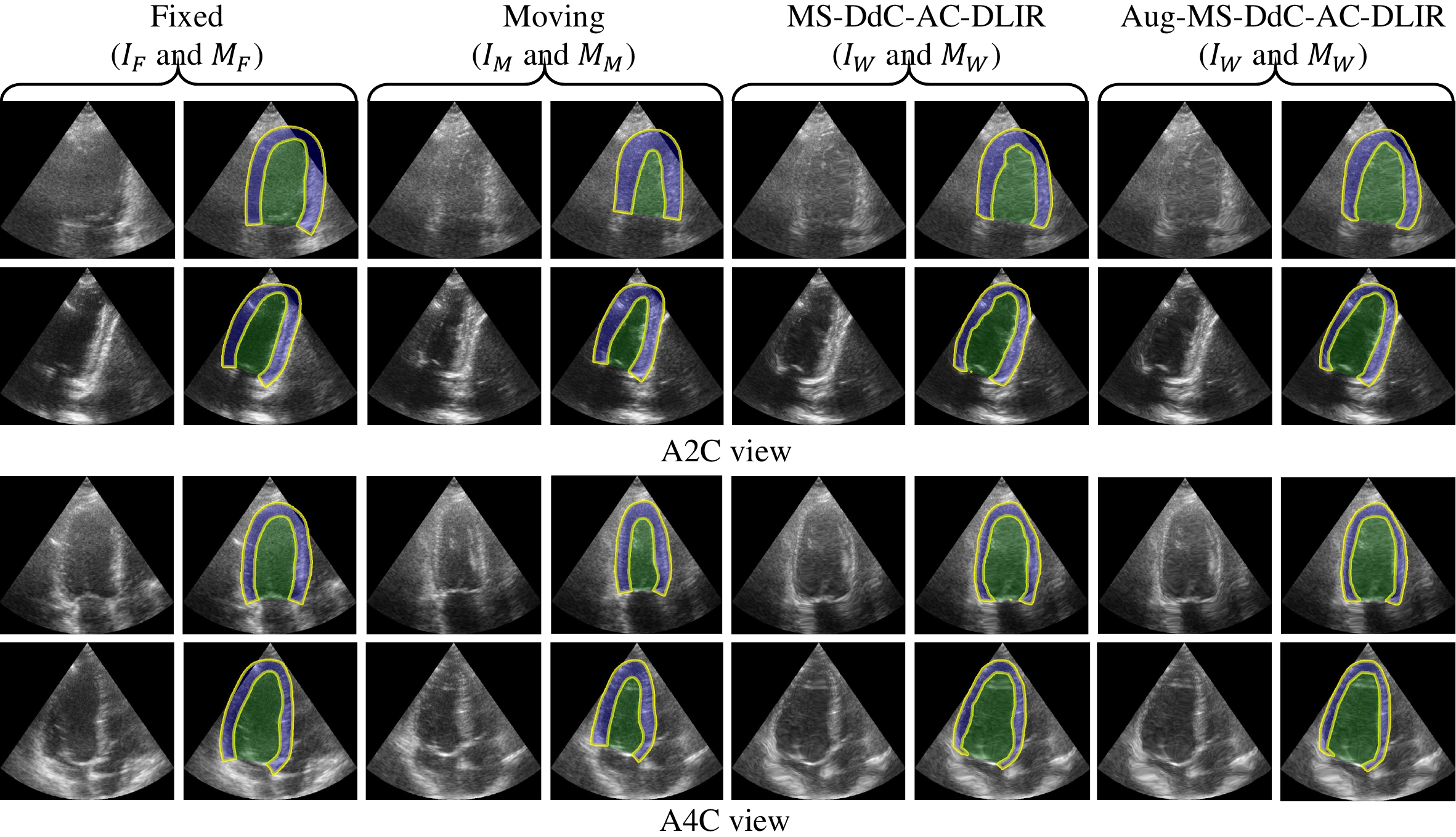}
\end{figure*}

\begin{figure*}[!ht]
\centering
\subfloat[For the first 25 patients (P1-P25) out of 50 testing patients (A4C view)]{\includegraphics[width=1.0\textwidth]{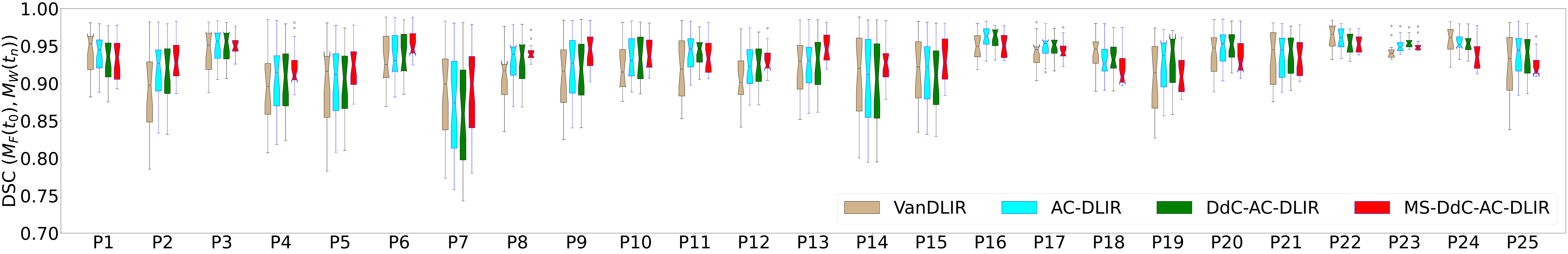}}\\
\subfloat[For the second 25 patients (P26-P50) out of 50 testing patients (A4C view)]{\includegraphics[width=1.0\textwidth]{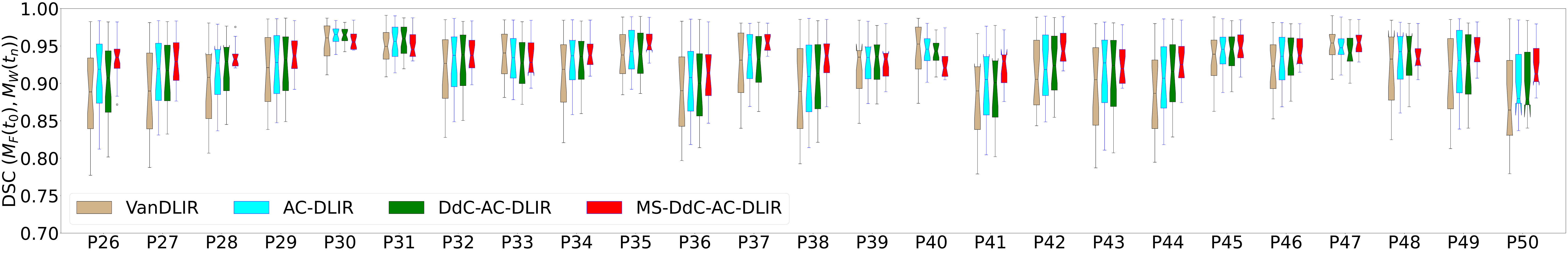}} \\
\subfloat[For the first 25 patients (P1-P25) out of 50 testing patients (A2C view)]{\includegraphics[width=1.0\textwidth]{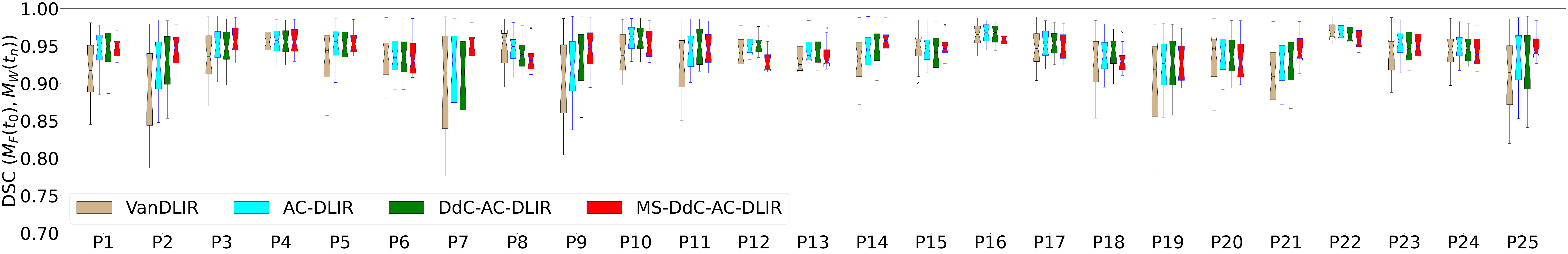}}\\
\subfloat[For the second 25 patients (P26-P50) out of 50 testing patients (A2C view)]{\includegraphics[width=1.0\textwidth]{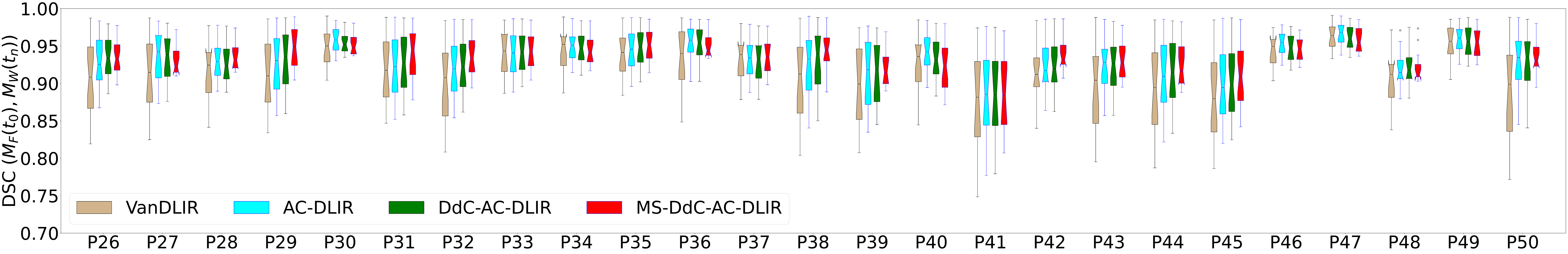}}
\caption{Demonstration of the non-overlapping benefits of data-driven and anatomical constraints in temporal echo image registration. The lower interquartile range in the box indicates better temporal consistency in the temporal echo registration. This figure was created using the CAMUS A2C and A4C views, where the first time point is set as a fixed sample, and other time points are warped to this fixed sample.}
\label{fig:DSC_Patients_A2C_A4C}
\end{figure*}

\end{document}